\documentclass[12pt]{elsarticle}

\usepackage{epstopdf}
\usepackage{subfigure}

\usepackage[utf8x]{inputenc}
\usepackage{url}
\usepackage{paralist}
\usepackage{csquotes}
\usepackage{textcomp}
\usepackage{longtable}

\usepackage{natbib,hyperref}
\journal{arXiv}

\newcommand{\eg}{e.g., }
\newcommand{\ie}{i.e., }

\begin{document}

\begin{frontmatter}

\title{Survey on Additive Manufacturing, Cloud 3D Printing and Services}

\author{Felix W. Baumann}
\ead{baumann@informatik.uni-stuttgart.de}
\author{Dieter Roller}
\ead{roller@informatik.uni-stuttgart.de}
\address{University of Stuttgart, Stuttgart, Germany}

\begin{abstract}
	Cloud Manufacturing (CM) is the concept of using manufacturing resources
	in a service oriented way over the Internet. Recent developments in
	Additive Manufacturing (AM) are making it possible to utilise resources
	ad-hoc as replacement for traditional manufacturing resources in
	case of spontaneous problems in the established manufacturing processes.
	In order to be of use in these scenarios the AM resources must adhere
	to a strict principle of transparency and service composition in adherence
	to the Cloud Computing (CC) paradigm. With this review we provide an
	overview over CM, AM and relevant domains as well as present the historical
	development of scientific research in these fields, starting from 2002. Part of
	this work is also a meta-review on the domain to further detail its development
	and structure.

\end{abstract}


\end{frontmatter}


\section{Introduction}
\label{sec:introduction}
	Cloud Manufacturing (here CM, in other works also CMfg) as a concept is
	not new and has been executed in enterprises for many
	years~\citep{Wu14}, under different terms,
	\eg Grid Manufacturing~\citep{Chen04} or
	Agile Manufacturing~\citep{Sanchez01}.

	The decision to have a globally distributed and with many
	contractors or partners interconnected production process and related
	supply chains is a luxurious one. Large global corporations and competitions
	makes \enquote{expensive} local production nearly impossible.

	CM is based on a strict service orientation of its constituent production
	resources and capabilities. 

	Manufacturing resources become compartmentalised and connected and worked
	with as service entities, that can be rented, swapped, expanded,
	dismantled or scaled up or down just by the use of software. This almost
	instantaneous and flexible model of resource usage is what made the
	Cloud Computing (CC) paradigm very successful for a number of companies.
	Here computing resources and data storage are all virtual, living in
	large data-centres around the globe, with the user only interfacing these
	resources through well-defined APIs (Application programming interface)
	and paying for only the resources utilised -- apart from the costs
	inflicted by the cloud service providers due to their business model
	and the surcharged or otherwise calculated requirement for profit.

	With this work we contribute to the dissemination of knowledge in
	the domain of Additive Manufacturing (AM) and the concept of CM.
	Cloud Manufacturing can be seen as having two aspects and applications,
	where the first application is within an industrial environment for
	which CM provides a concept to embed, connect and utilise existing
	manufacturing resources, \eg 3D printers, drilling-, milling- and
	other machines, \ie cloud manufacturing is not limited to AM but
	AM can be utilised within a CM concept.
	The second application is for end-users that use AM/3D resources
	over the Internet in lieu acquiring their own 3D printer.
	The usage in this second application is highly service oriented and
	has mainly end-users or consumers as target clients.
	The consumers can profit from online-based services without the
	requirement of having to own neither hard- nor software resources
	for 3D printing.

	We motivate this work by an overview of the historical development of
	scientific research in these domains starting from 2002. With this
	we show that the scientific output within these fields has increased
	by an average of 41.3 percent annually to about 20000 publications per year
	(see Sect.~\ref{subsec:development_in_scientific_publications}).

	To develop a better understanding of the topic at hand we discuss various
	terminological definitions found in literature and standards.
	We give critique on the common definitions of AM and propose a simpler,
	yet more accurate definition.

	For the reader to further grasp these domains we study existing
	journals catering for these communities and discuss reach
	and inter-connections.

	Cloud Manufacturing relies on a service oriented concept of
	production services or capabilities. We extend an existing study on
	cloud printing services as we see such services
	as integral components for future CM systems.

	Cloud manufacturing has two aspects which are detailed in this work.
	First CM is a methodology that is used within industrial settings for
	the connection of existing resources to form either a virtual assembly
	line or to acquire access to manufacturing resources in a service
	oriented manner.
	Due to the globalisation of the industry, manufacturers face increased
	challenges with shorter time-to-markets, requirements for
	mass customisation (MC) and increased involvement of
	customers within the product development process. In order to stay
	or become competitive, companies must utilise their resources more
	efficiently and where not available they must
	acquire those resources in an efficient and transparent way.
	These resources must then be integrated into the existing process
	environment and released when no longer required.
	The concepts of cloud computing, where resources are available
	as services from providers that can be leased, rented, acquired or
	utilised in other ways are proposed to be applied to the domain of
	manufacturing.

	Resources like machines and software, as well as capabilities/abilities
	become transparently available as services that customers or end-users can
	use through the respective providers and pay for only the services they
	require momentarily. Most often, no contractual obligations between
	the provider and the consumer exists (but it can exist, especially
	for high-value or high-volume usage) which
	gives the consumer great flexibility at the expense of possible
	unavailability of resources by the provider.

	In the end-user segment, or the consumer aspect of CM the user is
	interested in using AM resources like 3D printers through a web-based
	interface in order to be able to have objects produced that are designed
	to be 3D printed without the necessity to purchase and
	own a 3D printer themselves. The user commonly uses such services in
	a similar fashion that they would use a (online) photography 
	lab / printing service.
	The users' experience and knowledge of AM and 3D printing can
	vary significantly.

	Albeit these two aspects seem to be far apart, the commonality
	between them is, that the service operator must provide AM resources
	in both cases in a transparent and usable manner.
	Resources must be provided with clear definitions of the interface
	to the service, \ie the data consumed by the service and data rendered
	by the service. The description and provisioning of the service must
	be hardware agnostic as the consumer must be able to select the
	resources required, \eg have an object manufactured either on a FDM
	(Fused Deposition Modeling, also Fused Filament Fabrication FFF) machine
	or and SLA (Stereolithography) machine without the necessity to alter the underlying
	data and models but by selection.

	This work is structured as follows:
	Section~\ref{subsec:research_objective} provides information of the objective
	we accomplish with this review. Section~\ref{subsubsec:methodology}
	presents the research methodology applied for this work.
	In section~\ref{subsubsec:sources} we disseminate the sources that were used
	to gather information for this work.
	Section~\ref{subsec:development_in_scientific_publications} provides
	a dissemination of the scientific research in these fields with a
	discussion on its historical development.
	Chapter~\ref{sec:definition_and_terminology} contains sections on key
	terminology and their definition throughout literature and standards.
	We present these terms as well as synonyms and provide an alternative
	definition as a proposal.
	The Chapter~\ref{sec:journals_related_to_the_subject} is an exhaustive
	collection of scientific journals relevant to the domains discussed in
	this work. We provide an insight in their interconnectedness and their
	structure.
	Chapter~\ref{sec:reviews_on_the_subject} provides a meta-review on
	the subject for the reader to get a further reaching understanding of
	the subject and its relevant components.

	In Chapter~\ref{subsec:stakeholder_distinction} we discuss the audience
	or target group for CM 	and 3D printing related cloud services.
	Chapter~\ref{sec:3d_printing_services} extends the study by
	Rayna and Striukova~\citep{Rayna16} due to the
	importance of 3D printing related cloud services for the topic at hand.
	Section~\ref{sec:review} provides the information on the concepts,
	terminology, methods relevant to the subject as they are
	disseminated in literature.
	We conclude this work with a summary in Chapter~\ref{sec:summary}.

\begin{figure}
	\begin{center}

			\includegraphics[width=\textwidth]{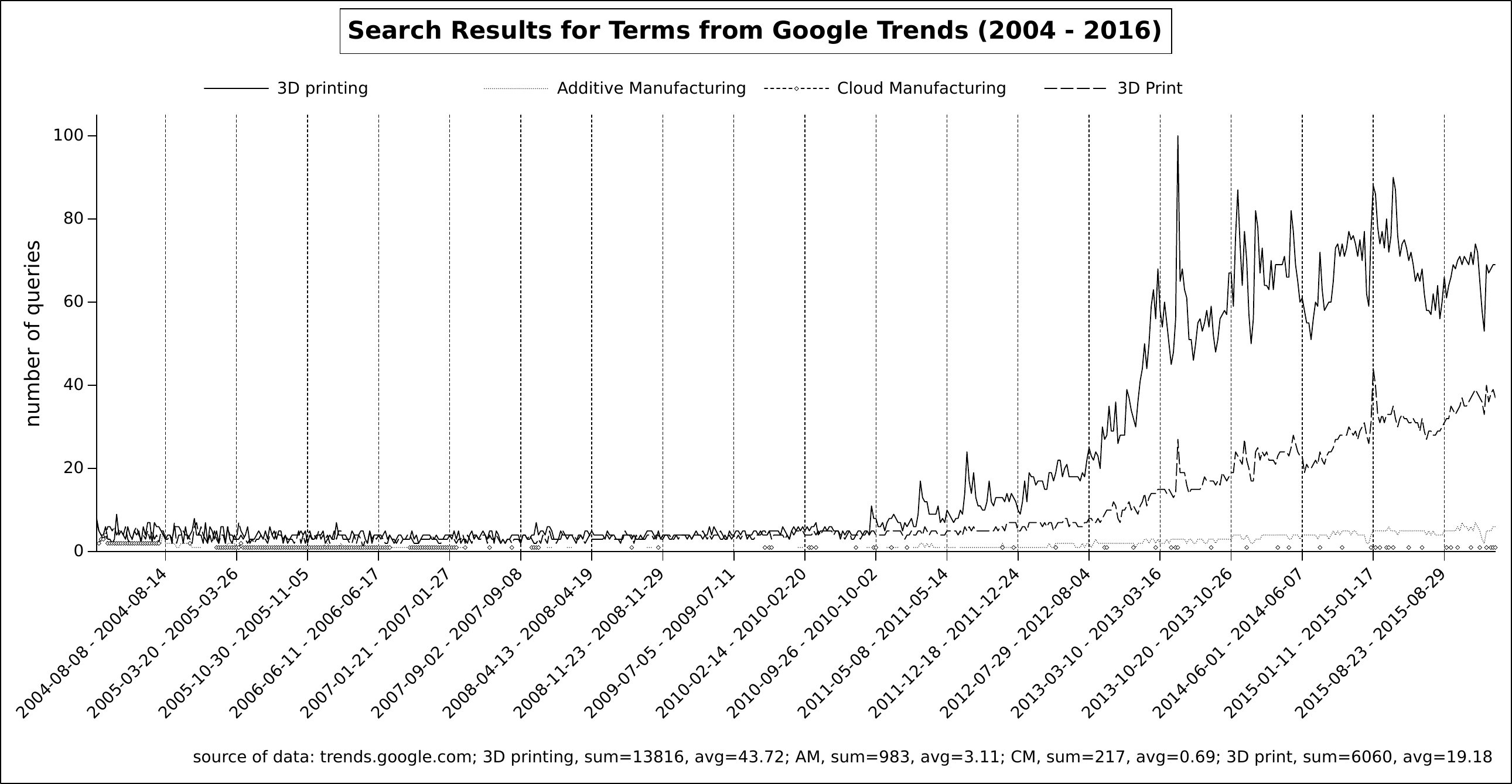}
			\caption{Queries for 3D Printing, AM, CM and 3D Print on google.com for 2004--2016 from trends.google.com}
			\label{fig:google_trends}
	\end{center}
\end{figure}

\subsection{Research Objective}
\label{subsec:research_objective}
	This review is performed to establish an overview on the concept and
	implementation of CM and the utilisation of Additive Manufacturing (AM)
	therein. For the understanding it is required to become familiar with 
	the various	definitions available and the problems arising from inconsistent
	usage of terminology. For this we compile differing definitions
	on key terminology. 

	With this work we aim to present an overview over the topic of CM,
	and its current research findings.
	We furthermore present a summary overview over existing
	online and cloud based 3D printing services that can either be regarded as
	implementations of CM or be utilised in CM scenarios. This
	part is to extend the knowledge on relevant online services and their
	orientation towards numerous services.
	With the presentation of the identified journals that cater for
	AM, DM, RP, RM and 3D printing research we want to provide other
	researchers with insight into possible publication venues and a
	starting point for the identification of relevant sources for their own
	work. The review work of this article has the objective to identify
	relevant literature and summarise the key and essential findings thereof.
	The review also is intended to provide a high level overview on
	identified research needs that are considered essential for the evolution
	of AM and CM.

	\subsubsection{Methodology}
	\label{subsubsec:methodology}
		The first part of this review is the analysis of other reviews in
		order to establish a foundation of the existing works and to have a
		baseline for the analysis of the journals catering to this domain.
		
		The journals are identified and presented in order to help researchers
		in finding a suitable publication venue and to present the recent
		development in this area. The journals are identified by literature
		research, web searching (see Sect. \ref{subsubsec:sources}), and as a
		result of the review analysis.

		This review identified its sources by web search for each of
		the identified topics depicted in the concept map
		(See Sect.~\ref{subsec:topological_map}), where the first 30 results
		from the search engines (see Sect.~\ref{subsubsec:sources}) each are
		scanned first by title, then by their abstract. For the creation
		of the topological map an iterative process is applied. The process
		starts with the analysis of the following
		works~\citep{Wu13c, Wang13d, Wu14, Wu13, He15} which we had prior
		knowledge of due to previous engagements in this research area.
		After the analysis a backward- and forward search is performed.

		The searches for the content of the review are
		sorted by relevance, according to the search engine
		operator. The articles are then analysed and its core concepts
		are presented in this work.
		
		The reviews for the meta-review are identified by a web search and
		data gathered during our review.
		
		For the compilation of the definitions an extraction process is employed
		where the identified literature for the review is basis for
		information extraction and dissemination. The compilation is
		expanded by literature and Internet research for the appropriate
		keywords and concepts.
		
		The extension to the study by Rayna and Striukova~\citep{Rayna16}
		is performed following the research methodology applied in the original
		work.

	\subsubsection{Sources}
	\label{subsubsec:sources}
		This review is based on scientific literature acquired through the
		respective publishers and searched for using the following
		search engines:
		\begin{itemize}
			\item Google Scholar\footnote{\url{https://scholar.google.com}}
			\item SemanticScholar\footnote{\url{https://semanticscholar.org}}
			\item dblp\footnote{\url{https://dblp.uni-trier.de}}
			\item Web of Science\footnote{\url{https://webofknowledge.com}}
			\item ProQuest\footnote{\url{https://proquest.com}}
		\end{itemize}
		Microsoft Academic
		Search\footnote{\url{http://academic.research.microsoft.com}} is not used
		for the search as the quality and number of results is unsatisfactory.
		Scopus\footnote{\url{https://www.scopus.com}} is not used for the
		research, as we have no subscription for it.
		The search engines differ in the handling of grouping
		and selection operators (\eg OR, +). For each search engine
		the appropriate operators where selected when not explicitly
		stated otherwise. As a search engine
		for scientific literature, Google Scholar,
		yields the most results but with a high degree
		of unrelated or otherwise unusable sources, like the
		Google search engine\footnote{\url{https://google.com}} itself.
		Furthermore, the search engine enforces strict usage rules
		thus hindering automated querying and usage.
		Results from patents and citations are excluded from the result set
		for Google Scholar.

		SemanticScholar offers a responsive interface, that allows for
		automated querying through JSON\footnote{JavaScript Object Notation}, to
		\enquote{millions} of
		articles\footnote{\url{https://www.semanticscholar.org/faq\#index-size}}
		from computer science - a statement that we can not verify as we have
		seen articles from other domains too.
		The dblp project indexes over
		3333333\footnote{News from 2016-05-03: \enquote{Today, dblp reached
		the wonderful "Schnapszahl" of 3,333,333 publications}}
		from computer science in a very high quality. Its interface allows
		for automated and scripted usage.
		Web of Science provides an index of a large number
		(over 56\footnote{A search for publications with its publication date
		between 1700 and 2100 yields 56998216 results}
		millions) of scientific works.
		The entries in the index are of high quality but the interface
		is rather restrictive. ProQuest also has a very restrictive and
		non-scriptable interface and contains over 54
		million~\footnote{A search for publications with its publication
		date after 1700 yields 54266680 results} entries in
		its corpus, among which are historical news articles and dissertations.
		The quality of the results is high. ProQuest and Web of Science
		are subscription based services.

	\subsection{Development in Scientific Publications}
	\label{subsec:development_in_scientific_publications}
		The significance and maturity of a research area is reflected in the
		number of publications available. We perform
		a keyword based analysis utilising the sources described in
		Section~\ref{subsubsec:sources}. The searches are performed with a
		number of relevant keywords (including various technologies and
		methods for AM) and a restriction of the time period starting from
		2002 to 2016. The queries are also restricted on the type of
		results with an exclusion to citations and patents, where applicable.
		For a study on the patents and the development of patent registrations
		for this domain we refer to Park et al.~\citep{Park16}.

		Caveat:\label{cav:clip} Searching on search engines for specific
		keywords like clip and lens in their abbreviated form will lead to
		a number of skewed results from works that are not significant
		for this body of work. For example the search for 
		\enquote{Additive Manufacturing} and LENS yield articles in
		the results that are either fabricating (optical) lenses
		using AM or are about lenses in lasers that are used in AM.
		In case the result sets are as large as in our case it is
		not feasible to remove those erroneous results and adjust the result
		set accordingly. We make the reader aware to only take the
		given numbers as an indication.

\begin{figure}
		\begin{center}
				\includegraphics[width=0.5\textwidth]{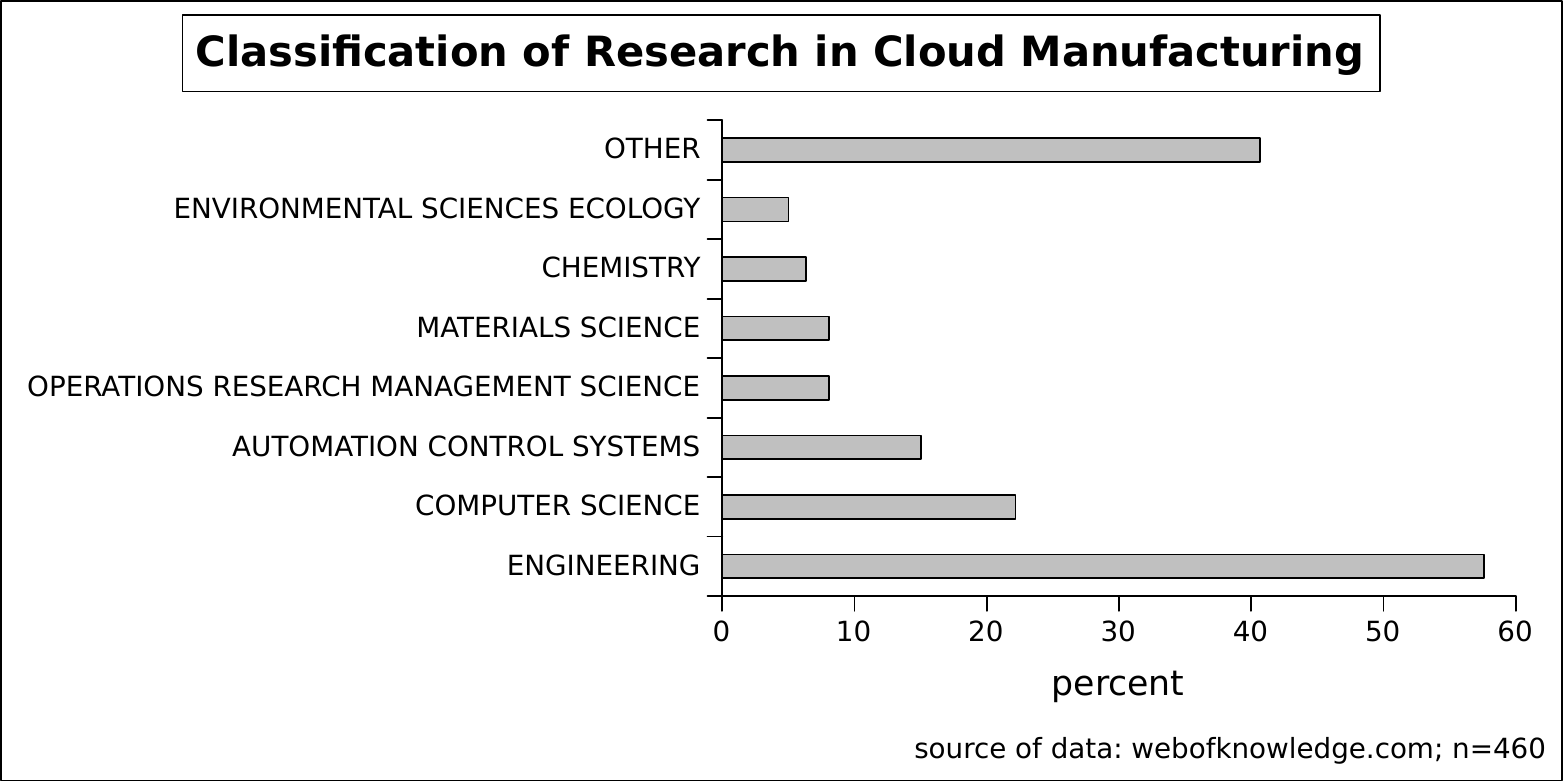}
				\caption{Classification of articles for Cloud Manufacturing; source of data: webofknowledge.com}
				\label{fig:wos_class_cm}
		\end{center}
	\end{figure}

\begin{figure}
		\begin{center}
				\includegraphics[width=0.5\textwidth]{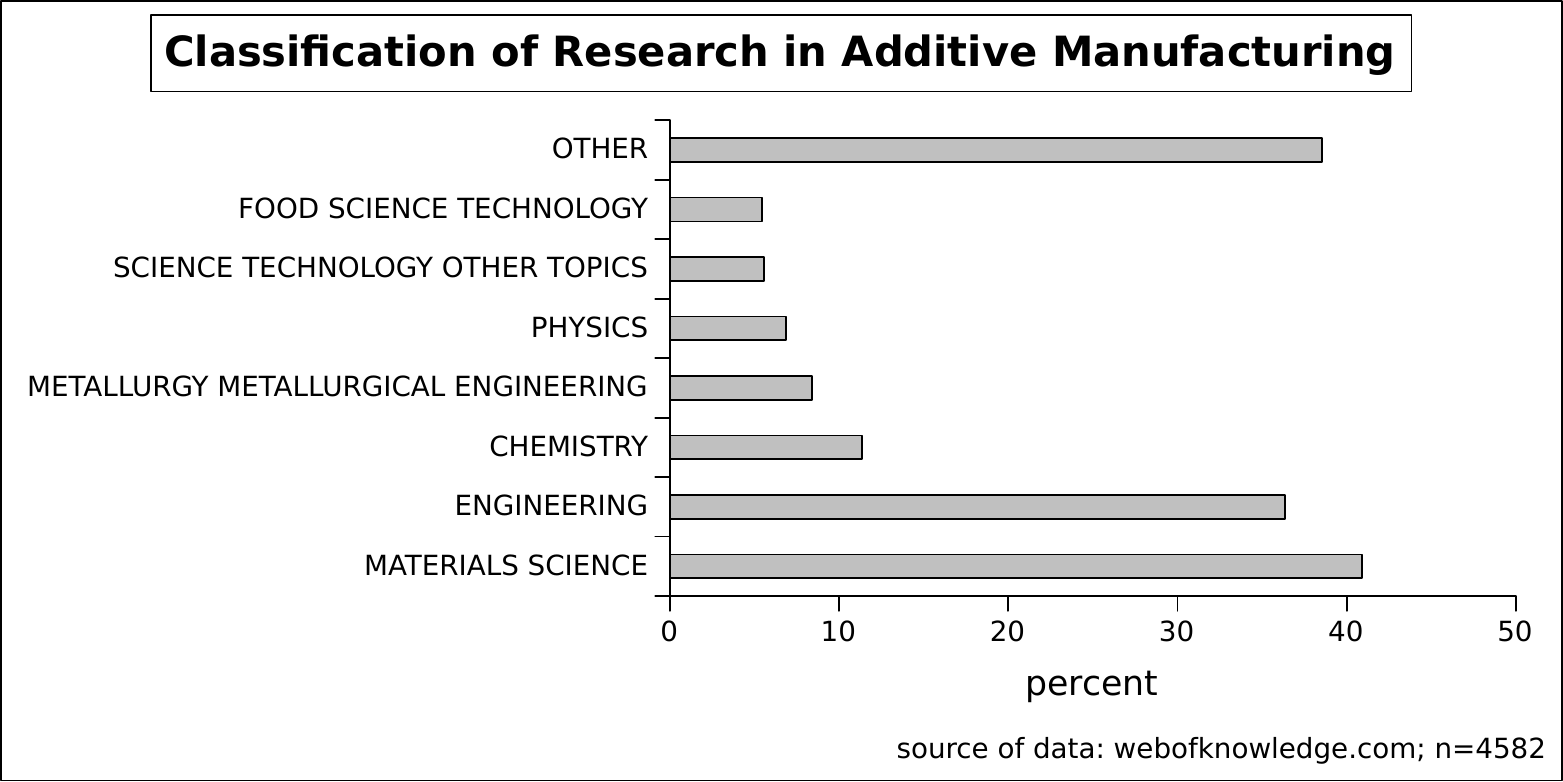}
				\caption{Classification of articles for Additive Manufacturing; source of data: webofknowledge.com}
				\label{fig:wos_class_am}
		\end{center}
	\end{figure}

\begin{figure}
		\begin{center}
				\includegraphics[width=0.5\textwidth]{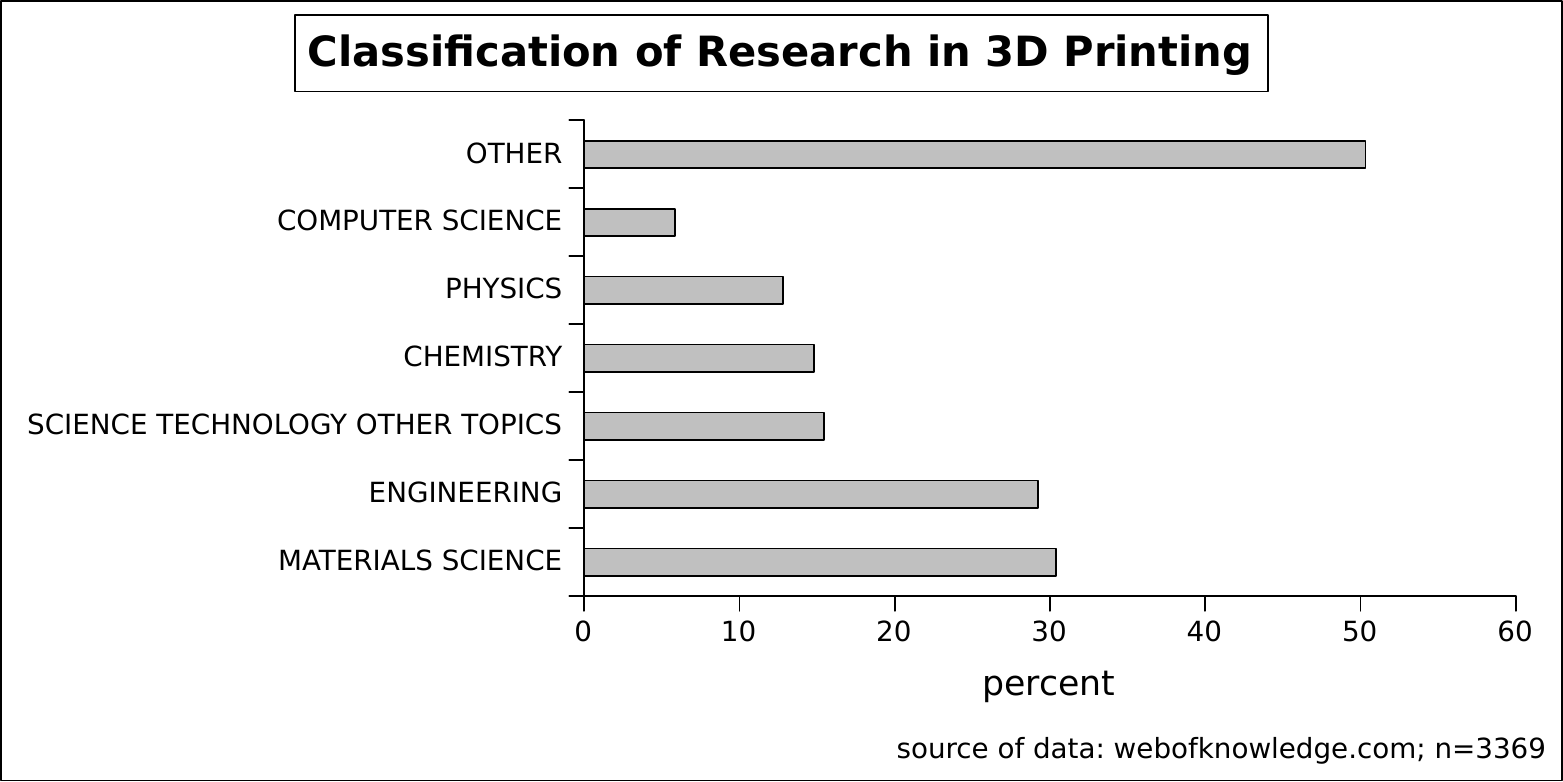}
				\caption{Classification of articles for 3D Printing; source of data: webofknowledge.com}
				\label{fig:wos_class_3d_printing}
		\end{center}
	\end{figure}

\begin{figure}
		\begin{center}
				\includegraphics[width=0.5\textwidth]{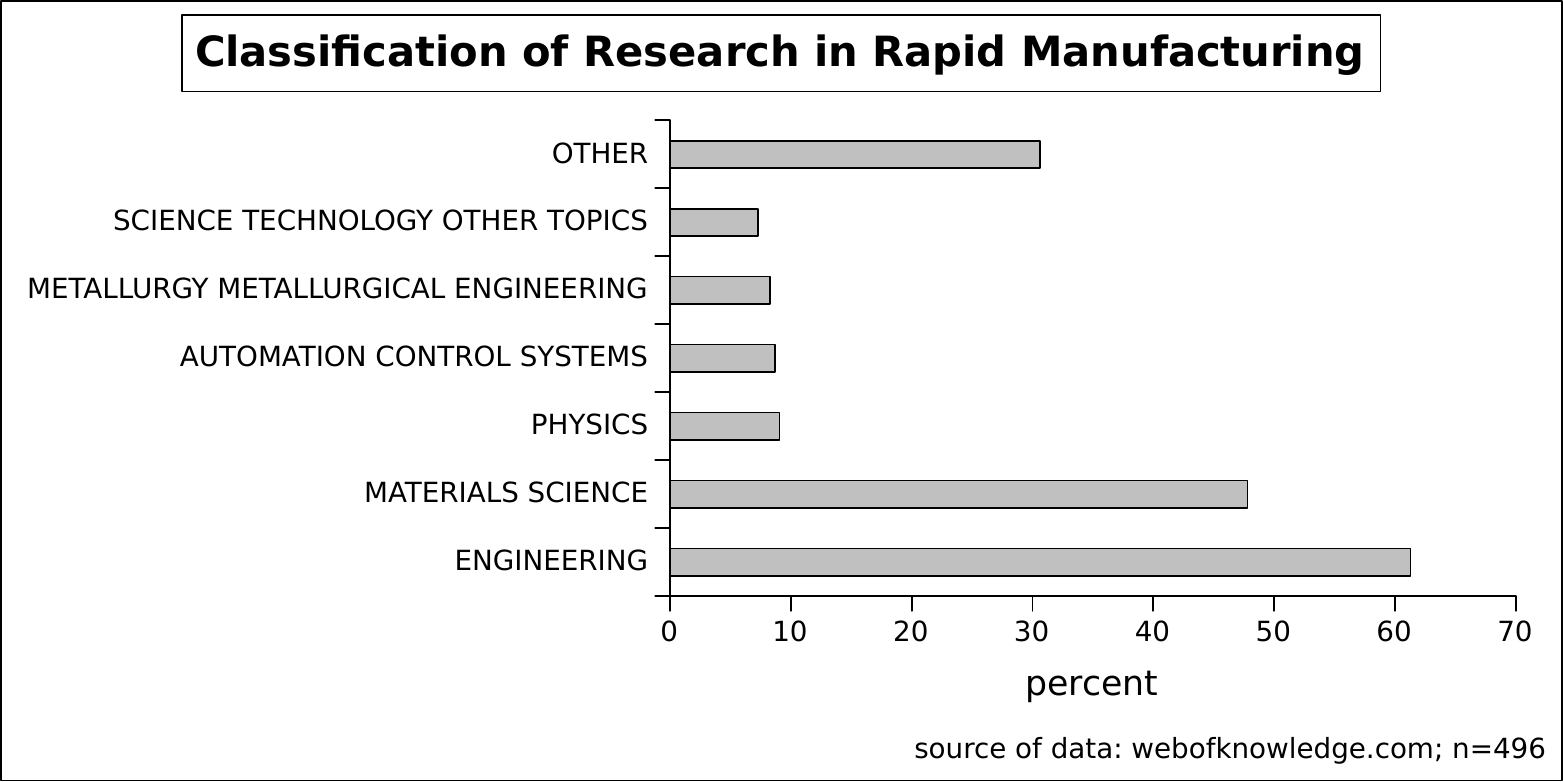}
				\caption{Classification of articles for Rapid Manufacturing; source of data: webofknowledge.com}
				\label{fig:wos_class_rm}
		\end{center}
	\end{figure}

\begin{figure}
		\begin{center}
				\includegraphics[width=0.5\textwidth]{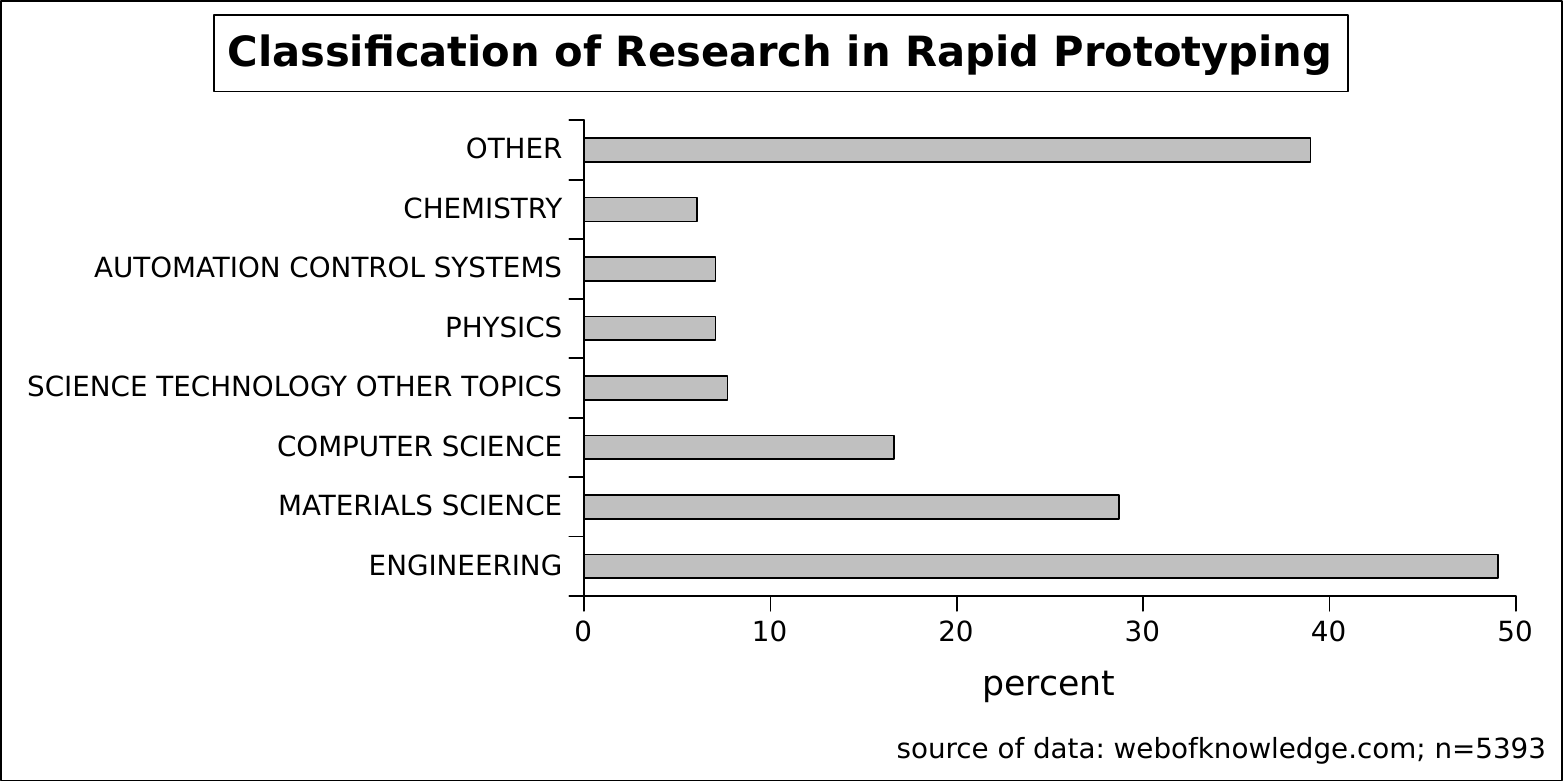}
				\caption{Classification of articles for Rapid Prototyping; source of data: webofknowledge.com}
				\label{fig:wos_class_rp}
		\end{center}
	\end{figure}

\begin{figure}
		\begin{center}
				\includegraphics[width=0.5\textwidth]{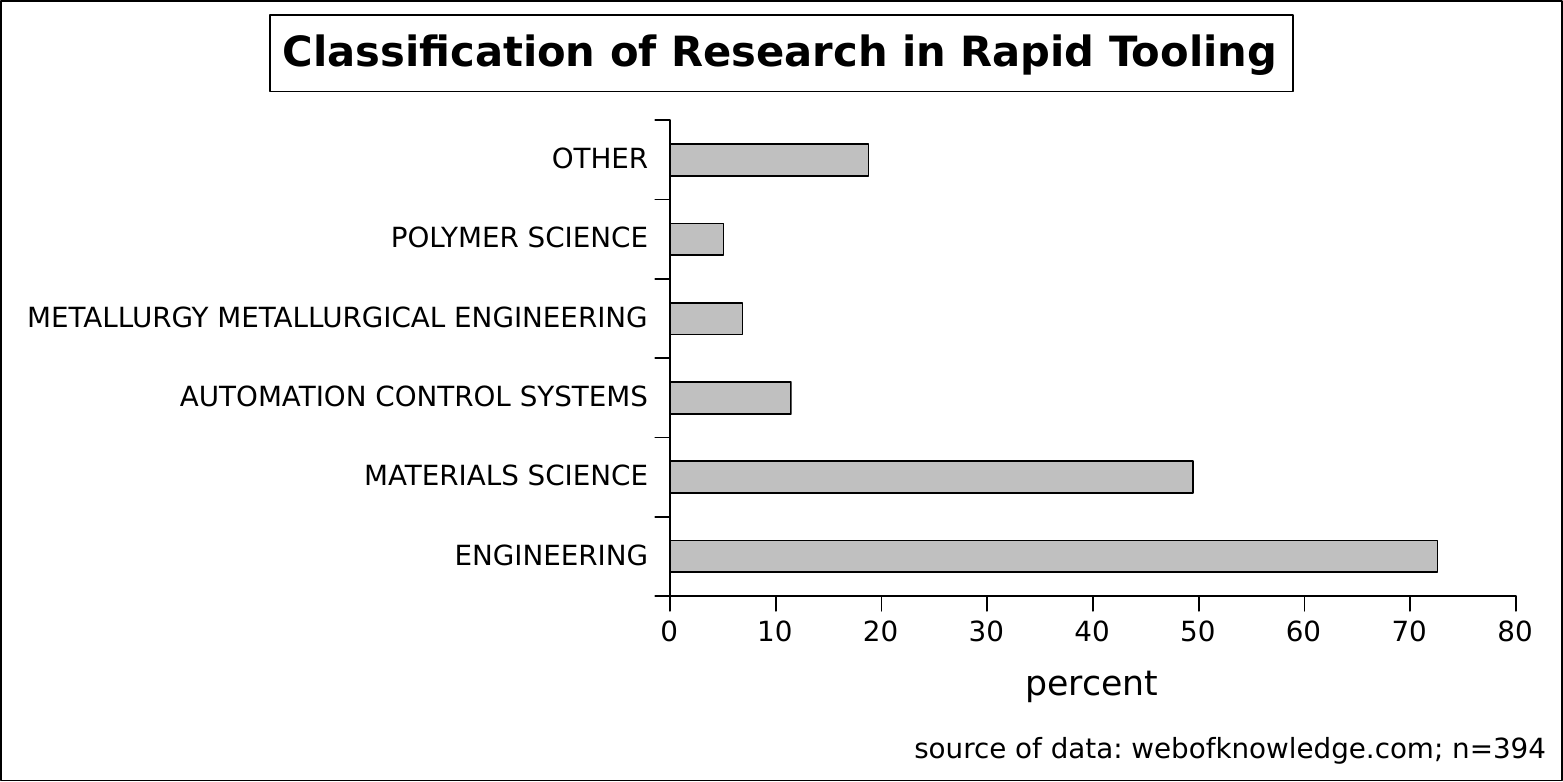}
				\caption{Classification of articles for Rapid Tooling; source of data: webofknowledge.com}
				\label{fig:wos_class_rt}
		\end{center}
	\end{figure}

		In Fig. \ref{fig:wos_class_cm} to Fig. \ref{fig:wos_class_rt} the
		classification of scientific articles according to Web of Science
		is shown. The classifications do not add up to 100 percent as
		the respective articles can be classified in more than one field.
		In the figures the number of results per search term is also listed.
		Domains with less than five percent aggregated classification are
		grouped together as \enquote{OTHER}.


\begin{figure}
		\begin{center}
				\includegraphics[width=\textwidth]{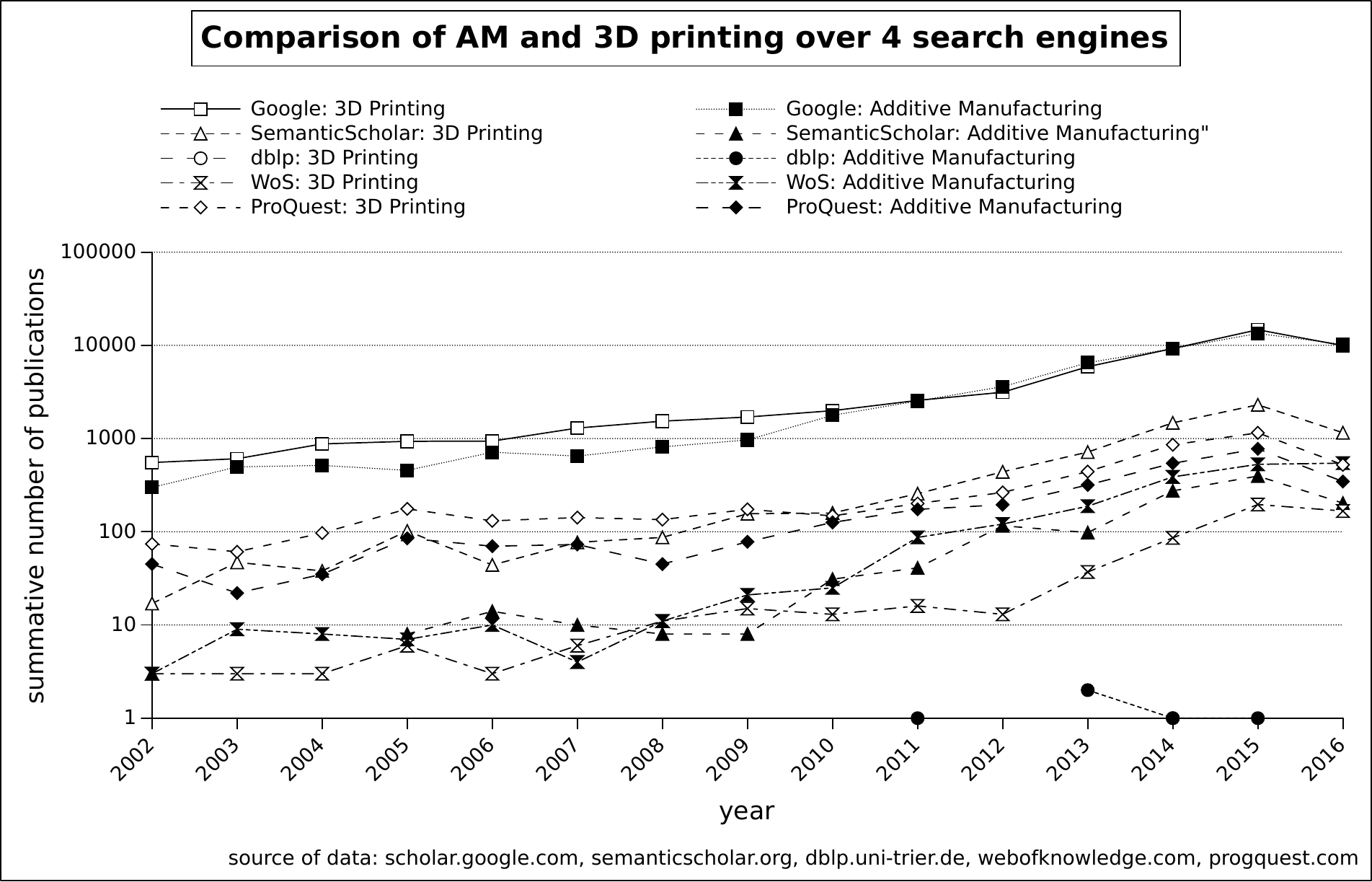}
				\caption{Comparison of AM and 3D Printing on selected search engines (2002--2016)}
				\label{fig:annual_results}
		\end{center}
	\end{figure}

\begin{figure}
		\begin{center}
				\includegraphics[width=\textwidth]{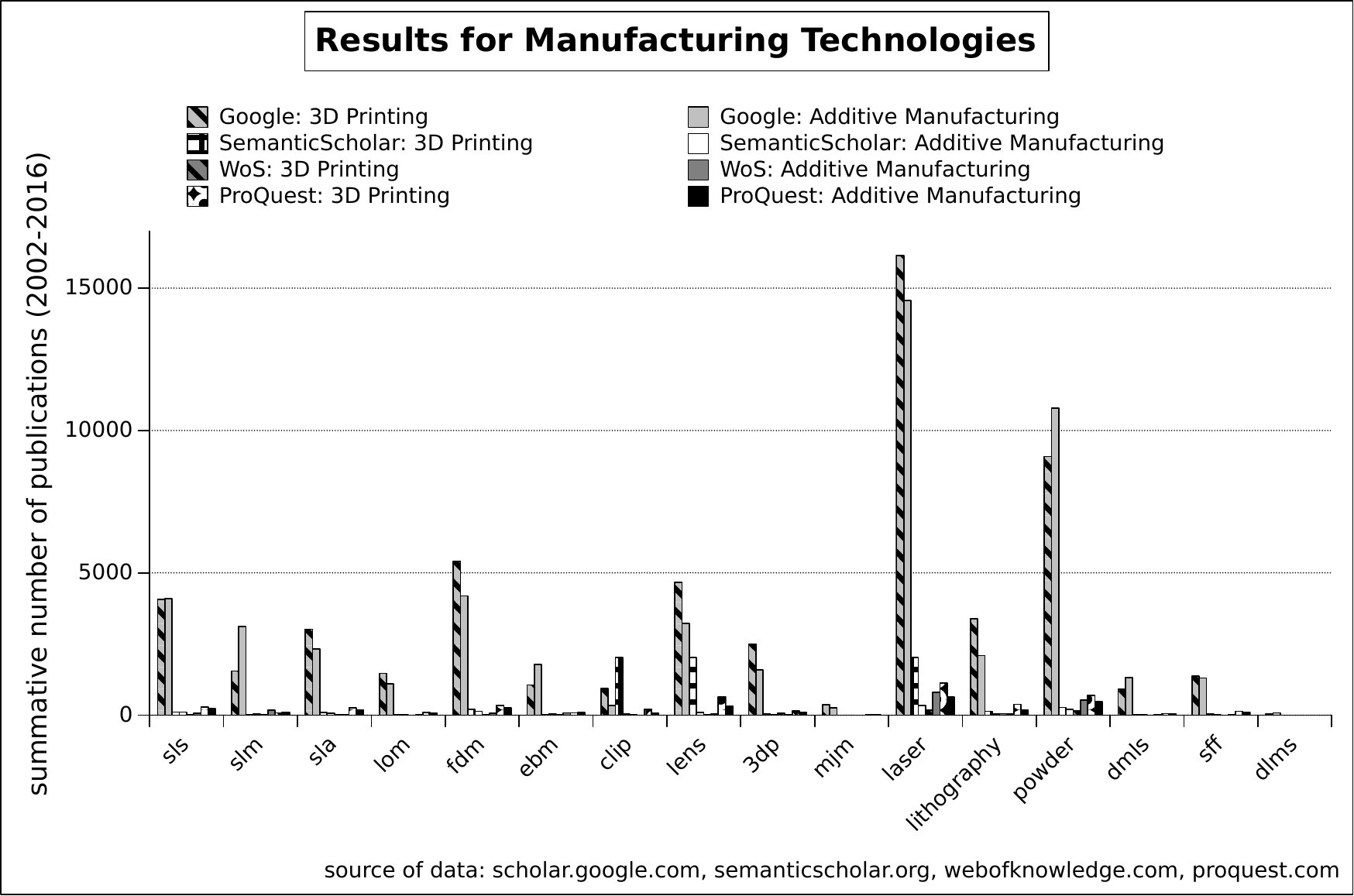}
				\caption{Comparison of 3D printing technologies on selected search engines (2002--2016)}
				\label{fig:technology_results}
		\end{center}
	\end{figure}

In Fig. \ref{fig:annual_results} the accumulated prevalence of the terms
3D printing versus Additive Manufacturing (AM) is displayed. For these
numbers queries are made for a combination of search terms and
restrictions on the time period. The scale of the Y-Axis is
logarithmic due to the large differences in the number of results per
search engine. The dblp database returned the lowest number of results
with results consistently less than 10. Google Scholar yielded the
largest number of results with the accumulated number of results for
the term AM gaining on the term 3D printing since 2009.
In Fig. \ref{fig:technology_results} the prevalence of certain AM or
3D Printing technologies is studied by the number of articles from
four different search engines for the respective combination of search terms.
The largest number of results are from Google Scholar for search term
combinations with \enquote{3D Printing}. Furthermore, a generalised
search is performed for the terminology \enquote{Laser, Lithography and Powder},
\eg summarising technologies like SLM (Selective Laser Melting),
SLS (Selective Laser Sintering), SLA, LOM (Laminated Object Manufacturing),
LENS (Laser Engineered Net Shaping) for
the term \enquote{Laser}.
The search for technologies like CLIP and LENS are problematic due to
the non-specificity of the terminology as described before
(See note \ref{cav:clip}).

\section{Definition and Terminology}
\label{sec:definition_and_terminology}
	In general the usage of the terminology within this field is very
	inconsistent. Commonly and colloquially the
	terms 3D printing and AM are used as synonyms.
	Analysing the prevalence of either of these terms we find that
	\textbf{3D printing} is slightly more prevalent for results of scientific
	literature with 68164 results for the sources described in
	Sect.~\ref{subsubsec:sources} during the period of 2002--2016. In the
	same period there are over 59506 results for the
	term \textbf{Additive Manufacturing}. SemanticScholar provided
	significantly more results (7072 over 1211) for	\textbf{3D printing}
	and Web of Science yielded almost four times the number of
	results for \textbf{Additive Manufacturing}	over \textbf{3D Printing}
	(1956 results to 578). There is also no clear trend in the usage
	of either terms. With this section we exemplify this situation and present
	common definitions throughout literature and standards. We furthermore add
	our point of view in the form of a critique at the end of the section.

\subsection{Additive Manufacturing and 3D Printing}
\label{subsec:additive_manufacturing_and_3d_printing}
	In this section we present established definitions for AM and related
	terminology as presented in literature and standards.

	\subsubsection{Definitions of Additive Manufacturing}
	\label{subsubsec:definitions_of_additive_manufacturing}
	AM is most often regarded as an umbrella term for technology and methods for the
	creation of objects from digital models from scratch. It is usually in contrast
	to subtractive and formative methods of manufacturing as defined in the
	standard~\citep{DIN_8580:03}. It is also commonly a synonym for 3D printing.

	Gibson et al.~\citep{Gibson15} define AM as:
	\enquote{Additive manufacturing is the formalised term for what used to be
	called rapid prototyping and what is popularly called 3D Printing. [...] Referred
	to in short as AM, the basic principle of this technology is that a model,
	initially generated using a three-dimensional Computer-Aided Design (3D CAD) system,
	can be fabricated directly without the need for process planning. [...]}

	Gebhardt~\citep{Gebhardt13} defines AM as:
	\enquote{Als Generative Fertigungsverfahren werden alle Fertigungsverfahren bezeichnet,
	die Bauteile durch Auf- oder Aneinanderfügen von Volumenelementen (Voxel’n),
	vorzugsweise schichtweise, automatisiert herstellen.}, which we
	translate as \enquote{As generative/additive manufacturing processes all production
	processes are referred that produce components automatically by
	depositioning of volume elements (Voxels), preferably layer-wise}.

	The VDI directives VDI 3404 (Version 2009~\citep{VDI-3404:09} and 2014~\citep{VDI-3404:14}) define additive fabrication as:
	\enquote{Additive fabrication refers to manufacturing processes
	which employ an additive technique whereby
	successive layers or units are built up to form a model.}.


	The 2009 directive \enquote{VDI-Richtlinie: VDI 3404 Generative Fertigungsverfahren - Rapid-Technologien (Rapid Prototyping) - Grundlagen, Begriffe, Qualitätskenngrößen, Liefervereinbarungen}
	and the 2014 directive \enquote{VDI-Richtlinie: VDI 3404 Additive Fertigung - Grundlagen, Begriffe, Verfahrensbeschreibungen} are both currently in retracted states.

	The also retracted ASTM standard F2792-12a \enquote{Standard terminology for additive manufacturing technologies} defines
	AM as \enquote{A process of joining materials to make
	objects from 3D model data, usually layer upon
	layer, as opposed to subtractive manufacturing methodologies.} with the following synonyms listed
	\enquote{additive fabrication, additive processes, additive techniques, additive layer manufacturing, layer manufacturing, and freeform fabrication.}.

	Bechthold et al.~\citep{Bechthold15} define AM as:
	\enquote{The terms additive manufacturing (AM) and 3D printing describe
	production processes in which a solid 3D structure is produced layer by layer
	by the deposition of suitable materials via an additive manufacturing machine.}

	Thomas and Gilbert~\citep{Thomas14} define AM as:
	\enquote{Additive manufacturing is
	the process of joining materials to make objects from three-dimensional (3D) models
	layer by layer as opposed to subtractive methods that remove material. The terms additive
	manufacturing and 3D printing tend to be used interchangeably to describe the same
	approach to fabricating parts. This technology is used to produce models, prototypes,
	patterns, components, and parts using a variety of materials including plastic, metal,
	ceramics, glass, and composites}

	Klocke~\citep{Klocke15} defines AM as:
	\enquote{Generative Verfahren: Diese Verfahrensgruppe umfasst alle Technologien,
	mit denen eine aufbauende, schichtweise Fertigung von Bauteilen realisiert wird.
	Sie werden auch als Additive Manufacturing Technologies oder als Layer based
	Manufacturing Technologies bezeichnet. Zum Herstellen der Schichten wird
	h\"aufig Laserstrahlung verwendet. [...].}

	translation \enquote{Generative Processes: This process group contains all technologies,
	with which an additive, layer-wise generation of parts is realised.
	They are also referred to as additive manufacturing technologies or layer based
	manufacturing technologies. For the creation of the layers oftentimes laser emission is used. [...]}

	In the ASTM F2792-12a~\citep{ASTM-F2792-12a} standard AM is defined as:
	\enquote{process of joining materials
	to make objects from 3D model data, usually layer upon
	layer, as opposed to subtractive manufacturing methodologies.
	Synonyms: additive fabrication, additive processes,
	additive techniques, additive layer manufacturing, layer
	manufacturing, and freeform fabrication.}

	Gao et al.~\citep{Gao15} use the term AM and 3D printing synonymously:
	\enquote{Additive manufacturing (AM), also referred to as 3D printing, [...]}.

	Sames et al.~\citep{Sames16} also use the term AM and 3D printing synonymously:
	\enquote{Additive manufacturing (AM), also known as three-dimensional (3D) printing, [...]}

	Lachmayer and Lippert~\citep{Lachmayer16} define AM as:
	\enquote{Das Additive Manufacturing (AM), als \"Uberbegriff für das Rapid Prototyping (RP), das
	Rapid Tooling (RT), das Direct Manufacturing (DM) und das Rapid Repair (RR) basiert
	auf dem Prinzip des additiven Schichtaufbaus in x-, y- und z-Richtung zur maschinellen
	Herstellung einer (Near-) Net-Shape Geometrie}
	which translates to:
	\enquote{Additive manufacturing as an umbrella term for Rapid Prototyping (RP), Rapid Tooling (RT),
	Direct Manufacturing (DM) and Rapid Repair (RR) is based on the principle of the additive
	layer fabrication in x-, y- and z-direction for the fabrication of a (near-) net-shape geometry by
	machines}

	The ISO/ASTM Standard 52900:2015(E)~\citep{ISO-ASTM_52900:15e} defines AM as:
	\enquote{process of joining materials to make parts (2.6.1) from 3D model data, usually layer (2.3.10) upon layer, as opposed to subtractive manufacturing
	and formative manufacturing methodologies}.

	\subsubsection{Definitions of 3D Printing}
	\label{subsubsec:definitions_of_3d_printing}

		According to Gebhardt~\citep{Gebhardt13} 3D Printing is a
		generic term that is synonymous to AM and is replacing the
		term AM in the future due to its simplicity.
		Bechtholdt et al.~\citep{Bechthold15} use the terms 3D Printing
		and AM synonymously as umbrella terms for technologies and applications.
		In the VDI directive~\citep{VDI-3405:14} the term 3D printing is
		used for a certain additive process but it is acknowledged that it
		is generally used as a synonym for AM.

		The  ASTM standard F2792-12a (retracted) defines 3D printing as
		\enquote{The fabrication of objects through the deposition
		of a material using a print head, nozzle, or another
		printer technology.} but also acknowledges the common synonymous use of this term
		for AM, mostly of low-end quality and price machines.

		Gibson~\citep{Gibson15} uses the term 3D Printing
		for the technology invented by researches
		at MIT~\citep{Sachs90} but also acknowledges that it is used synonymously for AM and
		will eventually replace the term AM due to media coverage.

		The ISO/ASTM Standard 52900:2015(E)~\citep{ISO-ASTM_52900:15e}
		defines 3D Printing as:
		\enquote{fabrication of objects through the deposition
		of a material using a print head,
		nozzle, or another printer technology}.

		It is also noted in this standard that the
		term 3D printing is often used as a synonym for
		AM, mostly in non-technical context.
		Furthermore, it is noted that 3D printing is associated
		with low price and capability machines.

	\subsubsection{Definitions of Rapid Prototyping}
	\label{subsubsec:definitions_of_rapid_prototyping}
		In Hopkinson and Dickens~\citep{Hopkinson01} Rapid Prototyping (RP) is defined as:
		\enquote{RP refers to a group of commercially available processes which are used to
		create solid 3D parts from CAD, from this point onwards these processes will be
		referred to as layer manufacturing techniques (LMTs)}

		The VDI directive 3405 defines RP as:
		\enquote{Additive fabrication of parts with limited functionality,
		but with sufficiently well-defined specific
		characteristics.}

		Weber et al.~\citep{Weber13} define RP as:
		\enquote{Early AM parts were created for the rapid prototyping
		market and were first employed as visual aids and presentation models. Many
		lower cost AM systems are still used in this way.}

	\subsubsection{Definitions of Rapid Manufacturing}
	\label{subsubsec:definitions_of_rapid_manufacturing}
		Hopkinson et al.~\citep{Hopkinson06} define Rapid Manufacturing (RM) as:
		\enquote{the use of a computer aided design (CAD)-based automated
		additive manufacturing process to construct that are used directly
		as finished products or components.}
		Previously Hopkinson and Dickens~\citep{Hopkinson01} defined RM as:
		\enquote{Rapid manufacturing uses LMTs for the
		direct manufacture of solid 3D products
		to be used by the end user either as parts
		of assemblies or as stand-alone products.}

		The VDI directive 3404 Version 2009~\citep{VDI-3404:09} defines RM as:
		\enquote{Additive fabrication of end products (often also described as production parts).
		Characteristics: Has all the characteristics of the
		end product or is accepted by the customer for
		“series production readiness”.
		Material is identical to that of the end product.
		Construction corresponds to that of the end product.}

		The VDI directive 3405~\citep{VDI-3405:14} defines RM as a
		synonym for direct manufacturing, which is defined as:
		\enquote{Additive fabrication of end products.}

	\subsubsection{Definitions of Rapid Tooling}
	\label{subsubsec:definitions_of_rapid_tooling}

		King and Tansey~\citep{King02} define Rapid Tooling (RT) as an extension of RP as such:
		\enquote{Rapid tooling is a progression from rapid prototyping. It is
		the  ability to build prototype tools directly as opposed to
		prototype products directly from the CAD model resulting in
		compressed time to market solutions.}

		The VDI directive 3405~\citep{VDI-3405:14} defines RT as:
		\enquote{The use of additive technologies and processes to
		fabricate end products which are used as tools,
		moulds and mould inserts.}

		Weber et al.~\citep{Weber13} define RT as:
		\enquote{Another class of applications for AM parts is patterns for tooling
		or tooling directly made by AM. AM processes can be used to significantly
		shorten tooling time and are especially useful for low-run production of
		products.}

	\subsubsection{Definitions of Cloud Manufacturing}
	\label{subsubsec:definitions_cloud_manufacturing}
		The work by Li et al.~\citep{Li10} appears to be the first to introduce the
		concept and definition of Cloud Manufacturing (CM), but unfortunately this article
		is only available in Chinese and could therefore not be considered.
		The article is cited by more than 450 publications according to
		Google Scholar.

		Wu and Yang~\citep{Wu10} define CM as such:
		\enquote{Cloud manufacturing is an
		integrated supporting environment both for the share and integration of resources in
		enterprise. It provides virtual manufacturing resources pools,
		which shields the heterogeneousness and the regional
		distribution of resources by the way of virtualisation.
		cloud manufacturing provides a cooperative work environment
		for manufacturing enterprises and individuals and enables the
		cooperation of enterprise.}

		Tao et al.~\citep{Tao11} define CM indirectly by the following description:
		\enquote{Cloud manufacturing is a computing and
		service-oriented manufacturing model developed from
		existing advanced manufacturing models (e.g. ASP, AM,
		NM, MGrid) and enterprise information technologies
		under the support of cloud computing, IoT, virtualisation
		and service-oriented technologies, and
		advanced computing technologies}

		Xu~\citep{Xu12} defines CM similar to the NIST definition of CC as:
		\enquote{a model for enabling ubiquitous, convenient, on-demand network
		access to a shared pool of configurable manufacturing
		resources (e.g., manufacturing software tools, manufacturing
		equipment, and manufacturing capabilities)
		that can be rapidly provisioned and released
		with minimal management effort or service provider interaction}.
		This definition is also used in the work by
		Wang and Xu~\citep{Wang13}.

		Zhang et al.~\citep{Zhang14} describe CM as:
		\enquote{Cloud manufacturing (CMfg) is a new manufacturing paradigm based on networks.
		It uses the network, cloud computing, service computing and manufacturing
		enabling technologies to transform manufacturing resources and manufacturing
		capabilities into manufacturing services, which can be managed and operated in an
		intelligent and unified way to enable the full sharing and circulating of
		manufacturing resources and manufacturing capabilities. CMfg can provide safe,
		reliable, high-quality, cheap and on-demand manufacturing services for the whole
		life cycle of manufacturing.}


	\subsubsection{Synonyms for AM}
	As with the previous definitions for AM, RP, RT, RM and 3D printing there is no consensus
	in the terminology for synonyms of AM in general. The following synonyms can
	be found in literature and are used in existing works.
		\begin{itemize}
			\item direct layer manufacturing or layer manufacturing or additive layer manufacturing
			\item direct digital manufacturing is a synonym for rapid manufacturing~\citep{Gibson15}
			\item solid freeform fabrication (SFF), three dimensional printing~\citep{Weber13}
			\item 3D printing, Additive Techniques, Layer Manufacturing, and Freeform fabrication~\citep{Muthu16}
			\item additive fabrication, additive processes, additive techniques, additive layer manufacturing, layer manufacturing, and freeform fabrication~\citep{ASTM-F2792-12a}\footnote{Also \url{https://wohlersassociates.com/additive-manufacturing.html}}
			\item \enquote{The technical name for 3D printing is additive manufacturing [...]}~\citep{Lipson13}
		\end{itemize}

	\subsubsection{Critique}
	\label{subsubsec:critique}
		The existing definitions fall short on their focus on the layer-wise
		creation of objects as technologies like LENS and multi-axis (n $>$ 3)
		are not bound and defined by a layer structure but can regarded
		as a form of AM as they	create objects based on 3D (CAD) models from
		scratch without any of
		the characteristics of traditional subtractive or formative
		fabrication methods.

		Through a systematic decomposition of the existing definitions of
		AM we conclude that the basic commonality of AM is described as
		the creation of a physical object from a digital model by a machine.

		Furthermore, we propose the term AM as an umbrella term that signifies
		industrial, commercial or professional application and usage whereas
		3D printing can be colloquially used for technologies and methods for
		the creation of physical objects from 3D (CAD) models in other
		situations.

		For the actual building machines of additively manufactured parts
		we recommend the synonymous use of AM fabricator or 3D printer. The
		first as it describes the functionality in a precise way and the second
		as it is commonly used and understood by a broad audience.


\section{Journals related to the Subject}
\label{sec:journals_related_to_the_subject}
	We have identified a number of journals specialising in the domain of AM. In
this section we explain their foci and their scientific scope.

The following journals cater partially or solely for the
academic dissemination of works based in or related to the domains
of AM, RM, RP and 3D Printing. These journals are identified using the
service of the Directory of
Open Access Journals\footnote{\url{https://doaj.org}},
Thomson Reuters Web of Science\footnote{\url{http://webofknowledge.com}}
and the articles used for this review.
Only journals with indication for AM, RM, RP or 3D Printing
in either the title or the scope are listed below.

In the following overview the abbreviations EiC for Editor in Chief,
ImpactF for Impact Factor and SJR for SCImago Journal Rank
Indicator\footnote{\enquote{It expresses the average number of weighted citations received in the selected year by the documents published in the selected journal in the three previous years}, see \url{http://www.scimagojr.com/SCImagoJournalRank.pdf} for more details} are used.
The Impact Factor is either acquired from the journal's home page directly when available or
looked up from Thomson Reuters InCites Journal Citation Reports\footnote{\url{https://jcr.incites.thomsonreuters.com}}. For a number of journals neither a SJR nor the IF could be found.
The numbers for the available volumes, issues and articles are directly extracted from the respective journal's website.
The listing contains a full list of all members of the board and editors per journal for an assessment of the interconnection between the various journals.
Editors and members of the board that are involved in more than one journal are indicated by italicised text and the indication in which other journal they are
involved. The journals are ordered by their number of articles published and if two or more journals have an equal number of publications the ordering is chronological. The journals without
publications and age available are sorted by their ISSN.

The 20 journals have an accumulated 22616 articles published
(respectively 17877 articles, when only considering articles
from Journal \ref{j:9} after it was renamed). The median of
the first publication date is 2014. Under the assumption
that the articles are published equally since the first
Journal (Journal \ref{j:2}) started in 1985, 31 years ago, this results in an
average number of 576 articles per year, which accounts for approximately
18~\% of the average accumulated results of
3197 scientific works indexed by \url{http://scholar.google.com}
for the time frame of 2002 to 2016 (See also Section~\ref{subsec:research_objective}).
The information on the journals is accurate as of 2016-08-10 according to
the respective websites.

\begin{enumerate}
	\item The International Journal of Advanced Manufacturing Technology\label{j:2}
		\begin{compactitem}
			\item[Publisher] Springer
			\item[ISSN] 1433-3015
			\item[URL] \url{http://www.springer.com/engineering/production+engineering/journal/170/PSE}
			\item[ImpactF] 1.568
			\item[H-Index] 71\footnote{\url{http://www.scimagojr.com/journalsearch.php?q=20428&tip=sid&clean=0}}
			\item[SJR] 0.91\footnote{\url{http://www.scimagojr.com/journalsearch.php?q=20428&tip=sid&clean=0}}
			\item[Since] 1985
			\item[Volumes] 85
			\item[Issues] 432
			\item[Articles] 11727
			\item[EiC] \textit{Andrew Y. C. Nee}\label{j:anee} (See also Journal \ref{j:18})
			\item[Board and Editors]
				\begin{inparaenum}[1)]
					\item Kai Cheng
					\item David W. Russell
					\item M. S. Shunmugam
					\item Erhan Budak
					\item D. Ben-Arieh
					\item C. Brecher
					\item H. van Brussel
					\item B. Çatay
					\item F. T. S. Chan\label{j:fchan} (See also Journal \ref{j:18})
					\item F. F. Chen
					\item G. Chryssolouris
					\item \textit{Chee Kai Chua}\label{j:ckchua} (See also Journals \ref{j:10}, \ref{j:19}, \ref{j:20})
					\item M. Combacau
					\item A. Crosnier
					\item S. S. Dimov
					\item L. Fratini
					\item M. W. Fu
					\item H. Huang
					\item V. K. Jain
					\item M. K. Jeong
					\item P. Ji
					\item W.-Y. Jywe
					\item R. T. Kumara
					\item A. Kusiak
					\item B. Lauwers
					\item W. B. Lee
					\item C. R. Nagarajah
					\item E. Niemi
					\item D. T. Pham
					\item S. G. Ponnambalam
					\item M. M. Ratnam
					\item V. R. Rao
					\item C. Saygin
					\item W. Steen
					\item D. J. Stephenson
					\item \textit{M. K. Tiwari}\label{j:mtiwari} (See also Journal \ref{j:6})
					\item E. Vezzetti
					\item G. Vosniakos
					\item X. Xu\label{j:xxu} (See also Journal \ref{j:18})
					\item Y. X. Yao
					\item A. R. Yildiz
					\item \textit{M. Zoe} (See also Journals \ref{j:1}, \ref{j:4}) 
					\item H.-C. Zhang
					\item L. Zhang
					\item A.G. Mamalis
				\end{inparaenum}
		\end{compactitem}

		\item Journal of Manufacturing Science and Engineering\label{j:9}
			\begin{compactitem}
				\item[Publisher] The American Society of Mechanical Engineers
				\item[ISSN] 1087-1357
				\item[URL] \url{http://manufacturingscience.asmedigitalcollection.asme.org/journal.aspx}
				\item[ImpactF] 1.087
				\item[H-Index] 68\footnote{\url{http://www.scimagojr.com/journalsearch.php?q=20966&tip=sid&clean=0}}
				\item[SJR] 0.8\footnote{\url{http://www.scimagojr.com/journalsearch.php?q=20966&tip=sid&clean=0}}
				\item[Since] 1996
				\item[Volumes] 138
				\item[Issues] 101 (Since \enquote{Journal of Engineering for Industry} was renamed to its current title)
				\item[Articles] 7066 (2327 since the renaming in May 1996)
				\item[EiC] Y. Lawrence Yao
				\item[Board and Editors]
					\begin{inparaenum}[1)]
						\item Sam Anand
						\item Wayne Cai\label{j:wcai} (See also Journal \ref{j:8})
						\item Jaime Camelio
						\item Hongqiang Chen
						\item Dragan Djurdjanovic
						\item Guillaume Fromentin
						\item Yuebin Guo
						\item \textit{Yong Huang} (See also Journal \ref{j:7}) 
						\item Yannis Korkolis
						\item Laine Mears
						\item \textit{Gracious Ngaile} (See also Journal \ref{j:8}) 
						\item Radu Pavel
						\item Zhijian Pei
						\item Xiaoping Qian
						\item Tony Schmitz
						\item Jianjun (Jan) Shi
						\item Daniel Walczyk
						\item Donggang Yao
						\item Allen Y. Yi
					\end{inparaenum}
		\end{compactitem}

		\item Robotics and Computer-Integrated Manufacturing\label{j:18}
			\begin{compactitem}
				\item[Publisher] Elsevier B.V.
				\item[ISSN] 0736-5845
				\item[URL] \url{http://www.journals.elsevier.com/robotics-and-computer-integrated-manufacturing}
				\item[ImpactF] 2.077
				\item[H-Index] 61\footnote{\url{http://www.scimagojr.com/journalsearch.php?q=18080&tip=sid&clean=0}}

				\item[SJR] 1.61\footnote{\url{http://www.scimagojr.com/journalsearch.php?q=18080&tip=sid&clean=0}}
				\item[Since] 1984--1994, 1996 ongoing
				\item[Volumes] 44
				\item[Issues] 145
				\item[Articles] 2191
				\item[EiC] Andre Sharon
				\item[Board and Editors]
					\begin{inparaenum}[1)]
						\item M. Haegele
						\item L. Wang
						\item M. M. Ahmad
						\item K. Akella
						\item H. Asada
						\item J. Baillieul
						\item T. Binford
						\item D. Bossi
						\item T. Broughton
						\item M. Caramanis
						\item \textit{F. T. S. Chan} (See also Journal \ref{j:2}) 
						\item G. Chryssolouris
						\item J. Deasley
						\item S. Dubowsky
						\item E. Eloranta
						\item K. C. Fan
						\item \textit{J. Y. H. Fuh} (See also Journals \ref{j:10}, \ref{j:16}) 
						\item J. X. Gao
						\item M. Gevelber
						\item Y. Ito
						\item K. Iwata
						\item T. Kanade
						\item F. Liu
						\item L. Luong
						\item K. L. Mak
						\item K. McKay
						\item A. Meng
						\item N. Nagel
						\item \textit{A. Y. C. Nee} (See also Journal \ref{j:2}) 
						\item G. Reinhardt
						\item R. D. Schraft
						\item W. P. Seering
						\item D. Spath
						\item H. C. G. Spur
						\item N. Suh
						\item M. K. Tiwari
						\item H. Van Brussel
						\item F. B. Vernadat
						\item A. Villa
						\item M. Weck
						\item H. Worn
						\item K. Wright
						\item C. Wu
						\item \textit{X. Xu} (See also Journal \ref{j:2}) 
					\end{inparaenum}
			\end{compactitem}

		\item Rapid Prototyping Journal\label{j:10}
			\begin{compactitem}
				\item[Publisher] Emerald Group Publishing, Ltd
				\item[ISSN] 1355-2546
				\item[URL] \url{http://www.emeraldinsight.com/loi/rpj}
				\item[ImpactF] 1.352
				\item[H-Index] 49\footnote{\url{http://www.scimagojr.com/journalsearch.php?q=21691&tip=sid&clean=0}}
				\item[SJR] 0.81\footnote{\url{http://www.scimagojr.com/journalsearch.php?q=21691&tip=sid&clean=0}}
				\item[Since] 1995
				\item[Volumes] 22
				\item[Issues] 113
				\item[Articles] 882
				\item[EiC] Ian Campbell
				\item[Board and Editors]
					\begin{inparaenum}[1)]
						\item \textit{David Bourell} (See also Journals \ref{j:5}, \ref{j:19}) 
						\item \textit{Ian Gibson}\label{j:igibson} (See also Journals \ref{j:15}, \ref{j:19}) %
						\item James Martin
						\item Sung-Hoon Ahn
						\item \textit{Paulo Jorge da Silva Bártolo}\label{j:sbartolo} (See also Journals \ref{j:12}, \ref{j:19}, \ref{j:20}) %
						\item Deon de Beer
						\item \textit{Alain Bernard}\label{j:abernard} (See also Journals \ref{j:11}, \ref{j:15}, \ref{j:19}) %
						\item \textit{Richard Bibb}\label{j:rbibb} (See also Journal \ref{j:20})
						\item U. Chandrasekhar
						\item \textit{Khershed Cooper} (See also Journal \ref{j:5}) 
						\item \textit{Denis Cormier} (See also Journals \ref{j:7}, \ref{j:15}) 
						\item \textit{Henrique de Amorim Almeida} (See also Journal \ref{j:4}) 
						\item Phill Dickens
						\item \textit{Olaf Diegel} (See also Journal \ref{j:5}) 
						\item \textit{Jerry Fuh}\label{j:jfuh} (See also Journals \ref{j:16}, \ref{j:18}) %
						\item Jorge Ramos Grez
						\item \textit{Chua Chee Kai} (See also Journals \ref{j:2}, \ref{j:19}, \ref{j:20}) 
						\item Jean-Pierre Kruth
						\item Gideon N. Levy
						\item Toshiki Niino
						\item \textit{Eujin Pei} (See also Journal \ref{j:4}) 
						\item B. Ravi
						\item \textit{David Rosen} (See also Journals \ref{j:5}, \ref{j:7}) 
						\item Monica Savalani
						\item \textit{Tim Sercombe}\label{j:tsercombe} (See also Journals \ref{j:15}, \ref{j:16}) %
						\item \textit{Brent Stucker} (See also Journals \ref{j:5}, \ref{j:7}) 
						\item \textit{Wei Sun} (See also Journal \ref{j:5}) 
						\item Jukka Tuomi
						\item \textit{Terry Wohlers} (See also Journal \ref{j:5}) 
					\end{inparaenum}
				\end{compactitem}

		\item Journal of Manufacturing Processes\label{j:8}
			\begin{compactitem}
				\item[Publisher] Elsevier B.V.
				\item[ISSN] 1526-6125
				\item[URL] \url{http://www.journals.elsevier.com/journal-of-manufacturing-processes}
				\item[ImpactF] 1.771
				\item[H-Index] 24\footnote{\url{http://www.scimagojr.com/journalsearch.php?q=27677&tip=sid&clean=0}}
				\item[SJR] 1.09\footnote{\url{http://www.scimagojr.com/journalsearch.php?q=27677&tip=sid&clean=0}}
				\item[Since] 1999
				\item[Volumes] 24
				\item[Issues] 47
				\item[Articles] 620
				\item[EiC] Shiv G. Kapoor
				\item[Board and Editors]
					\begin{inparaenum}[1)]
						\item M. Annoni
						\item \textit{W. Cai}\label{j:wcai} (See also Journal \ref{j:9})
						\item G. Cheng
						\item J. Dong
						\item Z. Feng
						\item G. Y. Kim
						\item A. S. Kumar
						\item X. Li
						\item \textit{G. Ngaile}\label{j:gngaile} (See also Journal \ref{j:9})
						\item S. S. Park
						\item M. Sundaram
						\item B. Wu
						\item H. Yamaguchi Greenslet
						\item Y. Zhang
					\end{inparaenum}
			\end{compactitem}

		\item Virtual and Physical Prototyping\label{j:19}
			\begin{compactitem}
				\item[Publisher] Taylor \& Francis
				\item[ISSN] 1745-2767
				\item[URL] \url{http://www.tandfonline.com/loi/nvpp20}
				\item[ImpactF] N/A
				\item[H-Index] 15\footnote{\url{http://www.scimagojr.com/journalsearch.php?q=5800173379&tip=sid&clean=0}}

				\item[SJR] 0.42\footnote{\url{http://www.scimagojr.com/journalsearch.php?q=5800173379&tip=sid&clean=0}}
				\item[Since] 2006

				\item[Volumes] 11
				\item[Issues] 42
				\item[Articles] 294
				\item[EiC] \textit{Paulo Jorge da Silva Bártolo} (See also Journals \ref{j:10}, \ref{j:12}, \ref{j:20}), \textit{Chee Kai Chua} (See also Journals \ref{j:2}, \ref{j:10}, \ref{j:20})
				\item[Board and Editors]
					\begin{inparaenum}[1)]
						\item \textit{Wai Yee Yeong}\label{j:wyeong} (See also Journal \ref{j:20})
						\item \textit{Alain Bernard} (See also Journals \ref{j:10}, \ref{j:11}, \ref{j:15})
						\item \textit{Anath Fischer} (See also Journal \ref{j:4})
						\item Bopaya Bidanda
						\item \textit{Cijun Shuai}\label{j:cshuai} (See also Journal \ref{j:20})
						\item \textit{David Bourell} (See also Journals \ref{j:5}, \ref{j:10})
						\item \textit{David Dean} (See also Journal \ref{j:4})
						\item \textit{Dongjin Yoo}\label{j:dyoo} (See also Journal \ref{j:20})
						\item Jack Zhou
						\item \textit{Ian Gibson} (See also Journals \ref{j:10}, \ref{j:15})
						\item \textit{Jiankang He}\label{j:jhe} (See also Journal \ref{j:20})
						\item John Lewandowski
						\item Martin Dunn
						\item Ming Leu
						\item Peifeng Li
						\item \textit{Shoufeng Yang} (See also Journals \ref{j:12}, \ref{j:20})
						\item Shlomo Magdassi
						\item \textit{Yong Chen} (See also Journal \ref{j:15})
					\end{inparaenum}
			\end{compactitem}

		\item RTejournal\label{j:1}
			\begin{compactitem}
				\item[Publisher] University Library of the FH-Aachen University of applied Science
				\item[ISSN] 1614-0923
				\item[URL] \url{http://www.rtejournal.de}
				\item[ImpactF] N/A
				\item[H-Index] N/A
				\item[SJR] N/A
				\item[Since] 2004
				\item[Volumes] 13
				\item[Issues] 13
				\item[Articles] 155
				\item[EiC] Andreas Gebhardt
				\item[Board and Editors]
					\begin{inparaenum}[1)]
						\item Ralf Eckhard Beyer
						\item Dietmar Drummer
						\item Karl-Heinrich Grote
						\item Sabine S\"andig
						\item \textit{Gerd Witt}\label{j:gwitt} (See also Journal \ref{j:4})
						\item \textit{Michael Z\"ah}\label{j:mzaeh} (See also Journals \ref{j:2}, \ref{j:4})
					\end{inparaenum}
			\end{compactitem}

		\item International Journal of Rapid Manufacturing\label{j:15}
			\begin{compactitem}
				\item[Publisher] Inderscience Enterprises Ltd.
				\item[ISSN] 1757-8825
				\item[URL] \url{http://www.inderscience.com/ijrapidm}
				\item[ImpactF] N/A
				\item[H-Index] N/A
				\item[SJR] N/A
				\item[Since] 2009
				\item[Volumes] 5
				\item[Issues] 20
				\item[Articles] 93
				\item[EiC] \textit{Bahram Asiabanpour} (See also Journal \ref{j:11}) 
				\item[Board and Editors]
					\begin{inparaenum}[1)]
						\item Ali K. Kamrani
						\item \textit{Denis Cormier} (See also Journals \ref{j:7}, \ref{j:10}) 
						\item Ismail Fidan
						\item \textit{Ian Gibson} (See also Journals \ref{j:10}, \ref{j:19}) 
						\item Wei Jun
						\item Allan Rennie
						\item \textit{Joseph J. Beaman Jr.} (See also Journal \ref{j:7}) 
						\item \textit{Alain Bernard}\label{j:abernard} (See also Journals \ref{j:10}, \ref{j:11}, \ref{j:19}) %
						\item Georges Fadel
						\item Mo Jamshidi
						\item \textit{Behrokh Khoshnevis} (See also Journal \ref{j:5}) 
						\item John M. Usher
						\item Richard A. Wysk
						\item Abe Zeid
						\item Abdulrahman M. Al-Ahmari
						\item Manfredi Bruccoleri
						\item Satish T. S. Bukkapatnam
						\item \textit{Yong Chen}\label{j:ychen} (See also Journal \ref{j:19})
						\item Fred Choobineh
						\item \textit{L. Jyothish Kumar} (See also Journals \ref{j:5}, \ref{j:11}) 
						\item Mehdi Mojdeh
						\item Benoit Montreuil
						\item Kamran Mumtaz
						\item Hossein Tehrani Niknejad
						\item \textit{Pulak Mohan Pandey} (See also Journal \ref{j:11}) 
						\item Prahalad K. Rao
						\item Sa'Ed M. Salhieh
						\item \textit{Tim Sercombe} (See also Journals \ref{j:10}, \ref{j:16}) 
						\item Kathryn E. Stecke
						\item Albert Chi To
						\item Shigeki Umeda
						\item Omid Fatahi Valilai
						\item Nina Vojdani
						\item Micky R. Wilhelm
						\item Stewart Williams
					\end{inparaenum}
			\end{compactitem}

		\item Additive Manufacturing\label{j:7}
			\begin{compactitem}
				\item[Publisher] Elsevier B.V.
				\item[ISSN] 2214-8604
				\item[URL] \url{http://www.journals.elsevier.com/additive-manufacturing}
				\item[ImpactF] N/A
				\item[H-Index] 5\footnote{\url{http://www.scimagojr.com/journalsearch.php?q=21100349533&tip=sid&clean=0}}
				\item[SJR] 1.04\footnote{\url{http://www.scimagojr.com/journalsearch.php?q=21100349533&tip=sid&clean=0}}
				\item[Since] 2014
				\item[Volumes] N/A
				\item[Issues] 12
				\item[Articles] 93
				\item[EiC] Ryan Wicker
				\item[Board and Editors]
					\begin{inparaenum}[1)]
						\item E. MacDonald
						\item M. Perez
						\item A. Bandyopadhyay
						\item \textit{J. Beaman}\label{j:jbeaman} (See also Journal \ref{j:15})
						\item J. Beuth
						\item S. Bose
						\item S. Chen
						\item J. W Choi
						\item K. Chou
						\item \textit{D. Cormier}\label{j:dcormier} (See also Journals \ref{j:10}, \ref{j:15})
						\item K. Creehan
						\item C. Elkins
						\item S. Fish
						\item D. D. Gu
						\item O. Harrysson
						\item D. Hofmann
						\item N. Hopkinson
						\item \textit{Y. Huang}\label{j:yhuang} (See also Journal \ref{j:9}) %
						\item K. Jurrens
						\item K. F. Leong
						\item \textit{J. Lewis} (See also Journal \ref{j:5}) 
						\item L. Love
						\item R. Martukanitz
						\item D. Mei
						\item \textit{R. Resnick} (See also Journal \ref{j:5}) 
						\item \textit{D. Rosen} (See also Journals \ref{j:5}, \ref{j:10}) 
						\item C. Spadaccini
						\item \textit{B. Stucker} (See also Journals \ref{j:5}, \ref{j:10}) 
						\item C. Tuck
						\item C. Williams
					\end{inparaenum}
			\end{compactitem}

		\item 3D Printing and Additive Manufacturing\label{j:5}
			\begin{compactitem}
				\item[Publisher] Mary Ann Liebert, Inc
				\item[ISSN] 2329-7662
				\item[URL] \url{http://www.liebertpub.com/overview/3d-printing-and-additive-manufacturing/621}
				\item[ImpactF] N/A
				\item[H-Index] N/A
				\item[SJR] N/A
				\item[Since] 2014
				\item[Volumes] 3
				\item[Issues] 10
				\item[Articles] 86

				\item[EiC] Skylar Tibbits
				\item[Board and Editors]
					\begin{inparaenum}[1)]
						\item Hod Lipson
						\item Craig Ryan
						\item Anthony Atala
						\item David Benjamin
						\item Lawrence J. Bonassar
						\item \textit{David Bourell}\label{j:dbourell} (See also Journals \ref{j:10}, \ref{j:19})
						\item Adrian Bowyer
						\item Glen Bull
						\item Adam Cohen
						\item \textit{Khershed P. Cooper}\label{j:kcooper} (See also Journal \ref{j:10})
						\item Scott Crump
						\item \textit{Olaf Diegel}\label{j:odiegel} (See also Journal \ref{j:10})
						\item Richard Hague
						\item John F. Hornick
						\item Weidong Huang
						\item Takeo Igarashi
						\item Bryan Kelly
						\item \textit{Behrokh Khoshnevis}\label{j:bkhoshnevis} (See also Journal \ref{j:15})
						\item Matthias Kohler
						\item \textit{L. Jyothish Kumar}\label{j:jkumar} (See also Journals \ref{j:11}, \ref{j:15})
						\item Melba Kurman
						\item \textit{Jennifer A. Lewis}\label{j:jlewis} (See also Journal \ref{j:7})
						\item Jos Malda
						\item Gonzalo Martinez
						\item Neri Oxman
						\item Bre Pettis
						\item Sharon Collins Presnell
						\item Phil Reeves
						\item Avi N. Reichental
						\item \textit{Ralph Resnick}\label{j:rresnick} (See also Journal \ref{j:7})
						\item \textit{David W. Rosen}\label{j:drosen} (See also Journals \ref{j:7}, \ref{j:10})
						\item Jenny Sabin
						\item Carolyn Conner Seepersad
						\item \textit{Brent Stucker}\label{j:bstucker} (See also Journals \ref{j:7}, \ref{j:10})
						\item \textit{Wei Sun}\label{j:wsun} (See also Journal \ref{j:10})
						\item Hiroya Tanaka
						\item Thomas Toeppel
						\item Peter Weijmarshausen
						\item \textit{Terry Wohlers}\label{j:twohlers} (See also Journal \ref{j:10})
					\end{inparaenum}
			\end{compactitem}

		\item International Journal of Bioprinting \label{j:20}
			\begin{compactitem}
				\item[Publisher] Whioce Publishing Pte Ltd
				\item[ISSN] 2424-8002
				\item[URL] \url{http://ijb.whioce.com/index.php/int-j-bioprinting}
				\item[ImpactF] N/A
				\item[H-Index] N/A
				\item[SJR] N/A
				\item[Since] 2015
				\item[Volumes] 2
				\item[Issues] 3
				\item[Articles] 31
				\item[EiC] \textit{Chee Kai Chua} (See also Journals \ref{j:2}, \ref{j:10}, \ref{j:19})
				\item[Board and Editors]
					\begin{inparaenum}[1)]
						\item \textit{Wai Yee Yeong} (See also Journal \ref{j:19})
						\item \textit{Aleksandr Ovsianikov} (See also Journal \ref{j:12})
						\item \textit{Ali Khademhosseini} (See also Journal \ref{j:16})
						\item \textit{Boris N. Chichkov} (See also Journal \ref{j:12})
						\item Charlotte Hauser
						\item \textit{Cijun Shuai} (See also Journal \ref{j:19})
						\item \textit{Dong Jin Yoo} (See also Journal \ref{j:19})
						\item Frederik Claeyssens
						\item Geun Hyung Kim
						\item \textit{Giovanni Vozzi} (See also Journals \ref{j:12}, \ref{j:16})
						\item Ibrahim Tarik Ozbolat
						\item \textit{Jiankang He} (See also Journal \ref{j:19})
						\item Lay Poh Tan
						\item Makoto Nakamura
						\item Martin Birchall
						\item \textit{Paulo Jorge Da Silva Bartolo} (See also Journals \ref{j:12}, \ref{j:10}, \ref{j:19})
						\item Peter Dubruel
						\item \textit{Richard Bibb} (See also Journal \ref{j:10})
						\item \textit{Roger Narayan} (See also Journals \ref{j:3}, \ref{j:16})
						\item \textit{Savas Tasoglu} (See also Journals \ref{j:12}, \ref{j:16})
						\item \textit{Shoufeng Yang} (See also Journals \ref{j:19}, \ref{j:12})
						\item Vladimir Mironov
						\item Xiaohong Wang
						\item Jia An
					\end{inparaenum}
			\end{compactitem}

		\item Progress in Additive Manufacturing\label{j:4}
			\begin{compactitem}
				\item[Publisher] Springer
				\item[ISSN] 2363-9520
				\item[URL] \url{http://www.springer.com/engineering/production+engineering/journal/40964}
				\item[ImpactF] N/A
				\item[H-Index] N/A
				\item[SJR] N/A
				\item[Since] 2016
				\item[Volumes] 1
				\item[Issues] 1
				\item[Articles] 14
				\item[EiC] Martin Sch\"afer, Cynthia Wirth
				\item[Board and Editors]
					\begin{inparaenum}[1)]
						\item \textit{Henrique A. Almeida}\label{j:halmeida} (See also Journal \ref{j:10})
						\item \textit{David Dean}\label{j:ddean} (See also Journal \ref{j:19})
						\item Fernando A. Lasagni
						\item \textit{Eujin Pei}\label{j:epei} (See also Journal \ref{j:10})
						\item Jan Sehrt
						\item Christian Seidel
						\item Adriaan Spierings
						\item Xiaoyong Tian
						\item Jorge Vilanova
						\item \textit{Anath Fischer}\label{j:afischer} (See also Journal \ref{j:19})
						\item Russell Harris
						\item Dachamir Hotza
						\item Bernhard M\"uller
						\item Nahum Travitzky
						\item \textit{Gerd Witt} (See also Journal \ref{j:1}) 
						\item \textit{Michael Friedrich Z\"ah} (See also Journals \ref{j:1}, \ref{j:2}) 
					\end{inparaenum}
			\end{compactitem}

		\item International Journal on Additive Manufacturing Technologies\label{j:11}
			\begin{compactitem}
				\item[Publisher] Additive Manufacturing Society of India
				\item[ISSN] 2395-4221
				\item[URL] \url{http://amsi.org.in/homejournal.html}
				\item[ImpactF] N/A
				\item[H-Index] N/A
				\item[SJR] N/A
				\item[Since] 2015
				\item[Volumes] 1
				\item[Issues] 1
				\item[Articles] 7
				\item[EiC] \textit{Pulak M. Pandey}\label{j:ppandey} (See also Journal \ref{j:15}), David Ian Wimpenny, Ravi Kumar Dwivedi
				\item[Board and Editors]
					\begin{inparaenum}[1)]
						\item \textit{L. Jyothish Kumar} (See also Journals \ref{j:5}, \ref{j:15}) 
						\item Keshavamurthy D. B.
						\item Khalid Abdelghany
						\item Suman Das
						\item \textit{Alain Bernard} (See also Journals \ref{j:10}, \ref{j:15}, \ref{j:19}) 
						\item C. S. Kumar
						\item \textit{Bahram Asiabanpour}\label{j:basiabanpour} (See also Journal \ref{j:15}) %
						\item K. P. Raju Rajurkar
						\item Ehsan Toyserkani
						\item Wan Abdul Rahman
						\item Sarat Singamneni
						\item Vijayavel Bagavath Singh
					\end{inparaenum}
			\end{compactitem}

		\item 3D Printing in Medicine\label{j:13}
			\begin{compactitem}
				\item[Publisher] Springer
				\item[ISSN]  2365-6271
				\item[URL] \url{http://www.springer.com/medicine/radiology/journal/41205}
				\item[ImpactF] N/A
				\item[H-Index] N/A
				\item[SJR] N/A
				\item[Since] 2015
				\item[Volumes] 2
				\item[Issues] 4
				\item[Articles] 3
				\item[EiC] Frank J. Rybicki
				\item[Board and Editors]
					\begin{inparaenum}[1)]
						\item Leonid L. Chepelev
						\item Andy Christensen
						\item Koen Engelborghs
						\item Andreas Giannopoulos
						\item Gerald T. Grant
						\item Ciprian N. Ionita
						\item Peter Liacouras
						\item Jane M. Matsumoto
						\item Dimitrios Mitsouras
						\item Jonathan M. Morris
						\item R. Scott Rader
						\item Adnan Sheikh
						\item Carlos Torres
						\item Shi-Joon Yoo
						\item Nicole Wake
						\item William Weadock
					\end{inparaenum}
			\end{compactitem}

		\item Bioprinting\label{j:16}
			\begin{compactitem}
				\item[Publisher] Elsevier B.V.
				\item[ISSN] 2405-8866
				\item[URL] \url{http://www.journals.elsevier.com/bioprinting}
				\item[ImpactF] N/A
				\item[H-Index] N/A
				\item[SJR] N/A
				\item[Since] 2016
				\item[Volumes] 1
				\item[Issues] N/A
				\item[Articles] 1
				\item[EiC] A. Atala
				\item[Board and Editors]
					\begin{inparaenum}[1)]
						\item S. V. Murphy
						\item T. Boland
						\item P. Campbell
						\item \textit{U. Demirci} (See also Journal \ref{j:12}) 
						\item B. Doyle
						\item J. Fisher
						\item \textit{J. Y. H. Fuh} (See also Journals \ref{j:10}, \ref{j:18}) 
						\item A. K. Gaharwar
						\item P. Gatenholm
						\item K. Jakab
						\item J. Jessop
						\item \textit{A. Khademhosseini}\label{j:akhademhosseini} (See also Journal \ref{j:20})
						\item S. J. Lee
						\item I. Lelkes
						\item J. Lim
						\item A. G. Mikos
						\item \textit{R. Narayan} (See also Journals \ref{j:3}, \ref{j:20}) 
						\item \textit{T. Sercombe} (See also Journals \ref{j:10}, \ref{j:15}) 
						\item A. Skardal
						\item \textit{S. Tasoglu} (See also Journals \ref{j:12}, \ref{j:20}) 
						\item D. J. Thomas
						\item \textit{G. Vozzi} (See also Journals \ref{j:12}, \ref{j:20}) 
						\item \textit{I. Whitaker} (See also Journal \ref{j:12}) 
						\item S. K. Williams II
					\end{inparaenum}
			\end{compactitem}

		\item 3D Printing – Science and Technology\label{j:14}
			\begin{compactitem}
				\item[Publisher] DE GRUYTER OPEN
				\item[ISSN] 1896-155X
				\item[URL] \url{http://www.degruyter.com/view/j/3dpst}
				\item[ImpactF] N/A
				\item[H-Index] N/A
				\item[SJR] N/A
				\item[Since] 2016
				\item[Volumes] 0
				\item[Issues] 0
				\item[Articles] 0
				\item[EiC] Haim Abramovich
				\item[Board and Editors]
					\begin{inparaenum}[1)]
						\item Christopher A. Brown
						\item Paolo Fino
						\item Amnon Shirizly
						\item Frank Walther
						\item Kaufui Wong
					\end{inparaenum}
			\end{compactitem}

		\item Journal of 3D Printing in Medicine\label{j:12}
			\begin{compactitem}
				\item[Publisher] Future Medicine Ltd
				\item[ISSN] 2059-4755
				\item[URL] \url{http://www.futuremedicine.com/page/journal/3dp/editors.jsp}
				\item[ImpactF] N/A
				\item[H-Index] N/A
				\item[SJR] N/A
				\item[Since] 2016
				\item[Volumes] 0
				\item[Issues] 0
				\item[Articles] 0
				\item[EiC] Dietmar W Hutmacher
				\item[Board and Editors]
					\begin{inparaenum}[1)]
						\item Peter Choong
						\item Michael Schuetz
						\item \textit{Iain S. Whitaker}\label{j:iwhitaker} (See also Journal \ref{j:16}) %
						\item \textit{Shoufeng Yang}\label{j:syang} (See also Journals \ref{j:19}, \ref{j:20})
						\item \textit{Paulo Jorge Bártolo} (See also Journals \ref{j:10}, \ref{j:19}, \ref{j:20}) 
						\item Luiz E. Bertassoni
						\item Faiz Y. Bhora
						\item \textit{Boris N. Chichkov}\label{j:bchichkov} (See also Journal \ref{j:20})
						\item \textit{Utkan Demirci}\label{j:udemirci} (See also Journal \ref{j:16}) %
						\item Michael Gelinsky
						\item Ruth Goodridge
						\item Robert E. Guldberg
						\item Scott J. Hollister
						\item Zita M. Jessop
						\item Jordan S. Miller
						\item Adrian Neagu
						\item \textit{Aleksandr Ovsianikov}\label{j:aovsianikov} (See also Journal \ref{j:20})
						\item Katja Schenke-Layland
						\item Ralf Schumacher
						\item Jorge Vicente Lopes da Silva
						\item Chris Sutcliffe
						\item \textit{Savas Tasoglu}\label{j:stasoglu} (See also Journals \ref{j:16}, \ref{j:20}) %
						\item Daniel Thomas
						\item Martijn van Griensven
						\item \textit{Giovanni Vozzi}\label{j:gvozzi} (See also Journals \ref{j:16}, \ref{j:20}) %
						\item David J. Williams
						\item Chris J. Wright
						\item Jing Yang
						\item Nizar Zein
					\end{inparaenum}
			\end{compactitem}

		\item Smart and Sustainable Manufacturing Systems\label{j:6}
			\begin{compactitem}
				\item[Publisher] ASTM
				\item[ISSN] N/A
				\item[URL] \url{http://www.astm.org/SSMS}
				\item[ImpactF] N/A
				\item[H-Index] N/A
				\item[SJR] N/A
				\item[Since] 2017
				\item[Volumes] 0
				\item[Issues] 0
				\item[Articles] 0
				\item[EiC] Sudarsan Rachuri
				\item[Board and Editors]
					\begin{inparaenum}[1)]
						\item Darek Ceglarek
						\item Karl R. Haapala
						\item Yinlun Huang
						\item Jacqueline Isaacs
						\item Sami Kara
						\item Soundar Kumara
						\item Sankaran Mahadevan
						\item Lihong Qiao
						\item Roberto Teti
						\item \textit{Manoj Kumar Tiwari} (See also Journal \ref{j:2}) 
						\item Shozo Takata
						\item Tetsuo Tomiyama
						\item Li Zheng
						\item Fazleena Badurdeen
						\item Abdelaziz Bouras
						\item Alexander Brodsky
						\item LiYing Cui
						\item Bryony DuPont
						\item Sebti Foufou
						\item Pasquale Franciosa
						\item Robert Gao
						\item Moneer Helu
						\item Sanjay Jain
						\item I. S. Jawahir
						\item Sagar V. Kamarthi
						\item Jay Kim
						\item Minna Lanz
						\item Kincho H. Law
						\item Mahesh Mani
						\item Raju Mattikalli
						\item Michael W. McKittrick
						\item Shreyes N. Melkote
						\item P. V. M. Rao
						\item Utpal Roy
						\item Christopher J. Saldana
						\item K. Senthilkumaran
						\item Gopalasamudram R. Sivaramakumar
						\item Eswaran Subrahmanian
						\item Dawn Tilbury
						\item Conrad S. Tucker
						\item Anahita Williamson
						\item Paul William Witherell
						\item Lang Yuan
						\item Rakesh Agrawal
						\item Dean Bartles
						\item Gahl Berkooz
						\item Jian Cao
						\item S. K. Gupta
						\item Timothy G. Gutowski
						\item Gregory A. Harris
						\item Rob Ivester
						\item Mark Johnson
						\item Thomas Kurfess
						\item Bahram Ravani
						\item William C. Regli
						\item S. Sadagopan
						\item Vijay Srinivasan
						\item Ram D. Sriram
						\item Fred van Houten
						\item Albert J. Wavering
					\end{inparaenum}
			\end{compactitem}

		\item Powder Metallurgy Progress\label{j:17}
			\begin{compactitem}
				\item[Publisher] DE GRUYTER OPEN
				\item[ISSN] 1339-4533
				\item[URL] \url{http://www.degruyter.com/view/j/pmp}
				\item[ImpactF] N/A
				\item[H-Index] N/A
				\item[SJR] N/A
				\item[Since] N/A
				\item[Volumes] 0
				\item[Issues] 0
				\item[Articles] 0
				\item[EiC] Beáta Ballóková
				\item[Board and Editors]
					\begin{inparaenum}[1)]
						\item Katarína Ondrejová
						\item Herbert Danninger
						\item Eva Dudrová
						\item Marco Actis Grande
						\item Abolghasem Arvand
						\item Csaba Balázsi
						\item Sergei M. Barinov
						\item Frank Baumg\"artner
						\item Paul Beiss
						\item Sigurd Berg
						\item Michal Besterci
						\item Jaroslav Briančin
						\item Francisco Castro
						\item Andrzej Cias
						\item Ján Dusza
						\item Juraj Ďurišin
						\item Štefan Emmer
						\item Sergei A. Firstov
						\item Christian Gierl-Mayer
						\item Eduard Hryha
						\item Pavol Hvizdoš
						\item Jan Kazior
						\item Jacob K\"ubarsepp
						\item Alberto Molinari
						\item John R. Moon
						\item Ľudovít Parilák
						\item Doan Dinh Phuong
						\item Raimund Ratzi
						\item Wolf D. Schubert
						\item František Simančík
						\item Marin Stoytchev
						\item Andrej \v{S}alak 
						\item José M. Torralba
						\item Andrew S. Wronski
						\item Timothy Martin
						\item Radovan Bureš
					\end{inparaenum}
			\end{compactitem}

		\item 3D-Printed Materials and Systems\label{j:3}
			\begin{compactitem}
				\item[Publisher] Springer
				\item[ISSN] 2363-8389
				\item[URL] \url{http://www.springer.com/materials/journal/40861}
				\item[ImpactF] N/A
				\item[H-Index] N/A
				\item[SJR] N/A
				\item[Since] N/A
				\item[Volumes] 0
				\item[Issues] 0
				\item[Articles] 0
				\item[EiC] \textit{Roger J. Narayan}\label{j:rnarayan} (See also Journals \ref{j:16}, \ref{j:20})
				\item[Board and Editors]
					\begin{inparaenum}[1)]
						\item Vipul Dave
						\item Mohan Edirisinghe
						\item Sungho Jin
						\item Soshu Kirihara
						\item Sanjay Mathur
						\item Mrityunjay Singh
						\item Pankaj Vadgama
					\end{inparaenum}
			\end{compactitem}
\end{enumerate}

Furthermore, the following journals are identified from the literature
relevant to this review. Journals catering specifically or
explicitly to AM, RM, RP and 3D printing are listed above.

The list contains only journals with more than 2 publications.
The goal for composing this list is to enable other researches to
identify possible publication venues for their work.
The list is sorted by the number of publications in our bibliography
for each identified journal. The number of each entry indicates
the number of publications for the journal.
	\begin{itemize}
		\item[11] The International Journal of Advanced Manufacturing Technology (See Journal \ref{j:2})
		\item[11] Rapid Prototyping Journal (See Journal \ref{j:10})
		\item[7] Computer-Aided Design\footnote{\url{http://www.journals.elsevier.com/computer-aided-design}}, ISSN: 0010-4485
		\item[6] Robotics and Computer-Integrated Manufacturing (See Journal \ref{j:18})
		\item[6] Journal of Manufacturing Science and Engineering (See Journal \ref{j:9})
		\item[5] Journal of Materials Processing Technology\footnote{\url{http://www.journals.elsevier.com/journal-of-materials-processing-technology}}, ISSN: 0924-0136
		\item[4] International Journal of Computer Integrated Manufacturing\footnote{\url{http://www.tandfonline.com/toc/tcim20/current}}, ISSN: 1362-3052
		\item[4] CIRP Annals - Manufacturing Technology\footnote{\url{http://www.journals.elsevier.com/cirp-annals-manufacturing-technology}}, ISSN: 0007-8506
		\item[3] Journal of Manufacturing Systems\footnote{\url{http://www.journals.elsevier.com/journal-of-manufacturing-systems}}, ISSN: 0278-6125
		\item[3] Computers in Industry\footnote{\url{http://www.journals.elsevier.com/computers-in-industry}}, ISSN: 0166-3615
		\item[2] Virtual and Physical Prototyping (See Journal \ref{j:19})
		\item[2] Proceedings of the Institution of Mechanical Engineers, Part B: Journal of Engineering Manufacture\footnote{\url{http://pib.sagepub.com}}, ISSN: 2041-2975
		\item[2] Journal of Intelligent Manufacturing\footnote{\url{http://link.springer.com/journal/10845}}, ISSN: 1572-8145
		\item[2] International Journal of Production Research\footnote{\url{http://www.tandfonline.com/toc/tprs20/current}}, ISSN: 1366-588X
		\item[2] International Journal of Machine Tools and Manufacture\footnote{\url{http://www.journals.elsevier.com/international-journal-of-machine-tools-and-manufacture}}, ISSN: 0890-6955
		\item[2] IEEE Transactions on Industrial Informatics\footnote{\url{http://ieeexplore.ieee.org/xpl/RecentIssue.jsp?punumber=9424}}, ISSN: 1551-3203
		\item[2] Enterprise Information Systems\footnote{\url{http://www.tandfonline.com/toc/teis20/current}}, ISSN: 1751-7583
		\item[2] Applied Mechanics and Materials\footnote{\url{http://www.scientific.net/AMM}}, ISSN: 1662-7482
		\item[2] Advanced Materials Research\footnote{\url{http://www.scientific.net/AMR}}, ISSN: 1662-8985
	\end{itemize}

\section{Reviews on the Subject}
\label{sec:reviews_on_the_subject}
	\begin{figure}
	\begin{center}
			\includegraphics[width=0.8\textwidth]{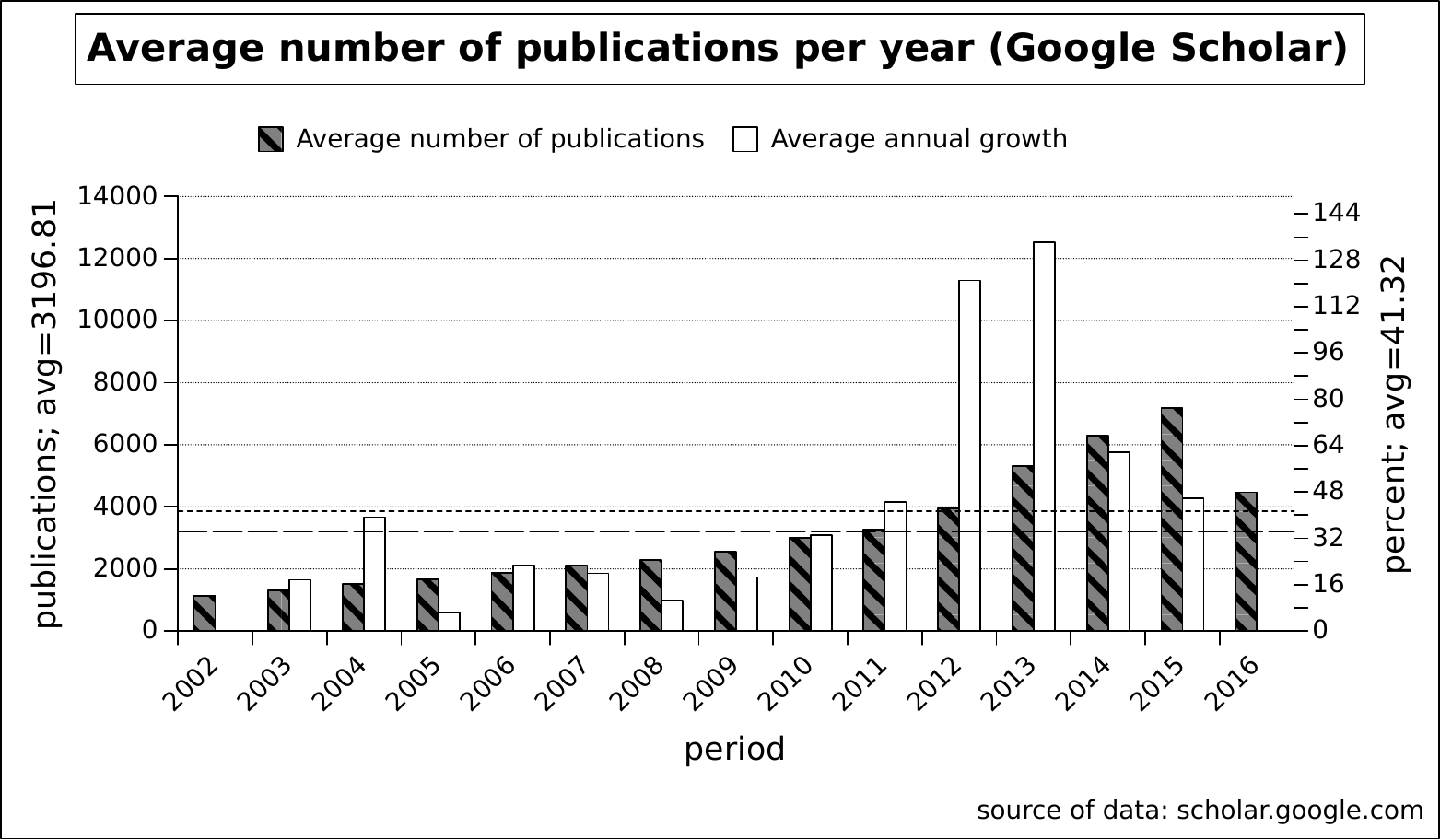}
			\caption{Average number of publications and annual average growth for the combined results from scholar.google.com for 2002--2016}
			\label{fig:annual_growth}
	\end{center}
\end{figure}

\begin{figure}
	\begin{center}
			\includegraphics[width=0.8\textwidth]{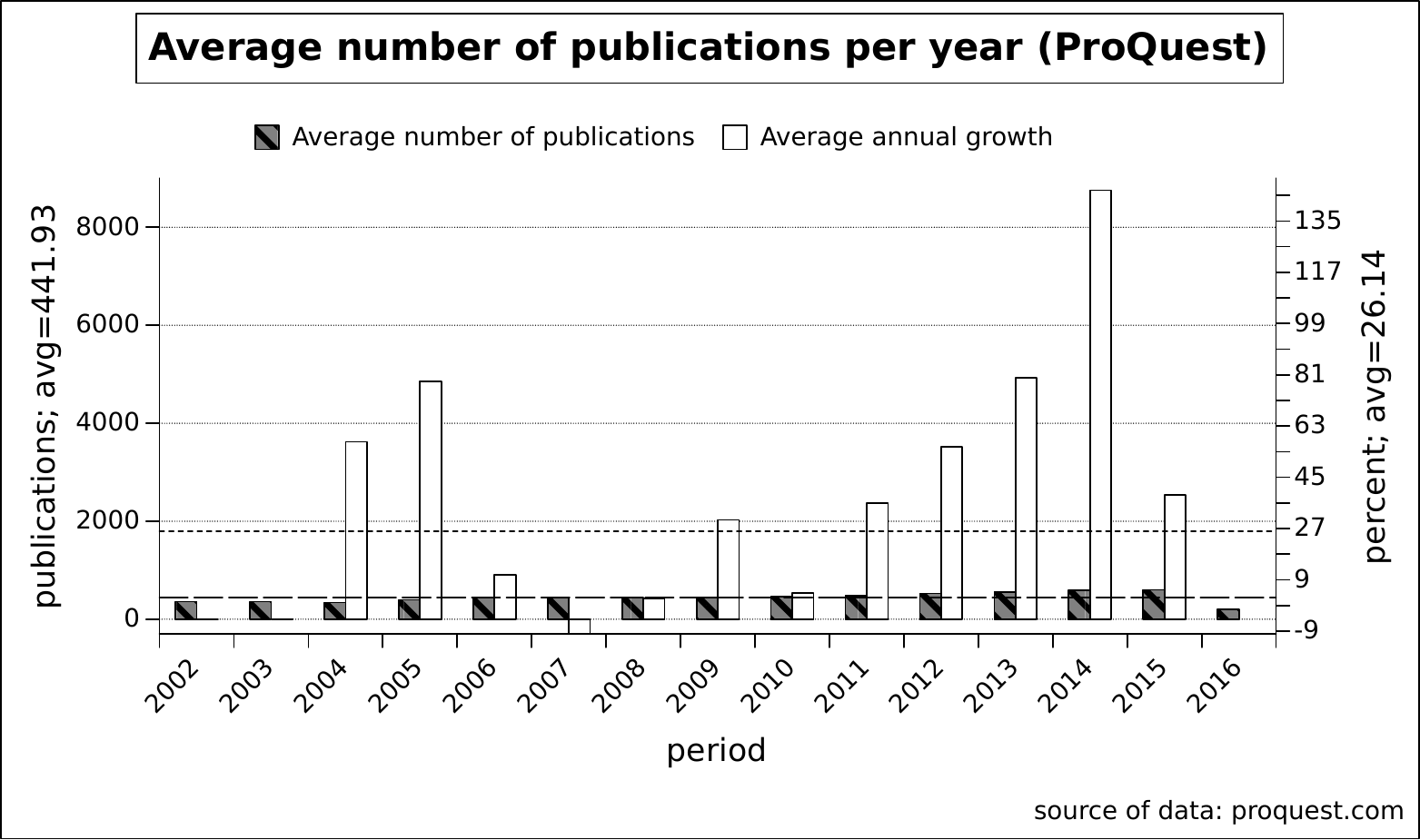}
			\caption{Average number of publications and annual average growth for the combined results from proquest.com for 2002--2016}
			\label{fig:annual_growth_proquest}
	\end{center}
\end{figure}

The topic of AM in general and its special applications,
technologies and directions is extensively researched and results published
in literature. The growth of the number of publications as found by
Google Scholar and Proquest is illustrated in the following figures
(See Fig.~\ref{fig:annual_growth} and 
Fig.~\ref{fig:annual_growth_proquest}).

An analysis of literature within this domain
from sources (See Sect.~\ref{subsubsec:sources}) for scientific literature
shows an increase in the number of
published works from 2002 to 2016 of 41.3~\% on average (See
Fig.~\ref{fig:annual_growth}), respectively 26.1~\% for the search
engine Proquest.
This number is from the average of the average growth of results found for
keywords related to specific AM topics and AM related literature
in general from \url{http://scholar.google.com}.

In this section we will present the findings of the analysis of available data on the scientific publications.

Specific aspects of AM, 3D printing and associated areas are
topic of a number of reviews listed below. The list
of reviews is compiled by searching on the previously mentioned
search engines (See sect. \ref{subsubsec:sources}) using a keyword search.
The keywords are \enquote{3D Printing} +Review/Survey/\enquote{State of the Art},
\enquote{Additive Manufacturing} +Review/Survey\enquote{State of the Art},
\enquote{Rapid Manufacturing} +Review/Survey\enquote{State of the Art}.

The time range for the search for reviews is restricted from 2005 to
2016. Following this literature search a backward search on the
results is performed.
From the 70 reviews identified we calculate the average number of authors
per review to be 3.3 with an average length of 15.2 pages.
The list is sorted chronologically with the general theme or
domain of the review provided.

\begin{enumerate}
	\item Dimitar Dimitrov, Kristiaan Schreve and N. de Beer~\citep{Dimitrov06}; General Introduction, Applications, Research Issues
	\item Vladimir Mironov, Nuno Reis and Brian Derby~\citep{Mironov06}; Bioprinting, Technology 
	\item Ben Utela et al.~\citep{Utela08}; New Material Development (Mainly Powders) 
	\item Abbas Azari and Sakineh Nikzad~\citep{Azari09}; Dentistry, Applications in Dentistry
	\item Hongbo Lan~\citep{Lan09}; Rapid Prototyping, Manufacturing Systems
	\item Daniel Eyers and Krassimir Dotchev~\citep{Eyers10}; Rapid Manufacturing, Mass Customisation
	\item Ferry P. W. Melchels, Jan Feijen and Dirk W. Grijpma~\citep{Melchels10}; Stereolithography, Biomedical Engineering 
	\item Fabian Rengier et al.~\citep{Rengier10}; Medicine, Data Acquisition (Reverse-Engineering) using Image Data 
	\item R. Sreenivasan, A. Goel and D. L. Bourell~\citep{Sreenivasan10}; Energy Consumption, Sustainability
	\item Rupinder Singh~\citep{Singh10}; Rapid Prototyping, Casting 
	\item R. Ian Campbell, Deon J. de Beer and Eujin Pei~\citep{Campbell11}; Application and Development of AM in South Africa
	\item Benjamin Vayre, Frédéric Vignat and François Villeneuve~\citep{Vayre12b}; Metal Components, Technology
	\item Dongdong Gu et al.~\citep{Gu12}; Metal Components, Technology, Terminology 
	\item Ferry P. W. Melchels et al.~\citep{Melchels12}; Medicine, Tissue and Organ Engineering 
	\item Kaufui V. Wong and Aldo Hernandez~\citep{Wong12}; General, Technology
	\item Lawrence E. Murr et al.~\citep{Murr12}; Metal Components, EBM, Laser Melting
	\item Shawn Moylan et al.~\citep{Moylan12}; Quality, Test Artifacts
	\item Timothy J. Horn and Ola L. A. Harrysson~\citep{Horn12}; General, Applications, Technology 
	\item Xibing Gong, Ted Anderson and Kevin Chou~\citep{Gong12}; EBM, Powder Based AM 
	\item Flavio S. Fogliatto, Giovani J.C. da Silveira and Denis Borenstein~\citep{Fogliatto12}; Mass-Customization
	\item K. P. Karunakaran et al.~\citep{Karunakaran12}; Rapid Manufacturing, Metal Object Manufacturing
	\item Carl Schubert, Mark C. van Langeveld and Larry A. Donoso~\citep{Schubert13}; General 
	\item Irene J. Petrick and Timothy W. Simpson~\citep{Petrick13}; Economics, Business 
	\item Iulia D. Ursan, Ligia Chiu and Andrea Pierce~\citep{Ursan13}; Pharmaceutical Drug Printing 
	\item Jasper Cerneels et al.~\citep{Cerneels13}; Thermoplastics
	\item Mohammad Vaezi, Hermann Seitz and Shoufeng Yang~\citep{Vaezi13}; Micro-Structure AM 
	\item Nannan Guo and Ming C. Leu~\citep{Guo13}; General, Technology, Materials, Applications  
	\item Olga Ivanova, Christopher Williams and Thomas Campbell~\citep{Ivanova13}; Nano-Structure AM 
	\item Robert Bogue~\citep{Bogue13}; General
	\item Samuel H. Huang et al.~\citep{Huang13}; Socio-Ecological and Economy 
	\item Zicheng Zhu et al.~\citep{Zhu13}; Hybrid Manufacturing 
	\item Dazhong Wu et al.~\citep{Wu13b}; Cloud Manufacturing
	\item Bethany C. Gross et al.~\citep{Gross14}; Biotech, Chemistry 
	\item Brett P. Conner et al.~\citep{Conner14}; Classification, Object Complexity 
	\item Brian N. Turner, Robert Strong and Scott A. Gold~\citep{Turner14}; Thermoplastics, Physical Properties 
	\item David W. Rosen~\citep{Rosen14}; Design for Additive Manufacturing
	\item Dimitris Mourtzis, Michael Doukas and Dimitra Bernidaki~\citep{Mourtzis14}; Simulation 
	\item Douglas S. Thomas and Stanley W. Gilbert~\citep{Thomas14}; Economy, Cost 
	\item Gustavo Tapia and Alaa Elwany~\citep{Tapia14}; Process Monitoring, Quality 
	\item Hae-Sung Yoon et al.~\citep{Yoon14}; Energy Consumption 
	\item Jan Deckers, Jef Vleugels and Jean-Pierre Kruth~\citep{Deckers14}; Ceramics AM 
	\item Rouhollah Dermanaki Farahani, Kambiz Chizari and Daniel Therriault~\citep{Farahani14}; Micro-Structure AM 
	\item Siavash H. Khajavi, Jouni Partanen and Jan Holmstr\"om~\citep{Khajavi14}; Supply Chain, Application
	\item William E. Frazier~\citep{Frazier14}; Metal Components 
	\item Wu He and Lida Xu~\citep{He15}; Cloud Manufacturing
	\item Syed Hasan Massod~\citep{Masood14}; Fused Deposition Modeling (FDM)
	\item Brian N. Turner and Scott A Gold~\citep{Turner15}; Thermoplastic AM, Material Properties 
	\item Carlos Mota et al.~\citep{Mota15}; Medicine, Tissue Engineering 
	\item C. Y. Yap et al.~\citep{Yap15}; SLM 
	\item Donghong Ding et al.~\citep{Ding15}; Metal Components, Wire Fed Processes 
	\item Adamson et al.\citep{Adamson15}; Cloud Manufacturing, Terminology 
	\item Jie Sun et al.~\citep{Sun15}; Food Printing, Technology 
	\item Jin Choi et al.~\citep{Choi15}; 4D Printing 
	\item K. A. Lorenz et al.~\citep{Lorenz15}; Hybrid Manufacturing 
	\item Merissa Piazza and Serena Alexander~\citep{Piazza15}; General, Terminology, Academic
	\item Omar A. Mohamed, Syed H. Masood and Jahar L. Bhowmik~\citep{Mohamed15}; Process Parameter Optimization (FDM) 
	\item Seyed Farid Seyed Shirazi et al.~\citep{Shirazi15}; Tissue Engineering, Powder Based AM
	\item Sheng Yang and Yaoyao Fiona Zhao~\citep{Yang15}; Design for AM, Complexity 
	\item Sofiane Guessasma et al.~\citep{Guessasma15}; Design for AM, Process Parameter Optimization
	\item Wei Gao et al.~\citep{Gao15}; General, Technology, Engineering
	\item Yong Huang et al.~\citep{Huang15}; General, Technology, Research Needs 
	\item Zhong Xun Khoo et al.~\citep{Khoo15}; Smart Materials, 4D Printing 
	\item Hammad H. Malik et al.~\citep{Malik15}; Medicine, Surgery
	\item Jie Sun et al.~\citep{Sun15b}; Food Printing
	\item Behzad Esmaeilian, Sara Behdad and Ben Wang~\citep{Esmaeilian16}; Manufacturing 
	\item H. Bikas, P. Stavropoulos and G. Chryssolouris~\citep{Bikas16}; General, Technology 
	\item Julien Gardan~\citep{Gardan16}; Technology, Engineering, Manufacturing
	\item Swee Leong Sing et al.~\citep{Sing16}; Metal Components, Medicine, Implants, Materials  
	\item William J. Sames et al.~\citep{Sames16}; Metal Components, Materials 
	\item Andrew J. Pinkerton~\citep{Pinkerton16}; Laser-technology
\end{enumerate}

\subsection{Stakeholder Distinction}
\label{subsec:stakeholder_distinction}
	Different 3D printing technologies, machines and manufacturers as well as
	services target	different clients for which we propose the following
	classification. Generally the discerning factors are
	\begin{inparaenum}
		\item cost per machine
		\item quality of print (\eg surface quality, physical properties of object)
		\item reliability of machine and
		\item materials available.
	\end{inparaenum}
	From literature the three classes of audience are apparent:
	\begin{itemize}
		\item consumer/end user
		\item professional user
		\item industrial application
	\end{itemize}
	For the consumer a very important factor is the cost of the
	printer itself with 45~\% of consumers are not willing to pay
	more than \$US~299 for a 3D printer~\citep{Matias15}.

	In recent years the price of entry level consumer 3D printers,
	especially for build-kits, decreased
	to about \$US~300\footnote{XYZPrinting da Vinci Jr. 1.0, \$US~297.97, \url{https://www.amazon.com/XYZprinting-Vinci-Jr-1-0-Printer}}.
	Open-source projects like RepRap have contributed to the decline
	of costs for these machine~\citep{Soderberg13}.

\begin{figure}[htp]
		\begin{center}
				\includegraphics[clip,trim={3cm 14cm 3cm 5cm},width=0.5\textwidth]{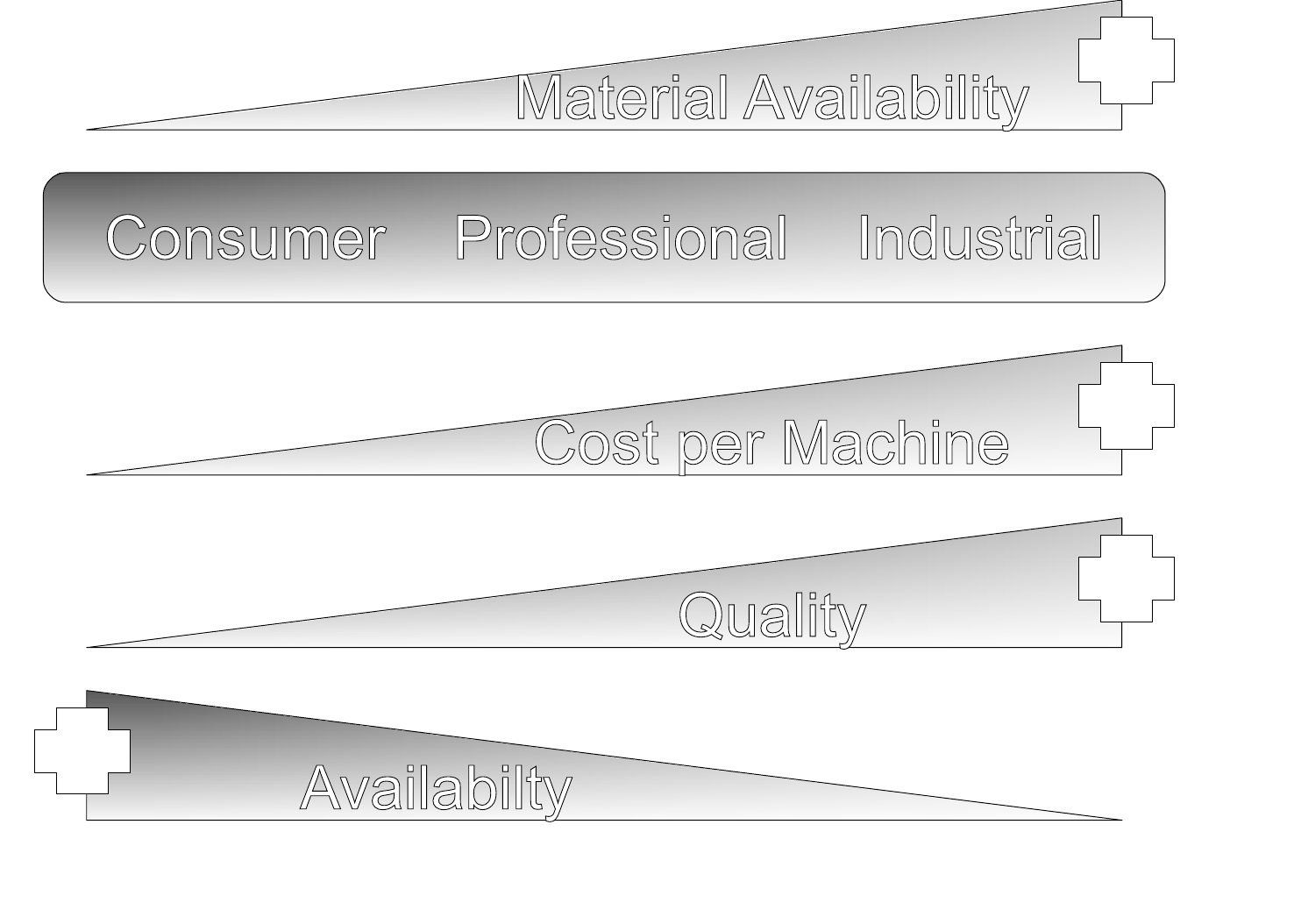}
				\caption{Audience classification and Expectations}
				\label{fig:category}
		\end{center}
	\end{figure}

	In Fig.~\ref{fig:category} we differentiate between the user groups
	of end-users/consumer, professional users and industrial users.
	Industrial users rely on high quality available with a large
	selection of processable materials.
	Machines for these users are expensive and out of reach of
	most end-users and professional users. The quality these
	machines produce is very high and the objects can be used for
	integration in a product or be a product themselves.
	Due to these restrictions the availability of such machines
	is not very wide spread but limited to
	highly specialised enterprises.

	On the other end of the spectrum the end-user/consumer has a large
	choice of 3D printers to select from, they are relatively inexpensive,
	produce objects of acceptable quality, work on a much lower
	number materials (typically thermoplastics) and have a reliability that
	is lower than the reliability of professional equipment.
	In the middle of the spectrum we see professional users, \eg from design
	bureaus or architects, that use such machines
	in a professional manner, draw benefits from the usage of such technology
	but it is mostly not their main concept of business.
	In an example, an architect makes use of a 3D printer for the creation of a
	high-quality model of a building he designed, which
	is faster and easier than making such a model by hand.

\section{3D Printing Services}
\label{sec:3d_printing_services}
	There are numerous dedicated 3D printing services available to end-users,
professionals and industrial users. They differ
in the clients they address, the services they offer,
the quality they can provide and the cost they charge. In this section
we give an overview of a selection of available 3D printing services.
The list is not conclusive as a number of enterprises does
offer 3D printing services in their portfolio but they are not necessarily
to be considered 3D printing services due to either their
local mode of operation or the number of 3D printers the user can chose from.
This overview is closely based on the work of~\citep{Rayna16}
and extends its findings.

We use the following list of properties to distinguish the services:
\begin{itemize}
	\item The target group (End-users, industrial users or professional users)
	\item The local reach (Local or global)
	\item Availability of an API
	\item Services rendered (Design, 3D printing, marketplace, other)
\end{itemize}

Rayna and Striukova~\citep{Rayna16} base their exploratory study on the following
list of services they have identified. For the original list of services
we add the following information.

\begin{itemize}
	\item 3D Burrito\footnote{\url{http://3dburrito.com}} - Pre-Launch Phase
	\item 3D Creation Lab\footnote{\url{http://www.3dcreationlab.co.uk}}
	\item 3DLT\footnote{\url{http://3dlt.com}} - Shut down on 2015-12-31
	\item 3DPrintUK\footnote{\url{https://www.3dprint-uk.co.uk}}
	\item Additer.com\footnote{\url{http://additer.com}} - Unreachable
	\item Cubify Cloud\footnote{\url{http://cubify.com}} - Acquired by 3D Systems, Service no longer available
	\item i.Materialise\footnote{\url{https://i.materialise.com/}}
	\item iMakr\footnote{\url{http://imakr.co.uk}}
	\item Kraftw\"urx.com\footnote{\url{http://www.kraftwurx.com}}
	\item MakerBot/Thingiverse\footnote{\url{http://thingiverse.com}}
	\item MakeXYZ\footnote{\url{https://www.makexyz.com}}
	\item Ponoko\footnote{\url{https://www.ponoko.com/}}
	\item Sculpteo\footnote{\url{https://www.sculpteo.com}}
	\item Shapeways\footnote{\url{http://www.shapeways.com/}}
\end{itemize}

For this study we extend the selection with the additional services
listed in Tabs.~\ref{tab:service_list} and .~\ref{tab:service_list_2}. Services
omitted in these two tables are described in the original study.

\begin{table}
	\caption{3D printing platforms and services included in this study -- Part 1}
	{\begin{tabular}[l]{llp{2.0cm}p{1.4cm}p{1.2cm}}
		\hline
			Company/Service Name & URL & Classification & Established & Location\\
		\hline
			3Faktur & \url{http://3faktur.com} & Modeling Service & 2014 & Germany \\
			3DaGoGo & \url{https://www.3dagogo.com} & Marketplace & 2013 & USA \\
			3DExport & \url{https://3dexport.com} & Marketplace, Repository & 2004 & USA \\
			3DHubs & \url{http://3dhubs.com} & Crowd Printing Provider & 2013 & USA \\
			3DPrinterOS & \url{https://www.3dprinteros.com} & Crowd Printing Provider & 2014 & USA \\
			3DShook & \url{http://www.3dshook.com} & Marketplace, Subscription Service & 2014 & Israel \\
			3D Warehouse & \url{https://3dwarehouse.sketchup.com} & Marketplace, Community, Repository & 2006 & USA \\
			Autodesk 123D & \url{http://www.123dapp.com} & Software, Marketplace, Repository & 2009 & USA \\
			Clara.io & \url{https://clara.io} & Repository, Modeling & 2013 & Canada \\
			CreateThis & \url{http://www.createthis.com} & Marketplace & 2013 & USA \\ 
			Cults & \url{https://cults3d.com} & Marketplace, Repository, Design Service & 2013 & France \\
			Grabcad & \url{https://grabcad.com} & Software, Marketplace, Repository & 2009 & USA \\
			La Poste & \url{http://impression3d.laposte.fr} & Print Provider, Marketplace & 2013 & France \\
			Libre3D & \url{http://libre3d.com} & Marketplace, Repository & 2014 & USA \\
			Makershop & \url{https://www.makershop.co} & Marketplace, Repository & 2013 & USA \\
			Materflow & \url{http://www.materflow.com} & Print Provider, Marketplace, Product Co-Creation/Support & 2013 & Finland \\
			MeltWerk & \url{https://www.meltwerk.com} & Print Provider & - & Subsidiary of trinckle \\
			MyMiniFactory & \url{https://www.myminifactory.com} & Crowd Printing Provider, Marketplace & 2013 & UK \\

		\hline
	\end{tabular}}
	\label{tab:service_list}
\end{table}

\begin{table}
	\caption{3D printing platforms and services included in this study -- Part 2}
	{\begin{tabular}[l]{llp{2.0cm}p{1.4cm}p{1.2cm}}
		\hline
			Company/Service Name & URL & Classification & Established & Location\\
		\hline
			NIH 3D Print Exchange & \url{http://3dprint.nih.gov} & Co-Creation, Repository & 2014 & USA \\
			p3d.in & \url{https://p3d.in} & Modeling & 2010 & Denmark \\
			Pinshape & \url{https://pinshape.com} & Marketplace & 2014 & Canada \\
			REPABLES & \url{http://repables.com} & Repository & 2013 & USA \\
			Rinkak & \url{https://www.rinkak.com} & Marketplace, Repository, Crowd Printing Provider & 2014 & Japan \\
			shapeking & \url{http://www.shapeking.com} & Marketplace, Repository & 2012 & Germany \\
			Shapetizer & \url{https://www.shapetizer.com} & Marketplace, Repository, Print Provider & 2015 & China \\
			Sketchfab & \url{https://sketchfab.com} & Marketplace, Repository & 2012 & France \\
			stlfinder & \url{http://www.stlfinder.com} & Search Engine & 2013 & Spain \\
			STLHive & \url{http://www.stlhive.com} & Marketplace, Repository, Design Service & 2015 & Canada \\
			Stratasys Direct Express & \url{https://express.stratasysdirect.com} & Print Provider & 2015 & USA \\
			Threeding & \url{https://www.threeding.com} & Marketplace, Print Provider & 2014 & Bulgaria \\
			Tinkercad & \url{https://www.tinkercad.com} & Design, Repository & 2011 & USA \\
			Treatstock & \url{https://www.treatstock.com} & Marketplace, Community, Crowd Printing Provider & 2016 & USA \\
			trinckle & \url{https://www.trinckle.com} & Print Provider & 2013 & Germany \\
			Trinpy & \url{https://www.trinpy.com} & Marketplace, Subscription Service & 2015 & Australia \\
			TurboSquid & \url{http://www.turbosquid.com} & Marketplace, Repository & 2000 & USA \\
			UPS & \url{https://www.theupsstore.com/print/3d-printing} & Print Provider & 2013 & USA \\			
			Watertight & \url{https://watertight.com} & Marketplace & 2015 & USA \\
			Yeggi & \url{http://www.yeggi.com} & Search Engine & 2013 & Germany \\
			YouMagine & \url{https://www.youmagine.com} & Community, Repository, Marketplace & 2013 & The Netherlands \\

		\hline
	\end{tabular}}
	\label{tab:service_list_2}
\end{table}

In contrast to the authors of the original work we think that an exhaustive list
of such services is impossible to compile
as a large number of local businesses do offer 3D printing services over the
Internet and would therefore qualify to be included in such a list.
These (local) businesses are hard to identify due to their limited size and reach.
Also, an exhaustive list would need to contain
3D printing services and repositories of which many similar and
derivative services exist.

Further, we extend the classification and study to the provisioning of an API
by the respective service.
An API should provide methods to use the service programmatically.
With an API such printing services can be used as a flexible
production means in CM settings.
The range of functionality of such APIs can vary significantly and range from
the possibility of having a widget displayed on a website with
a 3D model viewer, to upload and store digital models in a repository,
request quotes for manufacturing or digital fabrication.
A commonality for these APIs is the requirement for the third-party user
to have an account with the service, which is indicated in
Tabs.~\ref{tab:service_api} and \ref{tab:service_api_2} by
\textbf{Implementer} in the column
\textbf{Required for registration}. The indication \textbf{User} in this
column indicates that the user must be registered with this service too.

The implementer registration is intended for scenarios where the
API is embedded in a service or website
that a third party user then uses. The findings of this
study are presented in Tabs.~\ref{tab:service_api} and 
\ref{tab:service_api_2}, where we state whether
the service provides an API and if it is publicly available or only
accessible for business partners, who needs to be registered for
the usage of the API and what capabilities the API provides
(See Tab.~\ref{tab:service_category}).
\begin{table}
	\caption{3D printing platforms and services and their APIs - Part 1}
	{\begin{tabular}[l]{lp{1.4cm}p{1.4cm}p{1.4cm}p{1.4cm}c}
		\hline
			Company / Service Name & Provides an API & Required for registration & Capabilities & Reach & Target Group\\
		\hline
			3Faktur & No & N/A & N/A & Regional & Consumer \\
			3DaGoGo & No & N/A & N/A & Global & Consumer \\
			3DExport & No & N/A &  N/A & Global & Consumer + Professional \\
			3DHubs & Yes & Implementer + User & Upload & Global & Consumer \\
			3DPrinterOS & No & N/A & N/A & Global & Consumer \\
			3DPrintUK & No & N/A & Global & Consumer \\
			3DShook & No & N/A &  N/A & Global & Consumer \\
			3D Creation Lab & No & N/A & N/A & Global & Consumer \\
			3D Warehouse & No & N/A &  N/A & Global & Consumer \\
			Autodesk 123D & N/A &  No & N/A & Global & Consumer \\
			Clara.io & Yes & Implementer & Upload, Modify, Retrieve & Global & Consumer + Professional \\
			CreateThis & No & N/A &  N/A & Global & Consumer \\
			Cults & Yes (not public) & Implementer & View, Retrieve & Global & Consumer \\
			Grabcad & No & N/A & N/A & Global & Consumer + Professional\\
			iMakr & No & N/A & N/A & Global & Consumer \\
			i.Materialise & Yes & Implementer & Upload, Quoting, Order & Global & Consumer + Professional \\
			Kraftw\"urx.com & Yes (not public) & Implementer & Upload, Order & Global & Consumer \\
			La Poste & No & N/A & N/A & Regional & Consumer \\
			Libre3D & No & N/A &  N/A & Global & Consumer \\
			MakerBot / Thingiverse & Yes & Implementer & Upload, Retrieve & Global & Consumer \\
			Makershop & Yes & Implementer & Search, Retrieve & Global & Consumer + Professional \\
			MakeXYZ & Yes & Implementer + User & Order & Global & Consumer + Professional \\
			Materflow & No & N/A & N/A & Global & Consumer \\
			MeltWerk & Yes (not public) & Implementer & Upload, Quoting & Global & Consumer \\
			MyMiniFactory & No & N/A & N/A & Global & Consumer \\
			
		\hline
	\end{tabular}}
	\label{tab:service_api}
\end{table}

\begin{table}
	\caption{3D printing platforms and services and their APIs - Part 2}
	{\begin{tabular}[l]{lp{1.4cm}p{1.4cm}p{1.4cm}p{1.4cm}c}
		\hline
			Company / Service Name & Provides an API & Required for registration & Capabilities & Reach & Target Group\\
		\hline
			NIH 3D Print Exchange & Yes & Implementer & Upload, Retrieve & Global & Consumer \\
			p3d.in & No & N/A & N/A & Global & Consumer  \\
			Pinshape & No & N/A & N/A & Global & Consumer \\
			Ponoko & No & N/A & N/A & Global & Consumer \\
			REPABLES & No & N/A &  N/A & Global & Consumer \\
			Rinkak & Yes & Implementer & View, Order, Modeling & Global & Consumer \\
			Sculpteo & Yes & Implementer + User & Upload, Retrieve, Quoting, Order & Global & Consumer + Professional \\
			shapeking & No & N/A &  N/A & Global & Consumer \\
			Shapetizer & No N/A &  & N/A & Global & Consumer \\
			Shapeways & Yes & Implementer + User & Upload, Quoting, Order & Global & Consumer + Professional \\
			Sketchfab & Yes & Implementer & Upload, View & Global & Consumer \\
			stlfinder & No & N/A &  N/A & Global & Consumer \\
			STLHive & No & N/A &  N/A & Global & Consumer + Professional \\
			Stratsys Direct Express & No & N/A & N/A & Regional & Professional \\
			Threeding & No & N/A & N/A & Global & Consumer \\
			Tinkercad & No & N/A & N/A & Global & Consumer \\
			Treatstock & Yes & Implementer & Upload, Retrieve & Global & Consumer \\
			trinckle & No & N/A & N/A & Global & Consumer + Professional \\
			Trinpy & No & N/A &  N/A & Global & Consumer \\
			TurboSquid & No & N/A &  N/A & Global & Consumer + Professional \\
			UPS & No & N/A & N/A & Regional & Consumer \\
			Yeggi & Yes & Implementer & Search, Retrieve & Global & Consumer \\
			YouMagine & Yes & Implementer & Upload, Retrieve & Global & Consumer \\			
			Watertight & No & N/A & N/A & Global & Consumer \\
			
		\hline
	\end{tabular}}
	\label{tab:service_api_2}
\end{table}

This explorative extension study is performed as described by the original authors.

\begin{table}
	\caption{Categorising 3D printing online platforms}
	{\begin{tabular}[l]{p{2cm}|p{1.2cm}|p{1.4cm}|p{1.2cm}|p{1.4cm}|p{1.4cm}|p{1.2cm}|p{1.4cm}|p{1.4cm}l}
		\hline
			Company / Service Name & Design market place & Design repository & Design service & Printing market place & Printing service & Printer sale & Crowd sourcing platform & Editor \\
		\hline
			3Faktur & & & + & & + & & & \\
			3DaGoGo & + & + & & & & & & \\
			3DExport & + & + & & & & & & \\
			3DHubs & & & & + & + & & & \\
			3DPrinterOS & & & & + & + & & & \\
			3DPrintUK & & & + & & + & & & \\
			3DShook & + & + & & & & & & \\
			3D Creation Lab & & & & & + & & & \\
			3D Warehouse & + & + & & & & & & \\
			Autodesk 123D & + & + & & & + & & & \\
			Clara.io & + & + & & & & & & + \\
			CreateThis & + & + & & & & & & \\
			Cults & + & + & + & & & & & \\		
			Grabcad & + & + & & & & p & & + \\
			iMakr & & & & & + & + & & \\
			i.Materialise & + & + & + & & + & & + & \\
			Kraftw\"urx.com & + & + & + &  & + & & + & \\
			La Poste & + & + & & & + & & & + \\	
			Libre3D & + & + & & & & & & \\
			MakerBot / Thingiverse & + & + & & & & p & & + \\
			Makershop & + & + & & & & & & \\
			MakeXYZ & & & + & + & + & & & \\
			Materflow & + & + & + & & + & & & \\
			MeltWerk & & & & & + & & & \\
			MyMiniFactory & + & + & & & + & + & & \\
			NIH 3D Print Exchange & + & + & & & & & & \\
			p3d.in & + & + & & & & & & + \\
			Pinshape & + & + & & & & & & \\
			Ponoko & & & & & + & & & \\
			REPABLES & + & + & & & & & & \\
			Rinkak & + & + & & & + & & & \\
			Sculpteo & & & & & + & & & \\
			shapeking & + & + & & & & & & \\
			Shapetizer & + & + & & & + & & & \\
			Shapeways & + & + & & & + & & + & \\
			Sketchfab & + & + & & & & & & + \\
			stlfinder & & & & & & & & \\
			STLHive & + & + & + & & & & & \\
			Stratsys Direct Express & & & & & + & p & & \\
			Threeding & + & + & & & + & & & \\
			Tinkercad & + & + & & & o & & & + \\
			Treatstock & + & + & + & + & + & & & \\
			trinckle & & & & & + & & & \\
			Trinpy & + & + & & & & & & \\
			TurboSquid & + & + & & & & & & \\
			UPS & & & & & + & & & \\
			Watertight & + & + & & & & & & \\
			Yeggi & & & & + & + & & & \\
			YouMagine & + & + & & & o & p & & \\
		\hline
	\end{tabular}}
	\label{tab:service_category}
\end{table}
As analysed in Tab.~\ref{tab:service_category}, the services surveyed
offer a different range of services each. No provider could be
identified that offers a complete set of service for 3D printing and
related tasks. In the table, the indication of \textbf{p} marks
companies that do not themselves offer printers through this service
but their parental companies do. The \textbf{o} character in the
column for printing service for Tinkercad and YouMagine, indicates that
the service itself does not render printing services, but
has a cooperation with a third party for the provisioning of
this service. With the exception of La Poste, UPS and iMakr all
the services render their business completely on the Internet
without the requirement for physical interaction. La Poste and
UPS offer an Internet interface with the physical delivery of the
objects in certain shops of theirs. Services that offer a design market place can
offer designs and other files costless or for a fee, no distinction is made
for this study.
Yeggi and stlfinder are search engines for 3D model data that work on the
data from other sources. Albeit a search engine, Yeggi provides the integration
of printing services and cloud printing services for models available
from third party services, thus Yeggi can be classified as a service
of services. The service rendered by Trinpy is subscription based with various
membership options. Grabcab provides 3D printing planning and control services,
and integration with an online editor.

\section{Review}
\label{sec:review}

Cloud Manufacturing is mainly an overlapping manufacturing or
engineering concept with application and grounding the development of
parts or objects in \enquote{traditional} manufacturing.
With traditional manufacturing we denote all technologies and
methods to create or fabricate objects or parts other than AM. For a
distinction between manufacturing methods see Klocke~\citep{Klocke15},
Nee~\citep{Nee15} and the DIN Standard 8580~\citep{DIN_8580:03}.
In this sense all subtractive or formative manufacturing methods
are summarises as \enquote{traditional manufacturing} methods.
As AM offers a large degree of flexibility due to short lead times as
well as other beneficial properties, we see that AM is the ideal
technology to be considered within CM scenarios. Taking the
properties of AM into account we do not predict that AM will
replace other manufacturing methods, not even
within CM scenarios. Rather AM will fill niches for special
applications like mass-customisation, rapid replacement production
capabilities or RT, especially within CM scenarios. With this
work we aim to contribute to the development of AM
methodology and technology in the CM paradigm.

\subsection{Topological Map}
\label{subsec:topological_map}
\begin{figure}
		\begin{center}
				\includegraphics[width=0.8\textwidth]{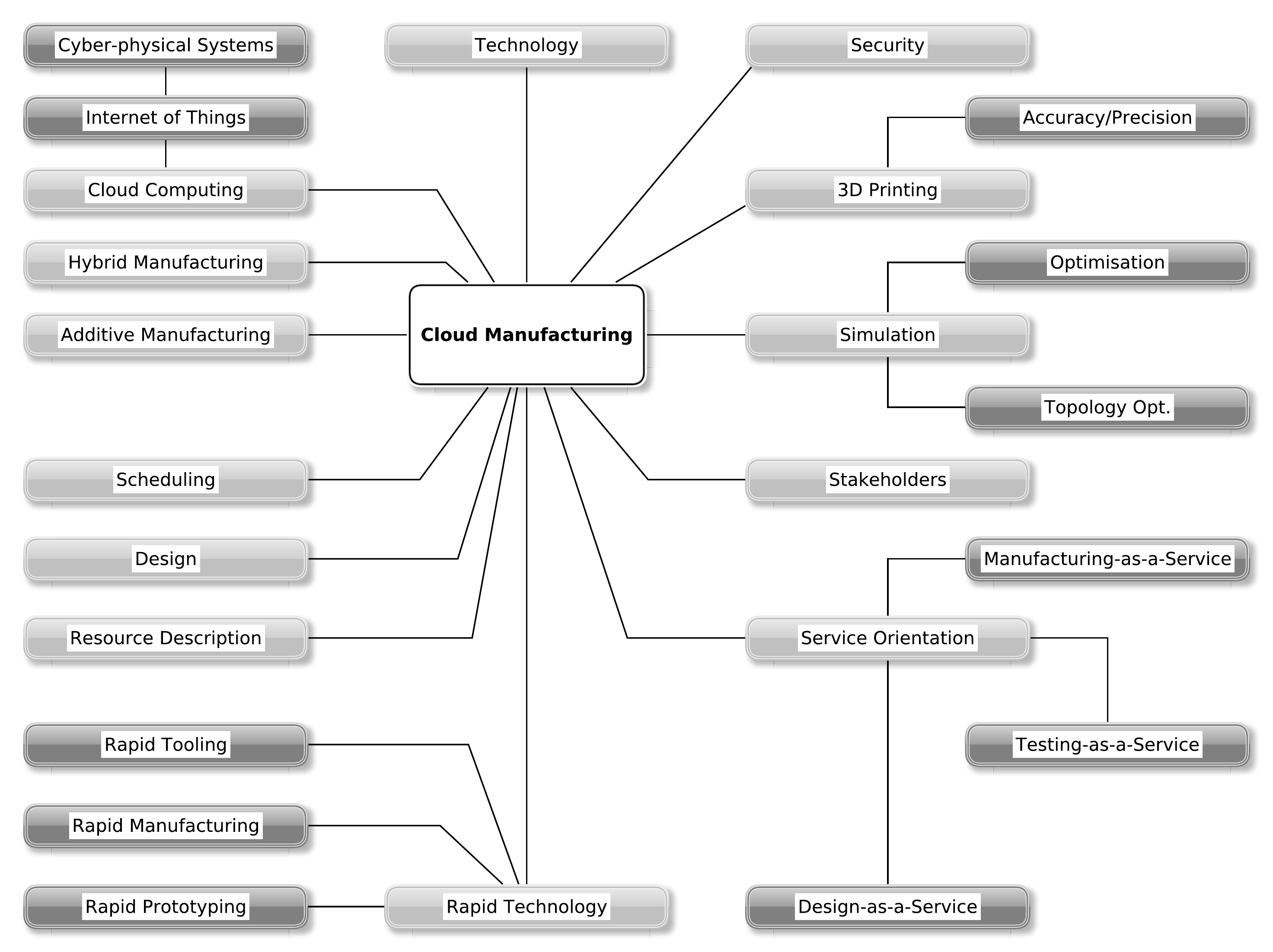}
				\caption{Topological Map for Cloud Manufacturing}
				\label{fig:topological_map}
		\end{center}
	\end{figure}

	In Fig.~\ref{fig:topological_map} the relationship and connection
	of various concepts relevant to	CM is described. This map forms
	the basis of the following review, where the nodes from the map
	represent sections from the review where we present the current
	state of research and elaborate on open research questions.
	The topics are extracted from literature.
	
	This topological map displays the relationship of CM with a
	variety of connected and enabling technologies and concepts.
	Additive Manufacturing (See Sect.~\ref{subsec:additive_manufacturing})
	enables CM to be more modular, flexible and
	offers new capabilities and business opportunities.
	The Rapid Technology (See Sect.~\ref{subsec:rapid_technology})
	and its composition Rapid Prototyping (RP, see
	Sect.~\ref{subsubsec:rapid_prototyping}),
	Rapid Manufacturing (RM, see
	Sect.~\ref{subsubsec:rapid_manufacturing}) and Rapid Tooling (RT,
	see Sect.~\ref{subsubsec:rapid_tooling}) are areas
	in which CM can be applied.
	The topic of Service Orientation
	(Sect.~\ref{subsec:service_orientation}) and its composition
	\enquote{as-a-Service} of which Design-as-a-Service (DaaS, 
	see Sect.~\ref{subsubsec:design_as_a_service}),
	Testing-as-a-Service (TaaS, see
	Sect.~\ref{subsubsec:testing_as_a_service}) and
	Manufacturing-as-a-Service (MaaS, see
	Sect.~\ref{subsubsec:manufacturing_as_a_service})
	are explored as examples, are concepts that enable the
	efficient application of CM. 
	For a broader understanding it is required to research the
	stakeholders involved in this technology which makes
	Sect.~\ref{subsec:stakeholder}.
	The topics of Scheduling (See Sect.~\ref{subsec:scheduling})
	and Resource Description (See
	Sect.~\ref{subsec:resource_description}) are to be discussed
	for the universal and efficient application
	of CM. The domain of Simulation (See Sect.~\ref{subsec:simulation})
	with its composition of Optimisation (See
	Sect.~\ref{subsubsec:optimization})
	and Topological Optimisation (See
	Sect.~\ref{subsubsec:topological_optimization}) enable a more rapid,
	more flexible and more robust usage of the technology. For AM 
	technology the application of Topology Optimisation enables
	the benefits of this technology. Similar to AM is 3D printing
	(See Sect.~\ref{subsec:3d_printing}) with its subtopic of
	Accuracy and Precision (See Sect.~\ref{subsubsec:accuracy_precision})
	as this technology is a appropriate basis for CM systems.
	The topic of Hybrid Manufacturing (See
	Sect.~\ref{subsec:hybrid_manufacturing}) gains importance in
	flexible and agile manufacturing systems which
	warrants and requires its research.
	In the topic of Technology (See Sect.~\ref{subsec:technology}) the 
	general principles and technologies of CM and AM are discussed
	as these are basic principles for the efficient implementation
	of these systems. The topic of Cloud Computing (CC, see
	Sect.~\ref{subsec:cloud_computing})
	with its sub-components Internet of Things (IoT, see
	Sect.~\ref{subsubsec:internet_of_things}) and Cyber-physical Systems
	(CPS, see Sect.~\ref{subsubsec:cyber-physical_systems}) is
	the conceptual progenitor of CM and therefore requires careful
	studying. IoT and CPS are key enabling technologies for CM.
	The topic of Security~\ref{subsec:security} is of increasing
	importance with the spreading application of AM and CM as
	attack surfaces grow and potential damage increases.
	
	\subsection{Technology}
\label{subsec:technology}
	A large number of technologies and technological advances have made
	AM possible to evolve from its origin as a RP method to
	its current state where it is used for end-part manufacturing (RM)
	and available to consumers~\citep{Gebhardt13, West14}.
	All 3D printed objects are based on a digital model.
	This model can either be created using CAD software, 3D sculpting software
	or acquired using reverse-engineering methods (\eg object scanning
	or photo reconstruction)~\citep{Gibson15}.

	Albeit direct slicing from a CAD model has been proposed
	by Jamieson~\citep{Jamieson95} in 1995 it is still rarely performed.
	Direct slicing requires implementation in the CAD software for
	each printer type and printer manufacturer, which is not feasible.
	Further shortcomings of the de-facto standard file format for
	AM, \ie STL, namely the possibility to contain mis-aligned facets,
	holes or to be non-watertight, as well as being to large in file
	size are reported by \citep{Wu06}.
	
	Besides a Steiner-patch based~\citep{Paul15} file format to replace
	the STL file format the ASTM Committee F42 has published an
	ISO Standard~\citep{ISO-ASTM_52915:16e} for the
	AMF (Additive Manufacturing File Format)
	with the same intention. Both file formats are created to increase
	the accuracy for the models described and therefore increase the
	quality of the resulting printed objects. STL seems to be the
	prevalent file format for AM with 25700 results on
	Google Scholar compared to 8230 results for AMF.
	Further investigation in the file support for different
	hard- and software vendors is warranted but out of the scope
	of this work.
	
	The review by Dimitrov et al.~\citep{Dimitrov06} presents further
	information on the technology that AM is based on with an overview
	of applications for it.
	
	In the review by Esmaeilian et al.~\citep{Esmaeilian16} the authors
	present the relationship of AM and Manufacturing in general
	as well its benefits.
	With the emergence of Internet or cloud based CAD Modelling software
	the creation of models for AM becomes easier as direct integration
	of 3D printing providers is possible.
	
	Furthermore, the collaborative aspect of 3D modelling is enhanced
	as studied by Jou and Wang~\citep{Jou13}. This study used
	a group of college students as a test group and investigated the
	adoption of an online CAD modelling (
	Autodesk AutoCAD~\footnote{\url{http://autodesk.com/products/autocad}})
	software in	the curriculum.
	
	The authors Andreadis et al.~\citep{Andreadis15} present a case
	study on the adoption of an unnamed cloud based CAD system in
	comparison to traditional software, as well as
	an exhaustive list of benefits of cloud based software.
	
	Wu et al.~\citep{Wu15c} present an economic analysis of cloud
	based services for design and manufacturing. This work also
	explores a number of cloud based services along with their pricing.

	Communities are of great importance to enterprises as shown in
	West and Kuk~\citep{West14}. One form of community is a repository
	for 3D printable digital models that collects and curates models
	supplied by users for collaboration, exchange, co-creation and sale.
	In this work the authors conduct a study to research the profit of
	catering for such a community/repository (Thingiverse) by a
	former open-source company (Makerbot).

	Wittbrodt et al.~\citep{Wittbrodt13} performed experiments to
	determine the ROI (Return on Investment) of 3D printers for
	common households and their	feasibility in application in
	end-user scenarios. With their experiment they concluded that
	an average household can achieve between 40 and 200 percent
	ROI on average usage of such machines.

\subsection{Security}
\label{subsec:security}
	Security for 3D Printing, AM or CM can be discussed from at least three
	perspectives. The first perspective would be the legal security of
	data and models processed within such a scenario.
	Discussions can range from whether it is legal to manufacture an
	existing object (replication) which might
	be protected by intellectual copyright laws to questions regarding
	product liability in case of company supplied model data.
	The second perspective is closely related to intellectual property (IP)
	as it is the technological discussion about the safeguarding of
	digital model files and data. The third perspective is about the data
	and process security itself in scenarios with malicious
	third-parties (\eg Hackers, Cyber-criminals).
	This third perspective is not limited to AM but shares many problems
	with CC and computing in general.

	Dolinksy~\citep{Dolinsky14} analyses the copyright and its application
	to 3D Printing for the jurisdiction of the USA. Because legal systems
	are different to each other such an analysis can not be exhaustive.

	Grimmelmann~\citep{Grimmelmann14} further exemplifies the legal status
	of 3D printing and model
	creation in the USA with fictitious characters from literature and theatre.
	He states that the creation of an object irregardless of the
	source of model for such a creation is infringing on copyright if the
	object that is replicated is protected by copyright.

	In \citep{Walther15} the author discusses the current situation of 3D
	printing in regard to gun laws. This discussion was
	started by the media in 2013 as models for a functional plastic gun
	were distributed and the gun manufactured. The author states
	that current gun control laws are adequate to control 3D printed
	weapons and that this is currently not a big issue.

	On a broader scope the authors McNulty et al.~\citep{McNulty12} research
	the implications of AM for the national security of the USA where the
	authors present the benefits
	of bio or tissue 3D printing for the treatment of battlefield wounds as
	well as the implications of AM technologies for criminal misconduct.

	For the analysis of data security the authors Wu et al.~\citep{Wu15}
	present the importance of such technologies within a CM environment.
	They propose the development of trust models for cyber-physical
	systems respectively the actors within such systems.

	The authors of Yampolskiy et al.~\citep{Yampolskiy14} provide a full
	risk analysis of a scenario for outsourcing AM under consideration of IP.
	The risk assessment does not include malicious behaviour other
	than IP infringement.
	
	To secure printed objects against counterfeiting the authors
	of \citep{Flank15} study and recommend the use of chemical
	components for authentication.
	Possible attacks on the 3D printing process by third parties is
	researched in~\citep{Zeltmann16} where one scenario is about the
	introduction of
	wilfully integrated material differences into an object in order to
	weaken the object under load. If the printing process itself is secured the
	question remains if a printed object is the original, a genuine replicate
	or a faked replicate. For the identification of genuine objects the
	authors of~\citep{Aliaga09}	research the applicability of physical
	signatures to 3D printed objects.
	
	In~\citep{Hou15} the authors present a watermarking technique for
	3D printed objects that is resilient against repeated digital
	scanning of the manufactured object.

	For a generalised discussion on security of cloud services and cloud
	computing we refer to~\citep{Subashini11} where the authors present
	issues ranging from data integrity to confidentiality. The concepts and
	terminology of CC security are also discussed in~\citep{Zissis12} of
	which the concept of confidentiality, trust and privacy are most relevant
	to scenarios of cloud based AM where users have physical objects created
	from digital models by third parties.

	Sturm et al.~\citep{Sturm14} present attack scenarios and
	mitigation strategies for attacks
	on AM system. The authors see rising CPS implementations in AM as
	potential intrusion vectors for attacks. The authors discuss various
	attacks for each of the manufacturing process phases. Furthermore, the
	authors identify the STL file format as a potential risk for tampering
	and attacking. Among the recommendations for mitigation is file hashing
	and improved process monitoring.
	
	Bridges et al.~\citep{Bridges15} briefly explore possible attacks 	
	on the cyber-physical systems that are used for AM. Among the
	attack scenarios the authors identify theft and tampering.


\subsection{3D Printing}
\label{subsec:3d_printing}
	Following the distinction between AM and 3D printing given in the
	definition of 3D printing (See 
	Sect.~\ref{subsubsec:definitions_of_3d_printing}) by some authors into
	high-quality professional or industrial usage and lower-quality
	end-user or semi-professional usage 3D printing could not be part of CM.
	As we relax the definition of AM and 3D printing and use the terms
	as synonyms, we survey technological developments within
	this chapter. Technological progress and development are essential to
	the widespread use and application of 3D Printing or AM in the CM paradigm.

	In the short article by Hansen et al.~\citep{Hansen14} the authors propose
	a measurement method for the correction or calibration of FDM printers. For
	this purpose the authors develop a measurement plate that is printed
	with specified parameters. In their experiment the authors recorded
	roundness errors of up to 100 $\mu$m. The calibration could not be
	applied due to the printer control software being closed-source.

	Anitha et al.~\citep{Anitha01} analyse the process variables layer
	thickness, bead width and deposition speed for their influence on the
	quality of objects manufactured using FDM. The
	authors find that the layer thickness is contributing with
	approximately 50~\% to the surface roughness of the manufactured objects.

	Balogun et al.~\citep{Balogun15} describe and an experiment on the energy
	consumption and carbon footprint of models printed using FDM technology.
	They define three specimens of 9000 $mm^3$ and 18000 $mm^3$ volume which are
	printed on a Stratasys dimension SST FDM.
	Their experiment also captures the energy consumption of the post processing
	with and ultrawave precision cleaning machine.
	The energy consumed for the print is approximately 1~kWh.
	Over 60~\% of the energy is consumed in non-productive
	states, \eg pre-heating.
	This energy consumption profile warrants high utilisation of 3D
	printers when aiming for a low ecological impact and penalises frequent
	and long idle times of the 3D printer.

	Brajlih et al.~\citep{Brajlih11} propose a comparison method for the speed
	and accuracy of 3D printers. As a basis the authors introduce
	properties and capabilities of 3D printers. A test-object designed
	by the authors is used to evaluate the average manufacturing speed of
	an Objet EDEN330 Polyjet and 3D Systems SLA3500 SLA manufacturing machine
	in an experiment.
	Furthermore, the experiment includes an EOS EOSINT P385 SLS and Stratasys
	Prodigy Plus FDM machine. The experiment concludes that the
	SLS machine is capable of the highest manufacturing speed
	(approx. 140~$cm^3/h$). In the experiment the angular and dimensional
	deviations are significant (up to 2.5\textdegree ~for a 90\textdegree
	~nominal, and 0.8 mm for a 10 mm nominal).
	
	Roberson et al.~\citep{Roberson13} develop a ranking model for the selection
	of 3D printers based on the accuracy, surface roughness and printing time.
	This decision making model is intended to enable consumers and buyers
	of such hardware to select the most appropriate device.

	Utela et al.~\citep{Utela08} provide a review on the literature related to
	the development of new materials for powder bed based 3D printing systems.
	They decompose the development into five steps, for which they
	provide information	on the relevant literature.

	Brooks et al.~\citep{Brooks14} perform a review on the history and
	business implications of 3D printing.
	They argue that the most promising approach for companies to benefit from
	3D printing	technology is to invest in and adapt current business
	models to support supplementary printing for the users.
	They also present the importance of the
	DMCA (Digital Millennium Copyright Act)
	in the USA under the aspect of 3D printing for current and upcoming
	businesses and services in the USA.

	Bogue~\citep{Bogue13} aims to provide an introduction into
	3D printing with this review.
	The historical development of the various printing technologies
	is presented and furthermore,
	applications with examples are explored.

	Petrick and Simpson~\citep{Petrick13} compare traditional manufacturing,
	which they classify as \enquote{economy of scale}, with AM.
	AM is classified by the authors as \enquote{economy of one}.
	They base their future hypotheses on the traditional design-build-deliver
	model and current patterns in supply chains from which they draw logical
	conclusion for future developments. These hypotheses are
	sparsely supported by literature.
	They predict that in the future the boundaries between the
	design-build-deliver paradigm will be less clear and that
	design and production will be closely coupled with experiments.
	One obvious prediction is that the supply chains will get
	shorter and the production will be more localised both
	geographically and in regard to time planning.
	
	Matias and Rao~\citep{Matias15} conduct an exploratory study
	on the business and consumer markets of 3D printing. This study
	consists of a survey based part for consumers within the
	area of 3D printing with a sample size of 66 participants conducted in 2014.
	One of their findings for the consumers is the willingness of
	45~\% of the participants to spend only \$US~299 on this technology.
	They also found out that a large number of consumers is not proficient
	with the technology and the required software.
	This finding was backed by five interviews conducted with business
	persons from five different companies.
	Their interviewees also expressed concerns that there will
	not be mass market for 3D printing within the next five to ten years.

\subsubsection{Accuracy/Precision}
\label{subsubsec:accuracy_precision}
	The accuracy, precision and geometrical fidelity of 3D printed objects has
	been researched in many works for over 20 years~\citep{Ippolito95,Fadel96}
	due to the necessity to produce objects that match their digital models
	closely. This topic is of general relevance for AM as only precise objects
	are usable for the various applications.
	Increased precision and accuracy enables AM and CM to be a valid
	manufacturing technology.
	
	Dimitrov et al.~\citep{Dimitrov06} conducted a study on the accuracy of
	the 3DP (3D-Printing) process with a benchmark model. Among the three
	influencing factors for the accuracy is the selected axis and the material
	involved.

	Turner and Gold~\citep{Turner15} provide a review on FDM with a
	discussion on the available process parameters and the resulting
	accuracy and resolution.

	Boschetto and Bottini~\citep{Boschetto14} develop a geometrical
	model for the prediction of the accuracy in the FDM process.
	They predict the accuracy based on process parameters for a
	case study for 92~\% of their specimens within 0.1 mm.

	Armillotta~\citep{Armillotta06} discusses the surface quality of
	FDM printed objects. The author utilises a non-contacting
	scanner with a resolution of 0.03 mm for the assessment of
	the surface quality. Furthermore,
	the work delivers a set of guidelines for the FDM process in
	respect to the achievable surface quality.

	Equbal et al.~\citep{Equbal11} present a Fuzzy classifier and
	neural-net implementations for the prediction of
	the accuracy within the FDM process under varying process
	parameters. They achieve a mean absolute relative error
	of 5.5~\% for the predictor based on Fuzzy logic.

	Sahu et al.~\citep{Sahu13} also predict the precision of FDM
	manufactured parts using a Fuzzy prediction, but with different
	input parameters (Signal to noise ratio of the width,
	length and height).

	Katatny et al.~\citep{Katatny10} present a study on the dimensional
	accuracy of FDM manufactured objects for the use as medical models.
	The authors captured the geometrical data with a 3D Laser scanner
	at a resolution of 0.2 mm in the vertical direction. In this
	work a standard deviation of 0.177 mm is calculated for a model
	of a mandible acquired from Computer Tomography (CT) data.

	To counter expected deviations of the object to the model,
	Tong et al.~\citep{Tong08} propose the adaption of slice files.
	For this adaption the authors present a mathematical error model
	for the FDM process	and compare the adaption of slice files to the
	adaption of STL (STereoLitography) files. Due to machine
	restrictions the corrections in either the slice file and the STL
	file are comparable, \ie control accuracy of the AM fabricator is not
	sufficient to distinguish between the two correction methods.

	Boschetto and Bottini~\citep{Boschetto16} discuss the implications of
	AM methods on the process of design. For this
	discussion they utilise digitally acquired images to compare to model files.

	Garg et al.~\citep{Garg16} present a study on the comparison of surface
	roughness of chemically treated and untreated specimens manufactured using
	FDM. They conclude that for minimal dimensional deviation from the
	model the objects should be manufactured either	parallel or perpendicular
	to the main axis of the part and the AM fabricator axis.

\subsection{Simulation}
\label{subsec:simulation}
	Simulation in the area of AM is of great importance even though the
	process of object manufacturing itself is relatively cheap and fast when
	compared to other means of production. But even 3D printed objects
	can take many hours to be manufactured in which the AM resource is occupied.
	Furthermore, with specialised and high value printing materials the
	costs can be prohibitively expensive for misprinted parts.

	In \citep{Hiller09} the authors describe a voxel based simulation for 3D
	printing for the estimation of the precision of AM objects.

	Pal et al.~\citep{Pal14} propose a finite-element based simulation for
	the heat-transfer of powder based AM methods. With this simulation the
	authors claim that the general quality of the printed objects can be
	enhanced and post-processing/quality-control can be reduced.

	The work of Zhou et al.~\citep{Zhou09} proposed a numerical simulation
	method for the packing of powder in AM. This research is conducted to
	provide a better understanding of the powder behaviour in methods like
	SLS or SLM.

	Alimardani~\citep{Alimardani07} propose another numerical simulation
	method for the prediction of heat distribution
	and stress in printed objects for Laser Solid Freeform Fabrication (LSFF),
	a powder based process similar to LENS. They compare their numerically
	computed predictions with experimental specimens, in one finding
	the maximum time-dependent stress could be reduced by eight percent	by
	improvements made by the simulation.
	
	Ding et al.~\citep{Ding11} discuss a FEM based simulation model for wire
	and arc based AM. Their simulation of the thermo-mechanical properties
	during this process is performed in	the 
	ABAQUS\footnote{\url{http://www.3ds.com/products-services/simulia/products/abaqus}}
	software.

	Chan~\citep{Chan03} presents graphical simulation models for the use in
	manufacturing scenarios not limited to AM. Such models must be adopted to
	contain virtual production entities like the ones provided by CM.
	
	Mourtzis et al.~\citep{Mourtzis14} present a review on the aspects of
	simulation within the domain of product development (PD).
	For this work they give introduction to concepts and technologies
	supporting and enabling PD.	The concepts are explained in sections of two
	paragraphs each and supported by existing literature.
	They link the concepts to simulation research within these
	areas, where applicable.
	The concepts and technology introduced includes Computer Aided Design (CAD),
	Computer Aided Manufacturing (CAM), Computer Aided Process Planning (CAPP),
	Augmented and Virtual Reality (AR and VR), Life Cycle Assessment (LCA),
	Product Data Management (PDM) and Knowledge Management (KM),
	Enterprise Resource Planning (ERP), Layout Planning, Process-,
	Supply Chain- and Material Flow Simulation,
	Supervisory Control and Data Acquisition (SCADA) and Manufacturing
	Systems and Networks Planning and Control.
	Their review is based on over 100 years of research in the area of
	simulation and 15954 scientific articles
	from 1960 to 2014. The articles are aligned with the product and
	production lifecycle.
	Furthermore, this review includes a comparison of commercially
	available simulation software.
	The authors conclude their work with a detailed analysis of research
	opportunities aligned to the concepts introduced within this work.


\subsubsection{Topological Optimisation}
\label{subsubsec:topological_optimization}
	One of the key benefits of AM is the ability to create almost any
	arbitrarily complex object which makes topological optimisation ideal
	for AM. In CM scenarios such optimisations can be embedded in the
	digital process chain and be offered and applied as services.
	In this section we present a number of research works on topological
	optimisation for AM.

	In~\citep{Chahine10} the authors discuss the application of topology
	optimisation for AM in a general manner giving an
	overview of the current state.

	Galantucci et al.~\citep{Galantucci08} present experimental results
	of topology optimised and FDM printed objects from compression tests.
	In their experiment	the reduction of filling material reduced the
	material consumption but also the maximum stress of the object.

	Almeida and Bártolo~\citep{Almeida10} propose a topology optimisation
	algorithm for the use in scaffold construction for bio-printing.
	This optimisation strategy is aimed	to create scaffolds that are
	more bio-compatible due to their porosity but
	yield high structural strength. The authors conducted an experiment for the
	comparison of the topological optimised structures and un-optimised
	structures with reduced infill. Their approach yields structurally
	stable scaffolds up to 60~\% porosity.

	The work by Leary et al.~\citep{Leary14} focuses on the topological
	optimisation to create objects without the requirement for additional
	support structures. For this approach the authors perform a general
	topological optimisation first, then orient the part optimally to reduce
	the required support material and apply a strategy
	to add part structures to remove the required support material.
	In an experiment conducted the authors create an object that requires
	significantly less material (89.7 $cm^3$ compared to 54.9 $cm^3$) and
	is manufactured in 2.6 hours compared to 5.7 hours for the optimised
	part with support structures.

	Tang et al.~\citep{Tang14} propose a design method for AM with the
	integration of topological optimisation. For this work the authors
	analyse existing design methods for AM.

	Bracket et al.~\citep{Brackett11} provide an overview and introduction of
	topology optimisation for AM. The authors identify constraints and
	restrictions for the usage of topology optimisation, \eg insufficient
	mesh-resolution or insufficient manufacturing capabilities. Among the
	identified opportunities of topology optimisation in AM is the ability
	create lattice structures and design for multi-material objects.

	Gardan~\citep{Gardan14} proposes a topology optimisation method
	for the use in RP and AM. The work is focused on the inner-part of
	the object. In non-optimised objects this is filled with a pre-defined
	infill pattern of a user-selectable density. The authors implement the
	method in a plugin for Rhinoceros 3D\footnote{\url{https://rhino3d.com}}
	 and provide an experiment with
	SLA and SLS. The article does not provide detailed information
	on the implementation of the software and the algorithm.

	Gardan and Schneider~\citep{Gardan15} expand on the prior work by Gardan
	by slightly expanding the previous article. In this work the authors
	apply the optimisation method to prosthetic hip implants and additionally
	experiment on a FDM 3D printer.

	Hiller and Lipson~\citep{Hiller09b} propose a genetic algorithm (GA) for
	the multi-material topological optimisation. With this approach
	the authors demonstrate the optimisation of varying degrees of stiffness
	within a part. They utilise a high-level description of the parts properties
	to design the desired object automatically in its optimised composition.

\subsubsection{Optimisation}
\label{subsubsec:optimization}
	For AM and CM as processes a number of optimisations is possible and
	necessary. The optimisations can relate to the optimisation of
	process parameters for quicker manufacturing, higher quality manufacturing
	or increased utilisation of hard- and software resources. The optimisation
	can furthermore, regard the embeddability and integration of AM and CM
	within existing	production processes.

	Optimisation for AM is a topic that is researched for a long time,
	as illustrated by the following two articles.
	Cheng et al.~\citep{Cheng95} propose an optimisation
	method for the criteria of manufacturing time, accuracy and stability.
	This optimisation is based on the calculation of the optimal part orientation.
	As a basis for the optimisation, the authors analyse the sources for
	errors in AM processes, \eg tessellation errors, distortion and shrinkage,
	overcuring or stair-stepping effects. For their model they weight
	input parameters according to the inflicting errors and perform a multi-objective
	optimisation.	
	
	Lin et al.~\citep{Lin01} propose a mathematical model to reduce the process
	error in AM. In the first part the authors analyse the different process
	errors for various types of AM, \eg under- and overfill, stair-stepping effects.
	Their model optimises the part orientation for minimal errors. Albeit this
	work is more than 15 years old, optimal orientation and placement of
	objects is not widely available in 3D printing control software.	
	
	More recently, Rayegani and Onwubolu~\citep{Rayegani14} present
	two methods for the process parameter
	prediction and optimisation for the FDM process. The authors provide an
	experiment to evaluate the tensile strength of specimens for the optimised
	process	parameters. For the optimised parameters the authors provide
	the solution of	0\textdegree ~part orientation, 50\textdegree ~raster angle,
	0.2034 mm raster width and -0.0025 mm air-gap. These parameters yield
	a mean tensile strength of 36.8603808 MPa.

	Paul and Anand~\citep{Paul12} develop a system for the calculation of the
	energy consumption in SLS processes and process parameter optimisation
	to minimise the laser energy. For the model the authors neglect the energy
	consumption of all elements (\eg heating bed) but the laser system.


	Paul and Anand~\citep{Paul15b} propose an optimisation for
	AM for the reduction of support material. Their approach is
	to optimise the part orientation for the minimum of support material.
	Furthermore, they provide optimisation for minimum cylindricity and
	flatness errors.

	Jin et al.~\citep{Jin14} propose a method to optimise the path generation
	of extrusion based AM, \eg FDM. The optimisation goals for this approach
	are machine utilisation and object precision. The authors perform a study
	with their approach but comparison data on the quality and time consumption
	of other algorithms is missing.
	
	Khajavi et al.~\citep{Khajavi14} present an optimisation for the spare-parts
	industry of fighter jets by the utilisation of AM. This work is on a systemic
	optimisation with AM being one strategy to achieve the optimum solution. 
	The authors analyse the current situation in this specific application
	and propose an optimised solution based on distributed manufacturing or AM.
	
	Ponche et al.~\citep{Ponche12} present an optimisation for the design
	for AM based on a decomposition of functional shapes and volumes. The authors
	argue that objects designed for traditional manufacturing are not necessarily
	suitable for AM but require partial or complete re-design to adjust
	for the specifics of a certain AM process, \eg in the inability to produce
	sharp corners and edges. In their work one optimisation goal is the reduction
	of material and therefore cost.
	
	Hsu and Lai~\citep{Hsu10} present the results of an experiment and
	the resulting process parameter optimisation for the 3DP manufacturing method.
	The authors improved the dimensional accuracy of each axis to under 0.1~mm.
	Furthermore, the authors improved on the building time by approximately 10~\%
	and on the flexural stress by approximately 30~\%. The authors experimented
	on the four process parameters that are layer height, object location, binder
	saturation and shrinkage.

\subsection{Stakeholder}
\label{subsec:stakeholder}
	In CM systems or cloud based printing systems naturally a number of
	stakeholders is involved. With this section we are presenting the current
	state in research on the identification of stakeholders in this domain as
	well as research regarding their agendas.

	In Rayna and Striukova~\citep{Rayna16} the authors identify the
	requirements of end-users for online 3D printing services. They base
	their study on concepts relevant to these stakeholders like
	user-participation and co-creation.

	Park et al.~\citep{Park16} provide a statistical analysis of patents
	and patent filings in the domain of 3D printing and bioprinting
	that can serve as a basis for decision making in the investment
	and R\&D in these fields. The stakeholders in this case are
	the investors and managers.

	H\"am\"al\"ainen and Ojala~\citep{Hamalainen15} applied the
	Stakeholder Theory by Freeman on the domain of AM and performed
	a study on eight companies with semi-structured interviews. They
	identified five companies that use AM for prototyping (RP). They further
	analysed the benefits of AM for the interviewed companies.

	Buehler et al.~\citep{Buehler16} created a software called GripFab for
	the use in special needs education. For this software and the
	use of 3D printing in special needs education they performed a
	stakeholder analysis. The analysis is based on observations and found
	beneficial uses for this technology, \eg in the form of assistive devices.

	Munguía et al.~\citep{Munguia08} analyse what influence missing standards
	have on the stakeholders of RM and develop a set of	best practises for RM
	scenarios. They identified the four main contributors to RM cost as
	Operation times, Machine costs, Labour costs and Material costs.

	Lehmhus et al.~\citep{Lehmhus15} analyse the usage of data acquisition
	technologies and sensors within AM from the perspectives
	of the identified stakeholders designer, producer, user and regulatory
	or public bodies. They argue that the producers
	in such scenarios might become obsolete if AM is utilised in a
	complete CM sense.

	Fox~\citep{Fox14} introduces the concept of virtual-social-physical (VSP)
	convergence for the application to product development. Within this
	concept he argues that AM can play an integral part to enhance
	product development. He identifies requirements from stakeholders in the
	product development process and addresses them in this work.

	Flatscher and Riel~\citep{Flatscher16} propose a method to integrate
	stakeholder in an integrated design-process for the scenarios of
	next-generation	manufacturing (Industry 4.0). In their study, a key
	challenge was the integration of all stakeholders in a team structure
	which they solved by integrating influential persons from different
	department in the joint operation.
	
	Maynard~\citep{Maynard15} discusses the risks and challenges, that come
	with the paradigm of Industry 4.0. Industry 4.0 as a concept is
	incorporating other concepts like AM and CM. The author briefly
	identifies the possible	stakeholders of this technology as consumers,
	CEOs and educators.

	In the report by Thomas~\citep{Thomas13} the author performs a very
	detailed and thorough analysis of stakeholders for AM technology in the USA.
	The list of identified stakeholders is 40 entries long and contains
	very specific entries (\eg Air Transport Providers or Natural
	Gas Suppliers) as well as generalised stakeholder groups
	(\eg Professional Societies or Consumers).



\subsection{Service Orientation}
\label{subsec:service_orientation}
	Service orientation denotes a paradigm from the domain of
	programming (Service Oriented Architecture, SOA).
	Within this paradigm the functionality or capability of a software is
	regarded and handled as a consumable service. The services offer
	encapsulated capabilities that can be consumed by users or other
	services in an easy to use, well-defined and transparent manner.
	The services are inter- and exchangeable within business processes as their
	inner-working is abstracted and the services act like black boxes
	with well-defined interfaces.
	With CM or AM in general, this service orientation can be expanded to the
	physical resources of manufacturing. Similar to service orientation from
	the programming	domain, it must be bound by the same stringency of
	well-defined interfaces and transparent or abstract execution of
	functionality or capability.

	In the review on service-oriented system engineering (SOSE)
	by Gu and Lago~\citep{Gu09}, the authors propose the hypotheses
	that the challenges in this domain can be classified by topic and
	by type. SOSE is the engineering discipline to develop service oriented
	systems. The authors identified 413 SOSE challenges from the reviewed
	set of 51 sources. The authors furthermore identified quality, service
	and data to be the three top challenges in this domain.
	
	Wang et al.~\citep{Wang10} provide a review on the CC paradigm.
	For this work the authors classify CC into the three layers (
	Hardware-as-a-Service - HaaS, Software-as-a-Service - SaaS
	and Data-as-a-Service - Daas). In this early work on CC the
	authors establish the importance of SoA for the CC paradigm.
	
	Tsai et al.~\citep{Tsai10} present an initial survey on
	CC architectures and propose a service architecture for
	CC (Service-Oriented Cloud Computing Architecture - SOCCA)
	for the interoperability between various cloud systems. Among
	the identified problems with CC architectures are
	tight coupling, lack of SLA (Service Level Agreement) support,
	lack of multi-tenancy support and lack of flexibility in user
	interfaces. The authors utilise SOA for the implementation
	of their prototype that is deployed on Google App
	Engine\footnote{\url{https://appengine.google.com}}.
	
	Alam et al.~\citep{Alam15} present a review on impact analysis
	in the domains of Business Process Management (BPM) and SOA. In
	their work the authors discuss the relationship and convergence of
	the two methods. From a set of 60 reviewed studies the authors
	conclude that BPM and SOA are becoming dominant technologies.
	
	
	Zhang et al.~\citep{Zhang10} propose a management architecture
	for resource service composition (RSC) in the domain of CM. For this
	work the authors analyse and define the flexibility of resources
	and their composition. The implementation by the authors supports
	resource selection based on QoS and flexibility.
	
	
	Shang et al.~\citep{Shang13} propose a social manufacturing system
	for the use in the apparel industry. Their implementation connects
	existing logistics and manufacturing systems with a strong focus
	on the consumer. For this architecture the authors rely heavily on
	SOA technology and describe the implementation of various layers
	required.
	
	Tao et al.~\citep{Tao15b} analyse the development of Advanced
	Manufacturing Systems (AMS) and its trend towards socialisation.
	In this work the authors establish the relationship between service
	orientation and manufacturing. The authors identify three phases
	for the implementation of service-oriented manufacturing (SOM), namely
	\enquote{Perception and Internet connection of Manufacturing resource
	and capability and gathering,
	Aggregation, management, and optimal allocation of Manufacturing
	resource and capability in the form of Manufacturing service and
	Use of Manufacturing service}.
		
	Thiesse et al.~\citep{Thiesse15} analyse economic implications
	of AM on MIS (Management Information Systems) and the service orientation of these systems.
	In this work the authors analyse the economic, ecological and
	technological potential of AM and its services.	The authors
	conclude that the services for the product development will
	be relocated upstream. 
	
	For the service composition of cloud applications standards and
	definitions are essential. In the work by Binz et al.~\citep{Binz14}
	the authors introduce the TOSCA (Topology and Orchestration Specification for Cloud Applications)
	standard. Albeit this standard is focused on the deployment and management
	of computing and other non-physical resources its architectural
	decisions and structures are of relevance to CM systems.
	Support for encapsulation and modelling as described in this work
	is sparse for other CM systems.
	
	As an extension to the previous work, the authors Soldani et al.~\citep{Soldani16}
	propose and implement a marketplace (TOSCAMART) for TOSCA for the distribution of
	cloud applications. Such a marketplace would be highly 
	beneficial to CM systems as it can foster innovation, collaboration, re-use
	and competition.
	
\subsubsection{Manufacturing-as-a-Service}
\label{subsubsec:manufacturing_as_a_service}
	Described in Sect.~\ref{subsec:service_orientation} the service orientation
	regards	capabilities as services that can be consumed. Such a class of
	services is the manufacturing of products. As a consumer of such a
	service, one is	not necessarily interested in the process of
	manufacturing (\eg what type of machine is used) or the location
	of manufacturing as long as a pre-agreed upon list of
	qualities of the end-product is complied with. As an example it can be
	said that a user wants two parts made from a certain metal, within
	a certain tolerance, certain properties regarding stress-resistance and
	within a defined time frame. The input of this service would then be
	the CAD model and the properties that must be fulfilled. The parts
	could then either be milled or 3D printed in any part of the world and then
	shipped to the user. The user must pay for the service rendered, \ie the
	manufacturing of objects, but is not involved with the manufacturing itself
	as this is performed by a service provider. In the seventh EU
	Framework Programme, the project
	ManuCloud\footnote{\url{http://www.manucloud-project.eu}}
	was funded that consolidated research on this topic.
	In this section we present current research articles
	on the subject in order to illustrate the concept of MaaS,
	its role for CM and applications.

	Tao et al.\citep{Tao13} propose an algorithm for a more efficient
	service composition optimal-selection (SCOS) in cloud manufacturing systems.
	Their proposed method is named FC-PACO-RM (full connection based parallel
	adaptive chaos optimisation with reflex migration)
	and it optimises the selection of manufacturing resources for the quality properties
	time, cost, energy, reliability, maintainability, trust and function similarity.
	In an experiment they proof that their implementation performs faster than 
	a genetic algorithm (GA), an adaptive chaos optimisation (ACO) algorithm for
	the objectives of time, energy and cost, but not for the objective of reliability.
	
	Veiga et al.~\citep{Veiga12} propose a design and implementation
	for the flexible reconfiguration of industrial robot cells with SMEs
	in mind. These robot cells are mostly reconfigurable by design
	but with high barriers for SMEs due to the requirement of experts.
	The system proposed enables an intuitive interface for reconfiguration
	of the cells in order to enhance the flexibility of manufacturing.
	The implementation draws heavily on SOA concepts. The implementation
	supports the flexible orchestration of robotic cells as services.
	
	Zhang et al.~\citep{Zhang14} provide an introduction into
	the paradigm of CM. Within this work the authors discuss issues
	arising from the implementation and the architecture itself.
	The authors present the decomposition of this paradigm into its
	service components, that are \enquote{design as a service (DaaS),
	manufacturing as a service (MFGaaS),
	experimentation as a service (EaaS), simulation as a service (SIMaaS),
	management as a service (MANaaS), maintain as a service (MAaaS),
	integration as a service (INTaaS)}. The authors implement
	such a CM system as a prototype for evaluation and discussion.
	
	Moghaddam et al.~\citep{Moghaddam15} present the development of
	MaaS and its relationship to the concepts of CM, Cloud Based Design
	and Manufacture (CBDM) and others. The authors propose SoftDiss~\citep{Silva14}
	as an implementation platform for CM systems.
	
	Van Moergestel et al.~\citep{VanMoergestel15} analyse the requirements for and propose
	an architecture for a manufacturing system that enables low-volume
	and low-cost manufacturing. The authors identify customer requirements
	for low-volume and flexible production of products as a driver
	for the development of the CM concept or other MaaS implementations. The
	architecture relies on cheap reconfigurable production machines (equiplet).
	For the implementation of the system the authors utilise open source
	software like Tomcat and have a strong focus on the end-user integration via
	Web technology.
	
	Sanderson et al.~\citep{Sanderson15} present a case study on distributed
	manufacturing systems which the authors call collective adaptive systems (CAS).
	The example in their case study is a manufacturing plant by Siemens in the
	UK which is part of the \enquote{Digital Factory} division.
	The authors present the division, structure and features of the company 
	which is compared to CAS features. Among the identified challenges
	the authors list, physical layouting, resource flow through supply chains
	and hierarchical distributed decision making.
	
	For the integration of MaaS (which is called
	Fabrication-as-a-Service, FaaS, in this work) 
	into CM, Ren et al.~\citep{Ren16} analyse the service provider
	cooperative relationship (CSPR). Such a cooperation of MaaS/FaaS
	providers within a CM system is essential for the task completion
	rate and the service utilisation as demonstrated by the authors in an experiment.
	
	Guo~\citep{Guo16} proposes a system design method for the implementation
	of CM systems. Within this work the MaaS layer of the CM system
	is further divided into \enquote{product design, process design,
	purchasing, material preparing, part processing and assembly and marketing
	process}. In the generalised five-layer architecture for the implementation
	of CM systems, the MaaS is located in the fifth layer.
	
	Yu and Xu~\citep{Yu15b} propose a cloud-based product configuration system (PCS)
	for the implementation within CM systems. Such systems interface
	with the customer enabling the customer to configure or create
	products for ordering. Within a CM such a system
	can be employed to directly prepare objects that can be manufactured
	directly utilising MaaS capabilities. In the implementation within an
	enterprise the authors utilise the STEP file format for information
	exchange.

\subsubsection{Design-as-a-Service}
\label{subsubsec:design_as_a_service}
	Besides physical and computational resources that are exposed
	and utilized as services the concept of CM allows for and requires
	traditional services to be integrated.
	Such a service is for example the design of an object, which is traditionally
	either acquired as a service from a third-party company or
	rendered in-house.
	
	As with the physical Manufacturing-as-a-Service the
	service rendered here must be well-defined and abstract. The service in
	this section is that of the design for AM or traditional manufacturing.
	
	This paradigm can lead to increased involvement of the user
	as described by Wu et al.~\citep{Wu13}. The authors
	provide an introduction to social product development - a new product
	development	paradigm rooted in Web 2.0 and social media technology.
	They conducted a study on their students in a graduate level course
	on product development.
	They structure the process in four phases beginning with acquisition of user
	requirement through social media.
	With the social product development process (PDP) the product development
	involves the users or customers more directly and more frequently
	than with traditional PDP. This increased degree of integration
	requires support through technology which is provided by social
	media and Web 2.0 technology for communication and management.
	
	Unfortunately the scientific literature on DaaS is sparse and
	mostly only mentioned as part of architectural or systematic descriptions
	or implementations of CM systems.
	In Tao et al.~\citep{Tao11} the authors place DaaS among other
	capabilities services that are part of the CM layer. Other 
	capabilities services are Manufacturing, Experimentation,
	Simulation, Management, Maintenance and Integration-as-a-Service.
	In Adamson et al.~\citep{Adamson13} the same classification is used
	(but without the Integration and Maintenance, and a combination
	of the Simulation and Experimentation service). The authors
	also briefly review literature in the domain of collaborative design
	for CM systems. 
	In Yu et al.~\citep{Yu15} DaaS is also identified as a capability of
	CM systems and part of its layered structure.

	Johanson et al.~\citep{Johanson09} discuss the requirements and implications
	of distributed collaborative engineering or design services. According to
	the authors the service orientation of design and its collaborative aspects
	will render enterprise more competitive due to reduced costs for software,
	decreased design times and innovative design. Furthermore, such services
	promote tighter integration and cooperation with customers.
		
	Laili et al.~\citep{Laili11} propose an algorithm for a more efficient
	scheduling of collaborative design tasks within CM systems. As collaborative
	design task scheduling is NP-hard, the authors propose a heuristic energy
	adaptive immune genetic algorithm (EAIGA). In an experiment the authors
	prove that their implementation is more stable with higher quality results
	than compared to an genetic algorithm (GA) and a immune GA (IGA).
	
	Duan et al.~\citep{Duan15} explore the servitization of capabilities and
	technologies in CC scenarios. The authors explore and discuss a variety
	of service offerings as described in literature. Among the identified
	as-a-Service offerings is Design-as-a-Service which is referenced to
	Tsai et al.~\citep{Tsai10}. Duan et al. provide a large collection
	of \enquote{aaS} literature. Contrary to the indication by Duan et al.
	the work of Tsai et al. does not cover DaaS. It however
	covers the architectural design of CC systems and service provisioning as
	well as an analysis of potential drawbacks and limitations of CC systems.

\subsubsection{Testing-as-a-Service}
\label{subsubsec:testing_as_a_service}
	Similar to the Design as a-Service (See
	Sect.~\ref{subsubsec:design_as_a_service}) this
	exposition of a capability as a service can play an
	important role within CM systems. In general the QA for
	AM is not sufficiently researched and conducted as the traceability
	of information from the original CAD model to the manufactured part
	is insufficient due to the number of conversion steps, file formats
	and systems involved.

	Albeit mentioned in a number of publications on the design
	and implementation of CM architectures, designs or systems, \eg
	Ren et al.~\citep{Ren15b} or Gao et al.~\citep{Gao15b}, 
	the research on Testing-as-a-Service (TaaS) in CM systems
	is sparse and the authors are not aware of any dedicated
	works on this topic.
	
	In contrast, TaaS as a concept for software testing in the
	cloud is researched by a number of authors, see \eg Gao et al.~\citep{Gao13}
	or Mishra and Tripathi~\citep{Mishra17}	for an introductory overview,
	Yan et al.~\citep{Yan12} for
	the special application of load testing, Tsai et al.~\citep{Tsai14} 
	for service design or Llamas et al.~\citep{Llamas14} for a
	software implementation.
	
	Extrapolating from the benefits that TaaS brings to software quality,
	\eg transparency, scalability, concurrency, cost-reduction
	and certification by third parties, research on this area
	in CM scenarios is warranted. In contrast to software QA, physical
	testing has an extended set of requirements and limitations, \eg 
	object under test must be physically available, higher likelihood that
	standardised test protocols exist, inability to scale without
	hardware investment or inability to scale beyond minimum time required
	for testing. With this section we want to motivate further
	discussion and research into this area.

\subsection{Rapid Technology}
\label{subsec:rapid_technology}
	As an umbrella term in accordance to the definition
	\enquote{General term to describe all process chains that manufacture
	parts using additive fabrication processes.} by \citep{VDI-3404:09} we examine
	the relevance of this technology for CM with this chapter. This technology is
	integral to the product development especially with its sub-technology that
	is RP (See Sect.~\ref{subsubsec:rapid_prototyping}). This and the following
	sections extend on the definitions provided in
	Sect.~\ref{sec:definition_and_terminology} by examples and research
	findings.
	
	For a brief introduction we refer to the following articles.	
	Li et al.~\citep{Li14} propose a method for rapid new product ramp-up
	within large multi-national companies relying on disperse
	supply chain networks and out-sourcing partners. In this work the
	authors consider large-volume product development. For the
	conceptual framework the authors identified critical members
	and defined a ramp-up process as a flowchart.
	
	Mavri~\citep{Mavri15} describes 3D printing itself as a rapid
	technology and analyses the impact of this technology on the production
	chain. The author performs an analysis on the influences
	on the phases of product design, production planning, product
	manufacturing as well as the topics material utilisation,
	inventory and retail market. The findings of the author
	include that AM enables companies to act more agile, 
	cater for smaller markets, limit potential inventory issues
	and can sustain smaller and slimmer supply chain networks.
	
	Muita et al.~\citep{Muita15} discuss the evolution of rapid production
	technologies and its implications for businesses. The authors
	investigate business models and processes, transitions as well
	as materials and logistics. A decomposition of rapid technology
	into the phases or layers (Rapid Prototyping, 3D Printing,
	Rapid Tooling, Rapid Product Development and Rapid Manufacturing) is provided
	and discussed. The authors recommend the adaption of AM by all companies.
	
	In the book by Bertsche and Bullinger~\citep{Bertsche07} the authors
	present the work of a research project on RP
	and the various problems addressed within the topic of Rapid
	Product Development (RPD).
	One aspect of this research is the development and integration
	of systems to efficiently store and retrieve information required throughout the
	process. Information required in the process is knowledge on construction,
	quality, manufacturing, cost and time.
	
	In Lachmayer et al.~\citep{Lachmayer16} the authors present
	current topics of AM and its application in the industry. In the
	chapter by Zghair~\citep{Zghair16} the concept of rapid repair
	is discussed. This concept is intended to prolong the life-time
	of high-investment parts as well as modification of parts in academic
	settings. The authors perform an experiment for this approach
	with three objects and conclude that there is no visible
	difference between additional object geometry in the case
	of previously SLM manufactured objects. Differences are visually
	detectable for cast objects that are repaired.
	
\subsubsection{Rapid Tooling}
\label{subsubsec:rapid_tooling}
	The use case of RT for AM is that the required tools or moulds
	for the (mass-) production of other parts or objects is
	supported by provisioning of said tools or moulds. See the definitions
	of RT in Sect.~\ref{subsubsec:definitions_of_rapid_tooling}.
	RT as a concept is researched and applied for at least 26 years~\citep{Sachs90}.
	Conceptually little has changed since the early publications
	but the number of available AM technologies, materials
	and support by other concepts like CC has increased. Since
	its start the idea of RT is to create tools or tooling
	directly from CAD models thus saving time and material. In this
	section we present articles from this research to give an overview
	to the reader and present its relevance and relationship to the concept 
	of CM.

	In the review by Boparai et al.~\citep{Boparai16} the authors
	thoroughly analyse the development of RT using FDM technology.
	FDM manufactured objects commonly require post-processing
	for higher-quality surfaces which is discussed by the authors in a
	separate section of their work. The authors present a variety
	of applications of RT with FDM which include casting and
	injection moulds and scaffolds for tissue engineering. Furthermore,
	the authors discuss material selection and manufacturing, as well
	as testing and inspection.
	
	The review by Levy et al.~\citep{Levy03} on RT and RM for
	layer oriented AM from 2003 already states that AM is not
	just for RP anymore. According to the definition of RT by the authors
	tools are supposed to last a few thousand to millions applications.
	The authors focus mainly on plastic injection moulds for tooling
	and survey a large number of different technologies and materials.
	
	Similarly, the definition of RT by King and Tansey~\citep{King02},
	is focused on injection moulds, a definition that has since
	been expanded to other tooling areas. In this work the authors
	present research on material selection for SLS manufactured moulds.
	In this work the authors analyse RapidSteel and copper polyamide
	for the use in time-critical RT scenarios.
	
	Lušić et al.~\citep{Lusic16} present a study on the applicability
	of FDM manufactured moulds for carbon fibre reinforced plastic (CFRP)
	objects. The authors achieved up to 84~\% material saving
	or 47~\% time saving for optimised structures compared to a solid
	mould at a comparable stiffness. The authors experimented with
	varying shell thicknesses and infill patterns.
	
	Nagel et al.~\citep{Nagel16} present the industrial application of
	RT in a company. The authors present at a high level the benefits
	and thoughts leading to the creation of flexible grippers for
	industrial robots utilising 3D printing. The authors also
	present a browser based design tool for the creation of the individual
	grippers with which the company is able to reduce the time
	required for product design by 97~\%.
	
	Chua et al.~\citep{Chua15} present a thorough introduction 
	to RT as well as a classification into soft- and hard tooling,
	with a further divide into direct soft tooling, indirect soft tooling,
	direct hard tooling and indirect hard tooling. Among the benefits
	of RT the authors see time and cost savings as well as profit
	enhancements. The authors discuss each of the classifications
	with examples and the relevant literature. Examples from
	industry given support the benefits proposed by the authors.
	
	Rajaguru et al.~\citep{Rajaguru15} propose a method for the creation
	of RT moulds for the production of low-volume plastic parts.
	With this indirect tooling method, the authors are able to produce
	low-cost moulds in less than 48 hours. The authors present an
	experiment where the mould is used for 600 repetitions. The method
	uses electroless plating of nickel and phosphorous alloy for the
	micro-pattern moulds.
	
	In the introduction to RT, Equbal et al.~\citep{Equbal15} start
	with the basics of various AM technologies. The authors provide
	a classification schema for RT and discuss each class with the
	appropriate examples. According to the authors, RT is a key
	technology for globally active companies in respect to
	flexibility and competitiveness.
	
	In the review by Malik et al.~\citep{Malik15} the authors
	investigate the use of 3D printing in the field of surgery.
	The authors discuss the fabrication of medical models for 
	education and operation planning as well as drill-guides and templates
	as RT technology. In contrast, the direct fabrication of implants or prosthetics
	as described by the authors is regarded RM.

\subsubsection{Rapid Manufacturing}
\label{subsubsec:rapid_manufacturing}
	In contrast to RP the goal of RM is the creation of parts and objects
	directly usable as end-products or part of end-products
	(See Sect.~\ref{subsubsec:definitions_of_rapid_manufacturing}).
	To achieve this usability the requirements on the quality of the parts
	is higher, therefore the quality control and quality assurance are stricter.
	
	Hopkinson and Dickens~\citep{Hopkinson03} provides findings on cost
	analysis for the manufacturing of parts for traditional manufacturing
	and AM. The authors identify the current and potential future benefits 
	for RM as the ability to manufacture with less lead time, increased
	geometric freedom, manufacture in distributed environments
	and potentially the use of graded material for production. The
	authors compared the costs incurred for the creation of two
	objects with injection moulding (IM), SLA, FDM and SLS. 
	For IM the tool costs are high (27360 and 32100 Euro) whereas
	the unit costs are low (0.23 and 0.21 Euro). In their calculation
	the equilibrium for the cost of IM and SLS for one of the objects
	is at about 14000 units and for the other part at around
	600 units. This finding validates RM for certain low-volume
	production scenarios.
	
	Ruffo et al.~\citep{Ruffo06} also present a cost estimation
	analysis and is an extension and update to the previous work.
	The authors calculated with a much lower utilisation of the
	machines (57~\% compared to 90~\%), higher labour cost as well
	as production and administrative overhead costs. Furthermore,
	the authors took other indirect costs like floor/building costs
	and software costs into consideration. The authors calculated a
	higher unit cost for the object (3.25 Euro compared to 2.20 Euro),
	and a non-linear costing function due to partial low-utilisation
	of the printing resources which is due to incomplete rows for unit counts
	not equal or multiple of maximum unit packing. The comparison of these
	two works illustrates the necessity to use the most up-to date
	and complete models for costing estimation.
	
	Ituarte et al.~\citep{Ituarte15} propose a methodology to
	characterise and asses AM technologies, here SLS, SLA and Polyjet.
	The methodology proposed is an experimental design for
	process parameter selection for object fabrication.
	The authors find that surface quality is the hardest
	quality to achieve with AM and might not suffice for RM
	usage with strict requirements. Such an analysis is of
	value in order to asses the feasibility of certain manufacturing
	methods in RM scenarios.

	In the review by Karunakaran et al.~\citep{Karunakaran12}, the authors
	survey and classify technologies capable of manufacturing metallic objects
	for RM. The technologies surveyed are CNC-machining, laminated manufacturing,
	powder bed processes, deposition processes, hybrid processes and rapid casting.
	The authors develop different classification schemes for RM processes
	based on various criteria, \eg material or application. Furthermore, the
	authors compile a list of RM process capabilities to be used for
	the selection of appropriate RM processes.
	
	Simhambhatla and Karunakaran~\citep{Simhambhatla15} survey build strategies
	for metallic objects for RM. The authors focus on the issues of overhangs
	and undercut structures in metallic AM. The work concludes with a comparative
	study on the fabrication of a part using CNC-machining and a hybrid layered
	manufacturing (HLM) method. With the hybrid approach the authors
	build the part in 177 minutes compared to 331 minutes at a cost of
	13.83 Euro compared to 24.32 Euro.
	
	Hasan et al.~\citep{Hasan13} present an analysis of the implications
	of RM on the supply chain from a business perspective. For this
	study the authors interviewed 10 business representatives and 6
	RP or RM service providers. The authors propose both reverse-auctioning
	as well as e-cataloguing as modes for business transactions.
	
	With rapid changing production the need arises for rapid fixture design
	and fabrication for the RM provider itself. This issue is discussed by
	Nelaturi et al.~\citep{Nelaturi14}, as they propose a mechanism
	to synthesise fixture designs. The method analyses the models to
	be manufactured and supported by fixtures as STL files for possible
	fixture application areas. The algorithm furthermore calculates
	possible fixture positions and inflicting forces. The authors
	select existing fixtures from in-house or online catalogues
	of fixtures for application.
	
	Gupta et al.~\citep{Gupta14} propose an adaptive method to slice model
	files of heterogeneous objects for the use with RM. For this the
	authors decompose the slicing process into three phases (Slicing set up,
	Slices generation and Retrieving data). The work also surveys
	other existing slicing techniques for various optimisation goals, \eg
	quality, computing resources or part manufacturing time. For the
	extraction of geometric and material information the authors utilise
	a relational database for efficient storage. The authors find that
	utilising the appropriate slicing technique the fabrication time can be
	reduced by up to 37~\%.
	
	Hern{\'{a}}ndez et al.~\citep{Hernandez13} present the KTRM (Knowledge
	Transfer of Rapid Manufacturing) initiative which is created to
	improve training and knowledge transfer regarding RM in the European Union.
	For the requirement analysis of such a project, the authors conducted
	a study with 136 participants of which the majority (70~\%) are
	SMEs. Such training initiatives are beneficial to the growth
	in application and increased process majority as the authors find
	that the knowledge of RM is low but the perceived benefits of this
	technology include higher quality parts, lower time to markets
	and increased competitiveness.
	
	With the chapter by Paul et al.~\citep{Paul13}, the authors provide
	a thorough overview over laser-based RM. The authors discuss
	classifications of such systems as well as composition of these
	systems in general. Process parameters are presented and located
	in literature. Furthermore, the authors discuss materials available
	for this class of RM and applications. This work is a comprehensive 
	overview, covering all relevant aspects of the technology, including monitoring
	and process control.
	

\subsubsection{Rapid Prototyping}
\label{subsubsec:rapid_prototyping}
	Following the definitions in
	Sect.~\ref{subsubsec:definitions_of_rapid_prototyping}
	Rapid Prototyping (RP) is the concept to speed-up the creation of
	prototypes in product development.
	These prototypes can be functional, visual, geared towards user-experience
	or of any other sort. RP was one of the first uses for AM and oftentimes
	the terms AM and RP are used synonymously. The quick or rapid creation of
	prototypes does not necessarily mean fast in absolute terms but rather a
	more rapid way to create prototypes than traditionally created using
	skilled or expert labour (\eg wooden models created by carpenters)
	or subtractive or formative manufacturing methods oftentimes requiring
	specialised tooling or moulds.

	Pham and Gault~\citep{Pham98} provide an overview of commonly used methods to
	rapidly create prototypes with information on the achievable accuracy,
	speed and incurred costs of each technology from a very early perspective.
	A number of technologies, \eg Beam Interference Solidification (BIS),
	has since been disused. The accuracy for Fused Deposition Modeling (FDM)
	stated with 127 $\mu$m has not been improved significantly since then.

	Masood~\citep{Masood14} reviews the technology of FDM and examines the
	usability of it for RP. Among the advantages of this technology
	is \enquote{Simplicity, Safety, and Ease of Use} as well as
	\enquote{Variety of Engineering Polymers} which makes it suitable for the
	creation of functional prototypes. A number of limitations,
	like \enquote{Surface Finish and Accuracy}, can diminish the suitability
	of this technology for certain aspects of prototyping.
	
	In their keynote paper Kruth et al.~\cite{Kruth98}
	survey the technologies used for RP and produce examples for the
	technologies. Furthermore, they briefly explain
	the developmental bridge from RP to RT and RM.

	The authors Yan et al.~\citep{Yan09} present
	the historical development of RP from its roots in the analogue and manual
	creation of prototypes to digital fabrication methods. The also present a
	list of current limitations for digital RP. Among the five limitations
	they place high-manufacturing cost, for the manufacturing
	resources, and the insufficient forming precision. The first argument
	of cost is often put forward in its reversed statement as RP is proposed
	as a low-cost production method, when compared to traditional prototyping.

	Azari and Nikzad~\citep{Azari09} present a review on RP in dentistry with
	a distinction of models in dentistry and its general meaning.
	They discuss the problems in data-capture for RP due to the nature
	of living patients. They further discuss the use of AM for drill-guides
	which is an application for RT.

	Liu et al.~\citep{Liu15} present a study on profit mechanisms associated
	with cloud 3D printing platforms predominantly in China. They argue that
	such services can enable small and medium sized enterprises (SME) to
	produce prototypes more rapidly and cheaper thus
	increasing their competitiveness.

	Roller et al.~\citep{Roller97} introduce the concept of Active Semantic
	Networks (ASN) as shared database systems for the storage of information
	for the product development process.

\subsection{Design}
\label{subsec:design}
	In traditional (subtractive or formative) manufacturing the design is driven
	by the capabilities provided by the manufacturing equipment. This
	is described as Design for Manufacturing or Design for Manufacturability (DFM)
	which means that the parts designed must be easy and cheap to manufacture.
	Especially in large volume production the parts must be machinable in
	a simple way as tooling, tool changes and complex operations are
	expensive. Furthermore, with traditional manufacturing certain operations
	like hollowed or meshed structures are not possible to produce or incur
	large costs. With AM the design of objects or parts is not strictly limited
	by these considerations as flexibility comes for free and a number
	of operations (\eg intersecting parts, hollowed structures) become possible.
	The designer can chose more freely from available designs and is less
	restricted. The design itself can concentrate on the functionality
	of the part, rather than its manufacturability.

	In the review by Rosen~\citep{Rosen14} the author proposes principles
	that are relevant for design for AM (DFAM) as they exist in literature.
	The suitability of AM is declared for parts of high complexity and
	low production volume, high production volume and high degree of
	customisation, parts with complex or custom geometries or parts
	with specialised properties or characteristics.
	Within this review the author proposes a prototypical design process
	for AM that is derived from a European Union standardisation project
	by the name of SASAM\footnote{\url{http://www.sasam.eu}}.

	Kerbrat et al.~\citep{Kerbrat11} propose a multi-material
	hybrid design method for the combination of traditional
	manufacturing and AM. For this method the object is decomposed
	based on the machining difficulty. The authors implemented
	their method in a CAD System (Dassault Systems
	SolidWorks\footnote{\url{http://www.solidworks.com}})
	for evaluation. This hybrid design method is not limited to
	a specific AM or manufacturing technology. The authors omit
	information on how the decomposed part or parts are fused together
	and how to compensate for inaccuracies within the manufacturing
	process.
	
	Throughout the design process and later for manufacturing
	it is necessary to convey and transport information on 
	design decisions and other specifications. Brecher et al.~\citep{Brecher09}
	provide an analysis of the STEP~\citep{ISO10303} and STEP-NC~\citep{ISO-10303-238:07}
	file formats. This analysis is used to propose extensions necessary
	for the use in an interconnected CAD-CAM-NC (CAx) platform.
	
	Buckner and Love~\citep{Buckner12} provide a brief presentation
	of their work on the automatic object creation using Multi-Objective
	Constrained Evolutionary Optimisation (MOCEO) on a high-performance
	computing (HPC) system. With their software,
	utilising Matlab\footnote{\url{http://mathworks.com/products/matlab}}
	and driving SolidWorks, the objects are created automatically
	following a given set of restrictions and rules.

	Cai et al.~\citep{Cai15} propose a design method for
	the personalisation of products in AM. Their work defines basic
	concepts ranging from Design Intent to Consistency Models. The
	design method is intended to convey design intentions from users
	or designers in a collaborative design or CAD environment. 
	
	Vayre et al.~\citep{Vayre12} propose a design method for AM with a focus
	on the constraints and capabilities. This design method consists
	of four steps (Initial Shape Design, Parameter
	Definition, Parametric Optimisation, Shape Validation). For the initial shape design,
	the authors propose the use of topological optimisation. The
	authors illustrate this process with an example of the re-design
	of an aluminium alloy square bracket.
	
	Diegel et al.~\citep{Diegel10} discuss the value of AM for a sustainable
	product design. The authors explore the benefits (\eg Mass customisation,
	freedom of design) and design considerations or restrictions for AM (\eg
	surface finish, strength and flexibility). The authors
	argue that AM offers create potential for the creation of long-lasting,
	high-quality objects and parts that can save resources throughout
	their lifetime by optimised design.

	Ding et al.~\citep{Ding16} analyse existing slicing strategies
	for the creation of objects with AM. Besides the analysis, the authors
	propose a strategy to create multi-directional slicing paths to be
	used with AM machines that support multi-axis deposition or
	fabrication. By the authors' analysis the existing software for
	slice creation is insufficient and leaves uncovered areas (hole
	or gap). This work is not on the design for AM but rather on the
	design of the resulting machine-paths for the manufacture with
	AM fabricators. 
	
	
	In Wu et al.~\citep{Wu14c} the authors discuss the concept
	of Cloud-Based Design and Manufacturing (CBDM, see also~\citep{Wu12}) in which the whole
	design and manufacturing process chain is executed in a cloud environment.
	CBDM is an extension to the CM concept as it expands the process chain
	horizontally into the collaborative and cooperative domain of the cloud.
	CBDM utilises Web 2.0 technology, service-oriented architecture (SOA) concepts,
	semantic web technologies and has an inherent connection to social networking
	applications. In this article concepts like collaboration, cooperation
	and crowdsourcing for design for AM are discussed and exemplified.

\subsection{Additive Manufacturing}
\label{subsec:additive_manufacturing}
	We see AM an integral component in CM and Industry 4.0 settings due
	to the benefits it provides. Among those benefits are flexibility,
	resource efficiency and the freedom in and of design. In this section
	we survey scientific literature regarding AM, especially works
	that provide an overview (\eg reviews, surveys), present important aspects
	or exhibit common characteristics of this domain.

	Le Bourhis et al.~\citep{LeBourhis14} develop the concept of design for
	sustainable AM (DFSAM) to minimise the yet unknown environmental impact
	of AM. According to the authors about 41~\% of the total energy
	consumption globally is attributed to industry. The authors
	further provide a division for the French industry in 2010 where
	about 12~\% percent are attributed to manufacturing. The authors
	claim that AM can reduce the energy required as it limits waste material.
	The authors experiment on the energy and resource consumption of the
	Additive Laser Manufacturing (ALM) process and present a method to calculate
	electricity, powder and gas consumption for an object based on the respective
	GCode.
	
	In their work Kim et al.~\citep{Kim14} present a federated information
	systems architecture for AM. This architecture is intended to facilitate
	an end-to-end digital implementation of AM, \ie \enquote{digital thread},
	design-to-product process. The authors analyse, for each phase
	(Part geometry/design, Raw/tessellated data, Tessellated 3D model,
	Build file, Machine data, Fabricated Part, Finished Part, Validated Part) the
	available and used data formats and supporting software.
	The focus of their conceptual architecture is interoperability by an open
	architecture.
	
	Balogun et al.~\citep{Balogun14} perform an experiment on the electricity consumption
	of the FDM process. The authors divide the manufacturing process into
	its components (Start-up, warm-up, ready-state, build-state).
	In an experiment they analyse three different FDM machines (Stratasys
	Dimension SST FDM, Dentford Inspire D290 and PP3DP) for their power consumption
	profile during manufacturing. The machines differ significantly in the
	energy demand with the Dentford machine requiring 1418 Wh and the PP3DP
	only requiring 66 Wh.
	Furthermore, the authors compare the energy consumption and manufacturing
	duration of a FDM machine to a milling machine. In the experiment
	the AM process consumed 685 Wh and the Mikron HSM 400 milling machine only 114 Wh.
	The AM cycle time was 3012 s (without 3600 s for support structure removal
	in an ultrasonic cleaning tank) and the milling machine cycle time was
	137 s.
	
	Weller et al.~\citep{Weller15} discuss the implications of AM on the company
	and industry level. Economic characteristics, \ie opportunities like acceleration
	and possible price premiums, lower market entry barriers and limitations like
	missing economy of scale, missing quality standards are discussed in this analysis.
	The authors perform modelling of various scenarios and propositions for
	the market under the influence of AM. Their prediction for first adoption is
	within markets with an overall lower economy of scale.
	
	Efthymiou et al.~\citep{Efthymiou14} present a systematic survey on the complexity
	in manufacturing systems. Albeit not directly referencing AM this
	study is relevant to understand the implications of AM on manufacturing systems.
	
	
	Turner et al.~\citep{Turner14} survey melt extrusion AM processes. This work
	is part of a two piece series (See also~\citep{Turner15}) with this part focusing
	on the design and process modelling. The authors provide a short market
	analysis in their introduction. The authors discuss literature relating to
	various processing steps  and problems, \eg die swelling, with melt extrusion processes.
	The authors provide a thorough overview on the literature for this topic.
	
	Mitev~\citep{Mitev15} approaches the topic of AM in a very uncommon manner,
	namely with a philosophical approach. This is the sole publication with this
	approach found by the authors. The author discusses AM for the question
	on what matter is and how 3D printing affects our concept of matter and material.
	
	In contrast to the previous author, Bayley et al.~\citep{Bayley14} present
	a model for the understanding of error generation in FDM. This work consists
	of two parts with experiments. The first part analyses actual errors in FDM
	manufactured parts (\eg Roundness error, geometrical deviation). In the
	second part the authors construct a framework for error characterisation and
	quantification.

	In the review by Kai et al. \citep{Kai16} the authors evaluate the
	relationship of manufacturing systems and AM briefly.
	The authors also provide an overview over one possible decomposition
	of AM and its academic relevance through numbers of published works from 1975 to 2015.

\subsection{Cloud Computing}
\label{subsec:cloud_computing}
	Cloud Computing (CC) is the concept of virtualized computing resources available
	to consumers and professionals as consumable services without physical restraints.
	Computing, storage and other related tasks are performed in a ubiquitous cloud
	which delivers all these capabilities through easy to use and interface
	front-ends or APIs.
	These concepts enable enterprises to acquire computing capacities
	as required while, often paying only for the resources consumed
	(Pay-as-you-go) in contrast to payment
	for equipment and resources in stock (\eg leasing, renting or acquisition).
	Concepts developed for this computing paradigm are of importance for
	the CM domain, as many problems stated or solved are interchangeable
	within domains. What CC is to computing resources (\eg storage,
	computing, analysis, databases) CM is to physical manufacturing
	resources (\eg Tools, 3D printer, drills).

	In the definition of Cloud Computing, Mell and Grance~\citep{NIST11} from NIST
	develop and present the characteristics and services models for CC.
	
	Truong and Dustdar~\citep{Truong10} present a service for
	estimating and monitoring costs for cloud resources from the domain
	of scientific computing. This model is also suitable for
	the monitoring of costs in other cloud based computing scenarios
	as CM with adaptions. The authors present an experiment where
	they analyse the cost of scientific workflows on with on-premise
	execution and deployment to the Amazon Web Service (AWS) cloud
	system.
	
	Stanik et al.~\citep{Stanik12} propose the cloud layer that is
	Hardware-as-a-Service for the remote integration of distinct
	hardware resource into the cloud. The authors argue from the point
	of embedded systems development and testing but the concepts described
	are universally applicable for any hardware that is intended to
	be exposed as a service.
	
	Mehrsai et al.~\citep{Mehrsai13} propose a cloud based framework
	for the integration of supply networks (SN) for manufacturing. The authors
	discuss the basics of supply networks and CC in order to
	develop a concept to integrate CC for the improvement of SNs. This
	modular approach is demonstrated in an experimental simulation.
	
	Oliveira et al.~\citep{Oliveira14} research the factors influencing
	the adoption of CC in general and for the manufacturing sector.
	The authors test their hypothesis on a survey of 369 companies
	from Portugal with 37.94~\% of the companies from the domain of manufacturing.
	The authors find that security concerns do not inhibit the
	adoption of CC in the manufacturing domain sub-sample of their 
	survey group.
	
	Ramisetty et al.~\citep{Ramisetty15} propose an ontology based
	architecture for the integration of CC or cloud platforms
	in advanced manufacturing. The authors claim that adoption of CC
	in manufacturing is less than in comparable industries due to
	the lack of social or collaborative engagement. The authors implement
	three services (Ontology, Resource Brokering and Accounting) for
	an evaluation in the WheelSim App. The authors propose
	an \enquote{App Marketplace} for manufacturing services to further
	the adoption of CC in the manufacturing industry.

	Um et al.~\citep{Um14} analyse the benefit of CC on the supply
	chain interactions in the aerospace industry. The authors
	propose a manufacturing network for contracting and subcontracting
	based on CC. In this architecture the basis for information exchange
	is the STEP-NC file format.
	
	Valilai and Houshmand~\citep{Valilai13} propose a service oriented
	distributed manufacturing system on the basis of CC. For their work
	the authors analyse the requirements and basics of globally distributed
	manufacturing systems. The proposed system (XMLAYMOD) utilises
	the STEP file format for the information exchange and enables a collaborative
	and distributed product development process as well as process
	planning and execution.

\subsubsection{Internet of Things}
\label{subsubsec:internet_of_things}
	Internet of Things (IoT) is a term used to describe a network consisting of
	physical objects connected to the Internet. These physical objects can
	be tools, parts, machines, actuators or sensors. The concept of IoT
	is integral to the CM paradigm as it is necessary to control
	the AM resources transparently and monitor the resources for efficient
	utilisation planning and scheduling. In IoT scenarios the use
	of open-standards helps to avoid vendor lock-in.

	Tao et al.~\citep{Tao14} present a very high-level description of the
	possible integration of IoT in CM scenarios. In four proposed
	layers (IoT, Service, Application, Bottom Support) of CM systems, they
	declare IoT and the corresponding layer as core enabling technology.

	Tao et al.~\citep{Tao14b} propose a five layer (Resource layer,
	perception layer, network layer, service layer and application layer)
	architecture for a CM system. The authors propose the utilisation of IoT
	technology as a method to interface the manufacturing resources into
	the architecture. This work is very similar to~\citep{Tao14}.

	Qu et al.~\citep{Qu16} present a case study on the integration of CM
	and IoT technology into an enterprise to synchronise the
	production logistics (PL) processes. For the implementation they propose
	a five tier (Physical resource layer, Smart object layer, Cloud
	manufacturing resource management layer, Cloud manufacturing core
	service layer, Cloud manufacturing application layer) decomposition.
	The system uses AUTOM~\citep{Zhang11b} as a backbone for the IoT integration.
	
	Baumann et al.~\citep{Baumann16d} propose the development of flexible
	sensor boards for the use in the monitoring of AM processes. The authors
	analyse existing sensors available and provide an architectural overview
	over a system for the incorporation of these sensor boards into
	a manufacturing control and monitoring system. With these sensors
	AM resources can be bridged to control systems or services thus
	enabling IoT functionality for the resources.
	
	Caputo et al.~\citep{Caputo16} perform a review on IoT from a managerial
	perspective with the application of AM. The authors develop a four
	staged (Radical, Modular, Architectural and Incremental)
	conceptual framework to classify innovation and research
	on the topic. Within this framework's description AM resource will
	become digitally represented by sensors and IoT technology.
	
	In the review by Kang et al.~\citep{Kang16}, the authors focus on
	global research on smart manufacturing and its enabling technologies and
	underlying concepts. In the section on IoT the authors link
	this concept to other technologies like SoA, CM and smart sensors.	
	
	Vukovic~\citep{Vukovic15} discusses the importance of APIs for the 
	IoT deployment and usage. The author discusses the common architectural
	patterns in IoT scenarios and the arising requirement for APIs
	to further this technology.
	
	In the work by Kubler et al.~\citep{Kubler16}, the authors discuss the
	evolution of manufacturing paradigms and the origins of CM along
	its relationship with IoT technology. The authors conclude that CM
	is not widely adopted because of security concerns but research
	in AM and IoT will drive CM forward.

\subsubsection{Cyber-physical Systems}
\label{subsubsec:cyber-physical_systems}
	Cyber-physical Systems (CPS) are one of the key enabling technologies
	for the Internet of Things.
	CPS is a term coined by the NSF (National Science Foundation) to describe
	systems, \eg machines, parts or tools
	that have capabilities to sense and interact with their physical
	environment while being connected to the Internet in order to relay
	state and environment information to an Internet based control system.
	The first occurrence in scientific literature can be found
	in Lee~\citep{Lee06}. In the domain of 3D Printing, AM and CM such
	systems are required to enable seamless integration of systems.
	With CPS, it is possible for an AM hardware resource to signal
	its current status or utilisation to a centralised or cloud-based
	control infrastructure in order to participate in scheduling endeavours
	and become part of a controllable system.

	In the work by Chen and Tsai~\citep{Chen16} the authors propose the concept of
	ubiquitous manufacturing. This concept is similar to CM but with a stronger
	focus on mobility of users and manufacturing resources. For this concept
	ubiquitous sensor and IoT technology are key enabling technologies.

	Lee et al.~\citep{Lee15} propose a five layer architecture for CM based on IoT
	technology. The layers in this architecture are from bottom to top:
	Smart Connection Layer, Data-to-Information Conversion Layer,
	Cyber Layer, Cognition Layer and Configuration Layer.
	The goal of this work is to provide a guideline for implementation of
	such a CPS backed manufacturing systems and to improve the product
	quality as well as the system's reliability.

	Sha et al.~\citep{Sha09} provide a general introduction into CPS and the related
	research challenges. The authors identify QoS composition, knowledge
	engineering, robustness and real-time system abstraction as the
	four main research questions for this technology.

	In the survey by Khaitan and McCalley~\citep{Khaitan14} the authors study the
	design, development and application of CPS. In the list of identified application
	scenarios (Surveillance, Networking Systems, Electric Power Grid and Energy Systems, 
	Data Centres, Social Networks and Gaming, Power and Thermal Management, Smart Homes,
	Medical and Health Care, Vehicular and Transportation Systems) manufacturing systems
	are missing. Despite this lack of mention, application in \eg transportation 
	and power management is relevant for CM systems.

	Kuehnle~\citep{Kuehnle15} proposes a theory of decomposition for manufacturing resources
	in distributed manufacturing (DM) systems. DM is similar in concept to CM in
	relation to the decomposition of manufacturing resources in vitalised services.
	According to the author IoT technology and CPS are among the enabling technologies
	for this smart manufacturing concept.

	Yao and Lin~\citep{Yao16} expand the concept of CPS into
	socio-cyber-physical-systems (SCPS) 
	with this study on smart
	manufacturing. This extension to social aspects of manufacturing
	(\eg collaboration and cooperation) is expected to be an integral
	part of the next industrial revolution (Industry 4.0).

	Turner et al.~\citep{Turner15b} discuss the risk of attacks and
	their implications on CPS in the domain of manufacturing. The
	authors present a number of attack vectors, \eg attack on
	the QA process and counter or mitigation strategies. According
	to the authors CPS provide an additional attack surface for
	malicious third parties.

\subsection{Scheduling}
\label{subsec:scheduling}
	In CM as in CC a number of resources must be provisioned on demand. In contrast to CC
	the requirements for the execution resource can be more complex than just a computing
	resource. With CM manufacturing resources must first be described in an abstract way
	(See Sect.~\ref{subsec:resource_description}) to be schedulable. In this section we
	present current research on the challenges that come from scheduling.

	Cheng et al.~\citep{Cheng13} introduce the concept of CM in their work
	and perform a brief review over possible criteria for scheduling in such
	scenarios. The authors provide four scheduling modes based on the three identified
	stakeholders (Operator, Consumer and Provider) and the system as a whole. The proposed
	modes consider energy consumption, cost and risk. The proposed system-centred
	cooperative scheduling mode yields the highest utilisation in their experiment.

	Liu et al.~\citep{Liu16} propose a scheduling method for CM systems for multiple enterprise and
	services scenarios. The authors use the criteria time, cost and pass-rate for the task
	selection. Based on these criteria constraints are constructed for the decomposition
	of tasks into subtasks and their distribution onto resources. The authors take
	geographical distance, respectively delivery times between CM locations into consideration.
	Their simulation concludes that for a 50 task scenario, with 10 enterprises offering 10 services in total,
	the utilisation is 49.88~\% compared to 10 tasks (17.07~\%). The authors provide
	no specific scheduling solution with their work.

	Laili et al.~\citep{Laili12} define the problem of optimal allocation of resources based
	on a 3-tier model (Manufacturing Task Level, Virtual Resource Layer and Computing Resource Layer).
	The authors prove that the optimum resource allocation is NP-complete.
	For the reason of NP-completeness of the scheduling problem this and other
	authors propose heuristics based algorithms to provide near-optimal
	scheduling. Heuristics based scheduling algorithms provide near-optimum
	solutions for most of the scheduling instances without the guarantee to
	achieve an optimum solution but at greater speed than exact computation.
	In this work the authors propose a heuristic algorithm inspired by the
	immune system (Immune Algorithm, IA). In an experiment they compare
	their algorithm against three other heuristic algorithms and it
	performed comparable.

	Wang~\citep{Wang13b} proposes a web-based distributed process
	planning (Web-DPP) system that performs process planning, machining
	job dispatching and job execution monitoring. The system is implemented
	as a prototype and connects to legacy machine controllers. The
	proposed system acts directly on the manufacturing resource and
	interfaces with the Wise-ShopFloor framework~\citep{Wang02}.
	The author does not provide information on scheduling
	algorithms or methods used.

	Huang et al.~\citep{Huang16} propose a scheduling algorithm based on
	Ant Colony Optimisation (ACO). In an experiment they compare the
	algorithm with and without a serial schedule generation scheme (SSGS)
	against another heuristic Genetic Algorithm (GA). Their algorithm for
	conflict resolution performs faster and with better quality results than
	the GA when used with the SSGS.

	Lartigau et al.~\citep{Lartigau12} present an 11-step framework
	for scheduling and order decomposition within a CM system. This scheduling
	is deadline oriented and implemented in
	a company environment for
	evaluation. The paper lacks validation
	and conclusive results for the proposed algorithm.

	In the work by Zhang et al.~\citep{Zhang15} the authors propose
	a queue optimisation algorithm for the criteria lowest cost,
	fastest finished time, cleanest environment and highest quality.
	The proposed CM system relies on active and real-time environment
	and machine-status sensing through heterogeneous sensors. Furthermore,
	they utilise semantic web (Ontology) technology for the system.

	Cao et al.~\citep{Cao16} refine an ACO algorithm for efficient
	scheduling within a CM. This algorithm optimises for time, quality,
	service or cost (TQSC). With the addition
	of a selection mechanism to ACO their ACOS algorithm performs with better
	quality results and faster convergence in comparison to
	Particle Swarm Optimisation (PSO), GA and Chaos Optimisation (CO).

	Jian and Wang~\citep{Jian14} propose an adapted PSO
	algorithm (Improved Cooperative Particle Swarm Optimisation, ICPSO)
	for the use in batch processing within a CM system.
	Batch tasks are indivisible units of work to be executed with
	manufacturing resources. The authors present an experiment for the
	comparison of the proposed algorithm with an PSO and a
	cooperative PSO scheduling algorithm in respect to the cost and
	time criteria. The algorithm performs better than the other two
	algorithms.

\subsection{Resource Description}
\label{subsec:resource_description}
	For the usage of manufacturing resources within CM there must be an abstract
	definition or description of the resources. Open-standards are preferable
	where available in order to avoid vendor lock in.

	Luo et al.~\citep{Luo13} propose a six step framework for the description
	of Manufacturing Capabilities (MC). The representation of this information
	utilises ontology and Fuzzy technology. Within the framework the
	authors represent information on the manufacturing equipment,
	computing resources, intellectual resources, software and other resources.

	Wang et al.~\citep{Wang14} also propose an ontology based representation
	for manufacturing resources. The information and ontology is derived
	from manufacturing task descriptions. The authors implement their algorithm
	in an enterprise setting in a medium-sized Chinese company for evaluation.

	As a more general approach to CC scheduling Li et al.~\citep{Li13} propose
	an ontology based scheduling algorithm with PSO. The authors motivate
	their work by an example in a logistics centre which is relevant to the
	domain of CM. For this algorithm the selection is restricted based
	on the Quality of Service (QoS) with time, cost, availability and
	reliability as criteria.

	Zhu et al.~\citep{Zhu13b} develop an XML based description for
	manufacturing resources oriented at the Web Service Description Language
	(WSDL) for web-services. The authors separate the resource description
	into two parts (Cloud End, CE and Cloud Manufacturing Platform, CMP).
	In their approach, they reflect static data, \eg physical structure or
	input data types, in the CE layer whereas
	the CMP layer reflects the dynamic data, \eg function parameters.

	Wu et al.~\citep{Wu16d} propose an ontology based capability description
	for industrial robot (IR) systems. IR are regarded as manufacturing resources
	and described as such. Besides manufacturing machines such IR systems 
	enable CM to perform as a flexible and agile manufacturing system.

\subsection{Hybrid Manufacturing}
\label{subsec:hybrid_manufacturing}
	Hybrid Manufacturing is a term used for the combination of AM and traditional manufacturing
	methods. The combination of these methods promises to provide
	benefits from both, \eg speed and accuracy of a milling machine with
	the low material input from AM.

	Lu et al.~\citep{Lu14} propose an architecture for a hybrid manufacturing
	cloud. Their definition of hybrid refers to cloud operation modes (private, community
	and public cloud). Besides the architecture they present a cloud management engine (CME)
	which is implemented for evaluation purposes on Amazon Web Service (AWS).

	In the work by Kenne et al.~\citep{Kenne12} a model for a hybrid
	manufacturing-remanufacturing system is proposed. The authors refer the term hybrid
	to manufacturing and remanufacturing in combination. Remanufacturing denotes
	an alternative use of products at the end of their product lifecycle for value
	generation. In an experiment the authors calculate the cost for a mixture of
	parameters, \eg return rates and come to the conclusion that the system
	is applicable with customisation to various industries.

	In the review by Chu et al.~\citep{Chu14} the authors discuss 57 hybrid manufacturing
	processes. These micro- and nanoscale processes are categorised in three different
	schemes (concurrent, main/assistive separate and main/main separate).
	The authors survey a combination of 118 processes in this work.

	The review by Zhu et al.~\citep{Zhu13} provides a classification of hybrid manufacturing
	processes. The authors present an extensive list of mainly two-process combination
	manufacturing processes. For this work the authors explore the existing definitions
	of manufacturing and hybrid manufacturing processes in literature.

	In another work by Zhu et al.~\citep{Zhu13c} the authors propose a build
	time estimation for the iAtractive~\citep{Zhu12} process that combines
	additive, subtractive and inspection
	processes. This process is based on FDM and the build time prediction
	is based on the same parameters as normal FDM build time prediction. The
	authors provide a discussion on an experiment for which their estimation
	ranged from approximately -12~\% -- 12~\% to the real build time. The authors
	only provide a build estimation method for the additively manufactured part of
	the process.

	Lauwers et al.~\citep{Lauwers14} propose a definition and classification of hybrid
	manufacturing processes with their work. They define these processes as
	acting simultaneously on the same work area or processing zone. This
	definition excludes processes that combine processing steps sequentially.

	Elmoselhy~\citep{Elmoselhy13} proposes a hybrid lean-agile manufacturing system (HLAMS).
	The author develops the system for the requirements in the automotive industry.
	The definition of hybridity in this work refers to the school of thinking for
	manufacturing.

	The work by Kendrick et al.~\citep{Kendrick15} proposes a solution
	to the problems associated with distributed manufacturing through
	the utilisation of hybrid manufacturing processes.
	The authors propose four options for the usage of distributed
	hybrid manufacturing systems (Local factories, manufacturing shops,
	community areas, personal fabrication). The described usage of
	hybrid MS can be further utilised in CM systems.


	Yang et al.~\citep{Yang16} propose a hybrid system for the integration of
	multiple manufacturing clouds. The definition of hybridity used in this
	work refers to the mixture of diverse manufacturing clouds and not on the
	manufacturing process itself. The architecture proposed links the various
	clouds together for a single point of interaction integration. The authors
	define adaptors and a broker system and implement these for evaluation purposes.

	In the overview by Zhang et al.~\citep{Zhang14b} the authors use the term hybrid
	to describe the cloud management. Their definition for the three cloud types used
	in CM is private/enterprise cloud, public/industry cloud and a mixture of both as hybrid
	cloud.

\subsection{Research Implications}
\label{subsec:research_implications}
	From the provided literature we have identified the following number of open
	research questions. The listing compiled is non-exhaustive due to the nature
	of scientific research. 

	Bourell et al.~\citep{Bourell09} provide a report on the \enquote{Roadmap for
	Additive Manufacturing} workshop that took place in 2009 and resulted
	in proposal for research of the coming 10 to 12 years. The recommendations
	are grouped into 
	\begin{inparaenum}
		\item Design
		\item Process Modelling and Control
		\item Materials, Processes and Machines
		\item Biomedical Applications
		\item Energy and Sustainability Applications
		\item Education
		\item Development and Community and
		\item National Testbed Center.
	\end{inparaenum}
	The recommendations include the proposal to create design methods
	for aiding designers with AM, creation of closed-loop printing systems and
	the design and implementation of open-architecture controllers for 
	AM fabricators.

	The authors reflect on their proposed roadmap in an
	article~\citep{Bourell14} five years later. In this analysis the authors
	state that the direct influence
	of the Roadmap is hard to quantify. The authors remark that the report is
	referenced about 50 times in scientific literature but only one project
	can be clearly attributed to the Roadmap.

	Lan~\citep{Lan09} identifies the following four tasks for future
	research in his review.
	\begin{inparaenum}
		\item Combination of Web services and software agents
		\item Collaborative network environment with the focus on integration and interoperability
		\item Focus on Web technology integration in RM systems and
		\item Collaborative product commerce and collaborative planning and control.
	\end{inparaenum}

	In the review by Fogliatto et al.~\citep{Fogliatto12} on Mass Customisation (MC), the
	authors identify the following research needs:
	\begin{inparaenum}
		\item Research on Rapid Manufacturing (RM) to support MC
		\item Research on the value of MC for consumers as well as environmental, economic and ethic value
		\item Research on Quality Control
		\item Research on Warranty for MC objects and
		\item Case Studies and empirical validation.
	\end{inparaenum}

	Khan and Turowski~\citep{Khan16} perform a survey on challenges in manufacturing
	for the evolution to Industry 4.0. The authors identify six current and future
	challenges which are the following topics 
	\begin{inparaenum}
		\item Data integration (IoT, Big-Data, real-time data, data management)
		\item Process flexibility (Adaption, Change management)
		\item Security (Connectivity, monitoring, compliance)
		\item Process integration within and across enterprise boundaries (Integrated processes, logistics, optimisation) 
		\item Real-time information access on hand-held devices (Web technology, ERP integration) and
		\item Predictive Maintenance (Machine data, sensors).
	\end{inparaenum}

	Among the research needs identified by Adamson et al.~\citep{Adamson15}
	in their review are the following
	\begin{inparaenum}
		\item Capabilities, information and knowledge integration and sharing as well as cloud architectures
		\item Definitions and standards for CM
		\item Intelligent, flexible and agile, distributed monitoring and control systems
		\item Business models
		\item Intellectual properties and
		\item Cost, security and adoption of CM systems.
	\end{inparaenum}
	Furthermore, the authors identify and predict 
	\begin{inparaenum}
		\item The emergence of cloud service providers
		\item Real world connectivity (IoT)
		\item New collaboration and cooperation scenarios (Customer-manufacturer and manufacturing collaboration)
		\item Increased competitiveness
		\item Cloud closed-loop manufacturing
		\item Manufacturing of feature function blocks
		\item Increased awareness and research on sustainable operations.
	\end{inparaenum}

	In the work by Oropallo and Piegl~\citep{Oropallo16} the authors
	specifically researched and compiled ten challenges in current
	AM systems that require research. The challenges are
	\begin{inparaenum}
		\item Shape optimisation (Cellular structures and topology optimisation)
		\item Design for 3D printing (Software support, design methodology)
		\item Pre- and Postprocessing (File formats, model preprocessing, part postprocessing)
		\item Printing methodologies (Layered manufacturing, voxel and digital material, non-layer oriented methods)
		\item Error control (Before and during printing)
		\item Multi material printing (Modelling and manufacturing support)
		\item Hardware and Maintenance issues (Process and material based issues)
		\item Part orientation
		\item Slicing (Adaptive and direct slicing)
		\item Speed
	\end{inparaenum}

	Wu et al.~\citep{Wu14c} explicitly identify the following research needs
	for the evolution of CM 
	\begin{inparaenum}
		\item Cloud-Based Manufacturing (Modeling and simulation of material flow,
			concurrency and synchronisation for scalability)
		\item Cloud-Based Design (Social media integration and leveraging,
			CAx convergence and cloud enablement)
		\item Information, Communication, and Cyber Security (IoT,
			Security-as-a-Service) and
		\item Business Model.
	\end{inparaenum}

	The work by Huang et al.~\citep{Huang15} examines the state of the art of AM
	and names the following research areas for future investigation:
	\begin{inparaenum}
		\item Materials
		\item Design (Methods and tools, complex geometries, lifecycle cost analysis)
		\item Modeling, Sensing, Control, and Process Innovation (Multi-scale modelling
			simulation, error and failure detection, optical geometry reconstruction,
			faster hardware, bioprinting)
		\item Characterization and Certification and
		\item System Integration and Cyber Implementation (Knowledge management
			integration, cloud based systems).
	\end{inparaenum}

\section{Summary}
\label{sec:summary}
	This article provides an overview over the topic that is
	CM and 3D Printing services.
	
	With the overview of the existing definitions
	(See Sect.~\ref{sec:definition_and_terminology}) and the extension
	of the definition proposed we create the foundation for the following
	work.
	
	The review is based on the topological map presented
	in Sect.~\ref{subsec:topological_map}. Concepts,
	techniques, methods and terminology is presented by exploring
	different authors work.
	We perform an explorative extension study
	to~\citep{Rayna16} due to relevance for this domain
	(See Sect.~\ref{sec:3d_printing_services}). In this study
	we cover and analyse 48 publicly available services.
	extension considers APIs of such services a further distinction
	to be made. This work also gives an overview on available
	journals in the domain of AM or 3D printing in general as to support
	other researchers' in finding suitable audiences for their work.
	One journal was established 31 years ago and provides a catalogue of
	over 21000 articles with no exclusivity to AM or 3D Printing.
	In recent years a number of new journals were established or
	are currently in the process of being established. Their focus
	is solely on AM or related domains like bioprinting.

	The domain of AM, CM, 3D Printing, RM and related domains
	is thoroughly presented in this work
	by means of literature analysis in scientometrical sense (See
	Sect.~\ref{subsubsec:sources} and
	Sect.~\ref{subsec:development_in_scientific_publications}).
	
	The results presented in this work illustrate the
	scientific development of various techniques and methods
	from these domains in a time period ranging from 2002 to 2016
	(See Sect.~\ref{sec:review}).

\section*{Disclosure}
\label{sec:disclosure}
This work is not funded by any third party. The authors do not have any conflicting interests
in the topic and provide an unbiased and neutral article on the topic contained within this
article.

\bibliographystyle{plainurl}
\bibliography{am-article}

\end{document}